\pdfoutput=1
\newcommand*{\ATLASLATEXPATH}{}

\newcommand{\AtlasCoordFootnote}{
ATLAS uses a right-handed coordinate system with its origin at the nominal interaction point (IP)
in the centre of the detector and the $z$-axis along the beam pipe.
The $x$-axis points from the IP to the centre of the LHC ring,
and the $y$-axis points upwards.
Cylindrical coordinates $(r,\phi)$ are used in the transverse plane,
$\phi$ being the azimuthal angle around the $z$-axis.
The pseudorapidity is defined in terms of the polar angle $\theta$ as $\eta = -\ln \tan(\theta/2)$.
Angular distance is measured in units of $\Delta R \equiv \sqrt{(\Delta\eta)^{2} + (\Delta\phi)^{2}}$.}
 
\newcommand{\Cline}[1]{
\noalign{\global\setlength\arrayrulewidth{1pt}}\cline{#1}
\noalign{\global\setlength\arrayrulewidth{0.4pt}}
}

\documentclass[cernpreprint, atlasdraft=true, texlive=2016, UKenglish, british]{\ATLASLATEXPATH atlasdoc}
 
\usepackage{\ATLASLATEXPATH atlaspackage}
\usepackage{\ATLASLATEXPATH atlasbiblatex}
 
\usepackage{\ATLASLATEXPATH atlasphysics}
\usepackage{amsfonts}
\usepackage{pdflscape}
\usepackage{longtable}
\usepackage{transparent}
 
\graphicspath{{logos/}{figures/}}
 
\usepackage{ANA-SUSY-2018-12-PAPER-defs}
\usepackage{multirow}

\AtlasTitle{Search for a scalar partner of the top quark in the all-hadronic $t\bar{t}$ plus missing transverse momentum final state at $\sqrt{s}=13$~TeV with the ATLAS detector}
 
\AtlasAbstract{
A search for direct pair production of scalar partners of the top quark (top squarks or scalar third-generation up-type leptoquarks) in the all-hadronic $t\bar{t}$ plus missing transverse momentum final state is presented. The analysis of 139~fb$^{-1}$ of ${\sqrt{s}=13}$~TeV proton--proton collision data collected using the ATLAS detector at the LHC yields no significant excess over the Standard Model background expectation. To interpret the results, a supersymmetric model is used where the top squark decays via $\tilde{t} \to t^{(*)} \tilde{\chi}^0_1$, with $t^{(*)}$ denoting an on-shell (off-shell) top quark and $\tilde{\chi}^0_1$ the lightest neutralino. Three specific event selections are optimised for the following scenarios. In the scenario where $m_{\tilde{t}}> m_t+m_{\tilde{\chi}^0_1}$, top squark masses are excluded in the range 400--1250~GeV for $\tilde{\chi}^0_1$ masses below $200$~GeV at 95\% confidence level. In the situation where $m_{\tilde{t}}\sim m_t+m_{\tilde{\chi}^0_1}$, top squark masses in the range 300--630~GeV are excluded, while in the case where $m_{\tilde{t}}< m_W+m_b+m_{\tilde{\chi}^0_1}$ (with $m_{\tilde{t}}-m_{\tilde{\chi}^0_1}\ge 5$~GeV), considered for the first time in an ATLAS all-hadronic search, top squark masses in the range 300--660~GeV are excluded. Limits are also set for scalar third-generation up-type leptoquarks, excluding leptoquarks with masses below $1240$~GeV when considering only leptoquark decays into a top quark and a neutrino.
}
 
\author{The ATLAS Collaboration}
 
\AtlasRefCode{SUSY-2018-12}
 
\AtlasJournal{EPJC}
\AtlasJournalRef{Eur. Phys. J. C 80 (2020) 737}
\AtlasDOI{10.1140/epjc/s10052-020-8102-8}

\PreprintIdNumber{CERN-EP-2020-044}
\hypersetup{pdftitle={ATLAS document},pdfauthor={The ATLAS Collaboration}}
\begin{document}
 
\maketitle

% The next lines are included from the .//tex/Introduction.tex input file
\section{Introduction}
 
Supersymmetry (SUSY)~\cite{Golfand:1971iw,Volkov:1973ix,Wess:1974tw,Wess:1974jb,Ferrara:1974pu,Salam:1974ig} is an extension of the Standard Model (SM) that can resolve the gauge hierarchy problem~\cite{Dimopoulos:1981zb,Sakai:1981gr,Dimopoulos:1981yj,Ibanez:1981yh} by introducing supersymmetric partners of the SM bosons and fermions. The SUSY partner to the top quark, the top squark,
plays an important role in cancelling out potentially large top-quark loop corrections to the Higgs boson mass~\cite{Inoue:1982pi,Ellis:1983ed}. Naturalness arguments suggest that the superpartners of the third-generation quarks
may be $\cal{O}(\TeV)$, and thus experimentally accessible at the Large Hadron Collider (LHC)~\cite{Barbieri:1987fn,deCarlos:1993yy}. The superpartners of the left- and right-handed top quarks, $\stopL$ and $\stopR$, mix to form two mass eigenstates, $\stopone$ and $\stoptwo$, where $\stopone$ is the lighter one. Throughout this paper, it is assumed that $\stoptwo$ has sufficiently high mass such that the analysis is sensitive to $\stopone$ only, which is labelled $\tilde{t}$ in the following.
 
$R$-parity-conserving SUSY models~\cite{Farrar:1978xj} may also provide a dark-matter candidate through the lightest supersymmetric particle (LSP), which is stable~\cite{Goldberg:1983nd,Ellis:1983ew}. In these models, the supersymmetric partners are produced in pairs. At the LHC, top squarks are produced mostly via gluon--gluon
fusion as well as quark--antiquark
annihilation. In a simplified scenario where the first- and second-generation squarks and gluinos are decoupled, the cross section of direct top squark pair production is largely decoupled from the specific choice of SUSY model parameters except for the top squark mass. This production cross section falls steeply with increasing top squark mass, ranging from $10.0\pm6.7$~pb for $\mstop=300$~GeV\ to $0.89\pm0.13$~fb for $\mstop=1300$~GeV~\cite{Beenakker:1997ut,Beenakker:2010nq,Beenakker:2011fu,Borschensky:2014cia}.
 
\begin{figure}[htbp!]
\begin{center}
\subfloat[]{\includegraphics[width=0.32\textwidth]{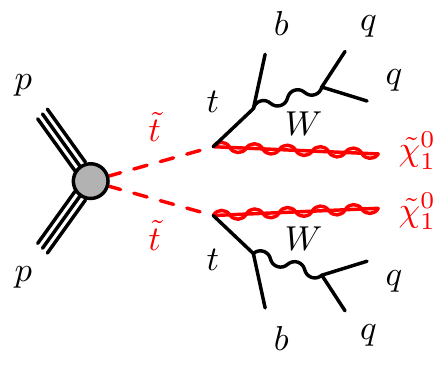}}
\subfloat[]{\includegraphics[width=0.32\textwidth]{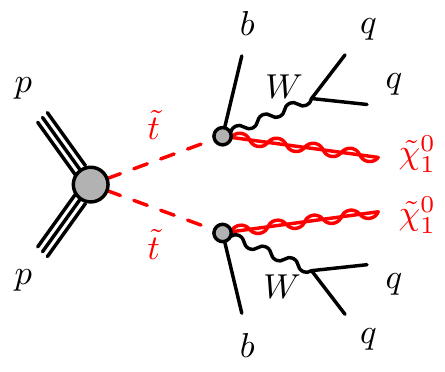}}
\subfloat[]{\includegraphics[width=0.32\textwidth]{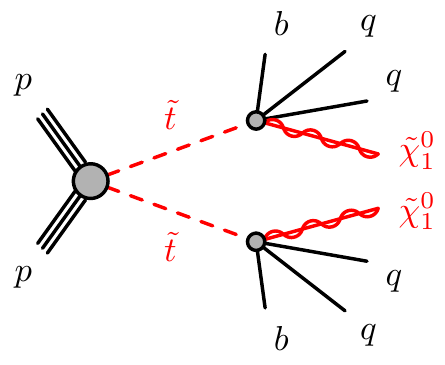}} \\
\subfloat[]{\includegraphics[width=0.32\textwidth]{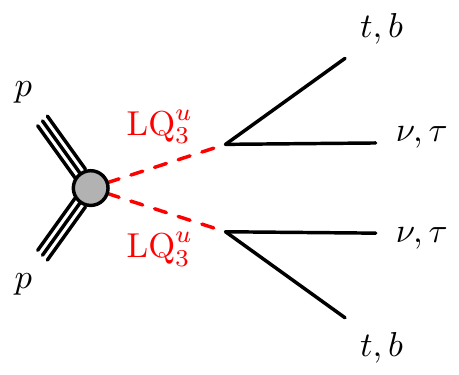}}
\end{center}
\caption{Decay topologies of the signal models considered in the analysis: (a) two-body, (b) three-body, (c) four-body top squark decays, the top quarks being produced in pairs, and (d) up-type, third-generation scalar leptoquark pair production, with both leptoquarks decaying into a top quark and a neutralino or a bottom quark and a $\tau$-lepton. For simplicity, no distinction is made between particles and antiparticles. Only hadronic $\Wboson$ boson decays are shown.
}
\label{fig:feynDiagModels}
\end{figure}

In this paper, each top squark is assumed to decay into a top quark (that may be either on-shell or off-shell) and the LSP, which is assumed to be the lightest neutral mass eigenstate of the partners of the electroweak gauge and Higgs bosons, i.e.\ the lightest neutralino, \ninoone. The degree to which the top quark is off-shell is directly related to the mass difference between $\tilde{t}$  and \ninoone. The top squark decay scenarios considered are shown in Figures~\ref{fig:feynDiagModels}(a)--\ref{fig:feynDiagModels}(c): the top quark is on-shell in two-body decays (${\stop\rightarrow t \ninoone}$), three-body decays contain an off-shell top quark but the $\Wboson$ boson is on-shell (${\stop\rightarrow t^{*}\ninoone \rightarrow bW \ninoone}$), and in four-body decays both the top quark and $\Wboson$ boson are off-shell (${\stop\rightarrow t^{*}\ninoone \rightarrow bW^{*}\ninoone \rightarrow bff' \ninoone}$, where $f$ and $f'$ are fermions originating from the off-shell $\Wboson$ boson decay). Only hadronic $\Wboson$ boson decays are considered in the following.
 
This paper presents a search for top squark pair production with an experimental signature of at least two jets, large missing transverse momentum, and no electrons or muons, using \intlumi\ of proton--proton ($pp$) collision data provided by the LHC at a centre-of-mass energy of $\rts = 13\tev$ and collected by the ATLAS detector in 2015--2018. Previous searches have been performed by both the ATLAS~\cite{SUSY-2014-07,SUSY-2016-15,SUSY-2016-16,SUSY-2016-17,SUSY-2016-26,EXOT-2016-27,TOPQ-2016-10} and CMS~\cite{CMS-SUS-15-004,CMS-SUS-15-005,CMS-SUS-16-008,CMS-SUS-16-049,CMS-SUS-16-051,CMS-SUS-17-001,CMS-SUS-19-005,CMS-SUS-19-006,CMS-SUS-19-009} collaborations. In this search, enhanced sensitivity to two-body top squark decays, where $m_{\stop}-m_{\ninoone}$ is greater than the top quark mass, $m_t$, is achieved by the analysis of the full LHC Run~2 dataset and the exploitation of techniques designed to efficiently reconstruct top quarks that are Lorentz-boosted in the laboratory frame. Sensitivity to compressed scenarios, where $m_{\stop}-m_{\ninoone} \sim m_t$, is extended compared with previous searches through the analysis of events in which high-transverse-momentum jets from initial-state radiation (ISR) boost the top squark system in the transverse plane. Finally, sensitivity to the four-body decay scenario where $m_{\stop}-m_{\ninoone}$ is less than the sum of the $\Wboson$ boson mass, $m_{\Wboson}$, and the $b$-quark mass, $m_{b}$, is achieved by extending the identification efficiency for low-transverse-momentum $b$-hadron decays through the use of charged-particle tracking information, adding sensitivity to the all-hadronic channel in comparison with previous searches. All sensitivities are also increased thanks to global enhancements in detector performance achieved by the end of LHC Run 2, including more precise estimates of the statistical significance of missing transverse momentum in an event~\cite{ATLAS-CONF-2018-038} and improved identification efficiencies for jets containing $b$-hadrons~\cite{FTAG-2018-01}. The interpretation of the results uses simplified models~\cite{Alwall:2008ve,Alwall:2008ag,Alves:2011wf}.
 
As has been demonstrated previously~\cite{SUSY-2016-15,SUSY-2016-16,SUSY-2016-17,EXOT-2017-32,EXOT-2017-30}, top squark searches are sensitive to a variety of additional signal models such as top squarks originating from gluino decays~\cite{Alwall:2008ve,Alwall:2008ag,Alves:2011wf}, top squark decays via charged electroweak SUSY partners~\cite{Alwall:2008ve,Alwall:2008ag,Alves:2011wf}, mediator-based dark-matter models~\cite{Fox:2008kb,Cassel:2009pu,Bai:2010hh,Abdallah:2015ter,Abercrombie:2015wmb,2HDMWGproxi}, scalar dark-energy models~\cite{Brax:2016did}, and third-generation scalar leptoquarks~\cite{dimop_suss,techni2, techni3, string, comp, pati_salam_colour,georgi_glashow_unification}. In this paper, the results are interpreted in models considering the pair production of up-type, third-generation scalar leptoquarks (\lqThreeUp), as shown in Figure~\ref{fig:feynDiagModels}(d), assuming that the \lqThreeUp\ only interact with leptons and quarks from the same generation~\cite{Buchmuller:1986zs}. Similar \lqThreeUp\ interpretations have been performed by both the ATLAS~\cite{EXOT-2017-30} and CMS~\cite{CMS-SUS-18-001} collaborations. The third-generation leptoquark production cross section is identical to that of top squark production and the $\lqThreeUp\to t\nu$ decay channel has the same experimental signature as heavy top squarks decaying into massless neutralinos, and thus additional sensitivity is achieved compared with previous \lqThreeUp\ results.
 
% End of text imported from the .//tex/Introduction.tex input file
\FloatBarrier

% The next lines are included from the .//tex/ATLASdetector.tex input file
\section{ATLAS detector}
 
The ATLAS experiment~\cite{PERF-2007-01,ATLAS-TDR-2010-19,PIX-2018-001} at the LHC is a multipurpose particle detector
with a cylindrical forward--backward- and $\phi$-symmetric
geometry
and an approximate $4\pi$ coverage in
solid angle.\footnote{\AtlasCoordFootnote}
It consists of an inner tracking detector surrounded by a thin superconducting solenoid
providing a \SI{2}{\tesla} axial magnetic field, electromagnetic and hadron calorimeters, and a muon spectrometer.
The inner tracking detector covers the pseudorapidity range $|\eta| < 2.5$.
It consists of silicon pixel, silicon microstrip, and transition
radiation tracking detectors. Lead/liquid-argon (LAr) sampling calorimeters provide electromagnetic (EM) energy measurements
with high granularity.
Hadronic calorimetry is provided by a steel/scintillator-tile calorimeter covering the central pseudorapidity range ($|\eta| < 1.7$).
The endcap and forward regions are instrumented with LAr calorimeters
for both the EM and hadronic energy measurements up to $|\eta| = 4.9$.
The muon spectrometer surrounds the calorimeters and features
three large air-core toroidal superconducting magnets with eight coils each,
providing coverage up to $|\eta| = 2.7$, as well as a system of precision tracking chambers and fast detectors for triggering.
The field integral of the toroids ranges between 2 and 6~T$\cdot$m across most of the detector.
% End of text imported from the .//tex/ATLASdetector.tex input file
\FloatBarrier

% The next lines are included from the .//tex/Simulation.tex input file
\section{Data collection and simulated event samples}
 
The data were collected from 2015 to 2018 at a $pp$ centre-of-mass energy of $13\tev$ with $25\ns$ bunch spacing, resulting in a time-integrated luminosity of \intlumiwithunc~\cite{ATLAS-CONF-2019-021}, measured using the LUCID-2 detector~\cite{LUCID2}. Multiple $pp$ interactions occur per bunch crossing (pile-up) and the average number of these interactions in the data was measured to be $\langle\mu\rangle = 34$. A two-level trigger system~\cite{TRIG-2016-01} is used to select events.
The first-level trigger is implemented in hardware and uses a subset of the detector information
to reduce the event rate to at most \SI{100}{\kilo\hertz}.
This is followed by a software-based trigger that reduces the accepted event rate to \SI{1}{\kilo\hertz} (on average) for offline storage.
 
Selected events are required to pass a missing transverse momentum (whose magnitude is denoted by \met) trigger~\cite{ATL-DAQ-PUB-2016-001}, which is fully efficient for events with reconstructed $\met>250$~GeV\ (the \met\ reconstruction is described in Section~\ref{sec:reconstruction}). In order to estimate the background originating from SM processes, events are also selected at lower values of \met\ using single-electron, single-muon, and single-jet triggers. Electron and muon triggers yield an approximately constant efficiency in the presence of a single isolated electron or muon with transverse momentum (\pt) above 27~GeV~(see Section~\ref{sec:reconstruction} for details of the electron, muon, and jet reconstruction); these triggers are needed for the estimation of $\Zboson\to \nu\bar{\nu}$ production in association with heavy-flavour jets (\Zjets) and top pair production in association with $\Zboson\to \nu\bar{\nu}$ (\ttZ) backgrounds. Triggers based on the presence of a single jet were used to collect data samples for the estimation of the multijet and all-hadronic \ttbar\ backgrounds. The jet \pt~thresholds after energy calibration ranged from $50$ to $400$~GeV. In order to stay within the bandwidth limits of the trigger system, only a fraction of the events passing the jet triggers were recorded to permanent storage.
 
Monte Carlo (MC) simulations are used to model the SUSY and leptoquark signals, as well as to aid in the description of the background processes.
SUSY signal models were all generated with \MGMCatNLOV{2.6.2}~\cite{Alwall:2014hca} at leading order (LO) in QCD, while leptoquark signals were generated with \MGMCatNLOV{2.4.3} at next-to-leading order (NLO) in QCD. All signal samples were interfaced to \PYTHIAV{8.230}~\cite{Sjostrand:2014zea} for the parton showering (PS) and hadronisation, and with {\scshape EvtGen} 1.6.0~\cite{Lange:2001uf} for the $b$- and $c$-hadron decays.

The parton distribution function (PDF) set used for the generation of the signal samples is NNPDF2.3 LO~\cite{Ball:2012cx} for SUSY signals and NNPDF3.0 NLO~\cite{Ball:2014uwa} for leptoquark signals, with the A14~\cite{ATL-PHYS-PUB-2014-021} set of tuned underlying-event and parton shower parameters (UE tune). Matching of the matrix element (ME) with parton showering was performed following the CKKW-L prescription~\cite{CKKW}, with a matching scale set to one quarter of the mass of the top squark or leptoquark. All signal cross sections are calculated to approximate next-to-next-to-leading order (NNLO) in the strong coupling constant, adding the resummation of soft gluon emission at next-to-next-to-leading-logarithm accuracy (approximate NNLO+NNLL)~\cite{Beenakker:2016lwe,Beenakker:1997ut,Beenakker:2010nq,Beenakker:2016gmf}.
 
The top squark mixing parameter between $\stopL$ and $\stopR$ was set to be maximal.\footnote{This refers to the Higgs--top-squark trilinear mixing term; the scenario of maximal mixing allows the top squark masses to be as light as possible, given a 125~GeV\ Higgs mass~\cite{Carena2003}.}
Finally, the top quark mass was set to $172.5$~GeV in all simulated samples.
 
SM background samples were generated with different MC event generators depending on the process. Details of the generators and parton showering used for the different processes are shown in Table~\ref{tab:mc_samples}.
 
\begin{table}[htbp]
\footnotesize
\centering
\caption{Overview of the simulated background samples.
}
\vspace{3mm}
\makebox[\textwidth][c]{
\begin{tabular}{ l | | lllll }
\hline\hline
Process            & ME event generator                    & PDF       & PS and        & UE tune & Cross-section                                                                                                 \\
&                                       &              & hadronisation &         & calculation                                                                                                   \\
\hline\hline
$V$+jets $(V=W/Z)$ & \SHERPA~2.2.1~\cite{Gleisberg:2008ta} & NNPDF3.0 NNLO & \SHERPA       & Default & NNLO~\cite{Catani:2009sm}                                                                                     \\
$\ttbar+V $        & \ \AMCatNLOV{2.3.3}                          & NNPDF3.0 NLO     & \PYTHIAV{8.210}  & A14     & NLO~\cite{Alwall:2014hca}                                                                                     \\
$\ttbar$           & \POWHEGBOXV{v2}~\cite{Alioli:2010xd}       & NNPDF3.0 NNLO   & \PYTHIAV{8.230}  & A14     & NNLO+NNLL~\cite{Czakon:2013goa,Czakon:2012pz,Czakon:2012zr,Baernreuther:2012ws,Cacciari:2011hy,Czakon:2011xx} \\
Single top         & \POWHEGBOXV{v2}                            & NNPDF3.0 NNLO   & \PYTHIAV{8.230}  & A14     & NNLO+NNLL~\cite{Kidonakis:2011wy,Kidonakis:2010ux,Kidonakis:2010tc}                                           \\
Diboson            & \SHERPA~2.2.1-2.2.2                   & NNPDF3.0 NNLO & \SHERPA       & Default & NLO                                                                                                           \\
$\ttbar H$         & \ \AMCatNLOV{2.2.3}                          & NNPDF3.0 NLO   & \PYTHIAV{8.230}  & A14     & NLO~\cite{Beenakker:2002nc,Dawson:2003zu,Yu:2014cka,Frixione:2015zaa}                                                   \\
$tWZ$, $tZ$         & \ \AMCatNLOV{2.3.3}                          & NNPDF3.0 NLO   & \PYTHIAV{8.212,8.230}  & A14     & NLO    \\
\hline\hline
\end{tabular}
}
\label{tab:mc_samples}
\end{table}
The detector simulation~\cite{SOFT-2010-01} was performed using either $\GEANT4$~\cite{Agostinelli:2002hh} or a fast simulation framework, where the showers in the electromagnetic and hadronic calorimeters are simulated with a parameterised description~\cite{ATL-PHYS-PUB-2010-013} and the rest of the detector is simulated with $\GEANT4$.
All signal samples were produced using the fast simulation, while SM background samples used the $\GEANT4$ set-up. All MC samples were produced with a varying number of simulated minimum-bias interactions overlaid on the hard-scattering event, to account for pile-up. These interactions were produced using \PYTHIAV{8.2}\ with the A3 tune~\cite{ATL-PHYS-PUB-2016-017} and NNPDF2.3 LO PDF set. The simulated events are reweighted to match the distribution of the number of $pp$ interactions per bunch crossing in data. Corrections are applied to the simulated events to account for differences between data and simulation for the lepton trigger, reconstruction, identification and isolation efficiencies, and for the lepton and jet momentum scale and energy resolution. Corrections are also applied to the efficiency of identifying jets containing $b$-hadrons ($b$-jets), the probability of mis-tagging jets containing only charm hadrons ($c$-jets) and only lighter hadrons (light-flavour jets), and the probability of mis-tagging jets originating from the hard $pp$ scattering as pile-up jets.
% End of text imported from the .//tex/Simulation.tex input file
\FloatBarrier

% The next lines are included from the .//tex/Reconstruction.tex input file
\section{Event reconstruction}
\label{sec:reconstruction}
 
Events are required to have a primary vertex~\cite{PERF-2015-01,ATL-PHYS-PUB-2015-026} reconstructed from at
least two tracks~\cite{PERF-2015-08} with $\pT>500~\MeV$. Among the vertices found, the vertex with the largest summed $\pT^2$ of the associated tracks is designated as the primary vertex.
 
Calorimeter jets are built from topological clusters of energy in the calorimeter~\cite{PERF-2014-07}, calibrated to the
electromagnetic scale, using the anti-$k_t$ algorithm with radius parameter
$R=0.4$~\cite{Cacciari:2008gp,Fastjet}. These types of jets are referred to as `jets'. Jet transverse momenta are further corrected to the corresponding particle-level jet \pt, based on the simulation~\cite{PERF-2016-04}. Remaining differences between data and simulated events are evaluated and corrected for using in situ techniques, which exploit the transverse momentum balance between a jet and a reference object such as a photon, $Z$ boson, or multijet system in data. After these calibrations, all jets in the event with $\pt > 20$~GeV\ and $|\eta| < 4.5$ must satisfy a set of loose jet-quality requirements~\cite{ATLAS-CONF-2015-029}. In the four-body analysis, the leading jet in \pt\ must satisfy a set of tighter jet-quality requirements. These requirements are designed to reject jets originating from sporadic bursts of detector noise, large coherent noise or isolated pathological cells in the calorimeter system, hardware issues, beam-induced background or cosmic-ray muons~\cite{ATLAS-CONF-2015-029}. If these jet requirements are not met, the event is discarded. All jets are required to have $\pt > 20$~GeV\ and $|\eta| < 2.8$ to be considered in this analysis.
In addition, the `medium' working point of the track-based jet vertex tagger~\cite{PERF-2014-03,ATLAS-CONF-2014-018} is required for
jets with $\pT<120$~GeV\ and $|\eta|<2.5$, to reject jets that originate from
pile-up interactions.
 
Jets which contain $b$-hadrons and are within the inner-detector
acceptance ($|\eta|<2.5$) are identified as `$b$-tagged' using a multivariate algorithm that exploits
the impact parameters\footnote{The transverse impact parameter, $d_{0}$, is defined as the distance of closest approach of a track to the beam-line, measured in the transverse plane. The longitudinal impact parameter, $z_{0}$, corresponds to the $z$-coordinate distance between the point along the track at which the transverse impact parameter is defined and the primary vertex.} of the charged-particle tracks, the presence of secondary
vertices, and the reconstructed flight paths of $b$- and $c$-hadrons inside
the jet~\cite{FTAG-2018-01}. The output of the multivariate algorithm is a single $b$-tagging output score, which signifies the likelihood of a jet to contain $b$-hadrons. The average identification efficiency of jets containing $b$-hadrons is $77\%$ as determined in simulated \ttbar\ events. Using the same simulated sample, a rejection factor of approximately 110 (5) is reached for jets initiated by light quarks and gluons (charm quarks).
 
In order to identify low-\pt\ $b$-hadrons that are not contained in jets passing the $\pt>20$~GeV\ requirement, `track-jets' are reconstructed from inner-detector tracks using the anti-\kt algorithm with radius parameter $R=0.4$. Tracks considered for inclusion in track-jets are required to have $\pt>500$~\MeV, $|\eta|<2.5$, at least seven hits in the silicon microstrip and pixel detectors, no more than one hit shared by multiple tracks in the pixel detector, no more than one missing hit in the pixel detector, and no more than two missing hits in the silicon microstrip detector. Additional requirements on the longitudinal impact parameter projected along the beam direction ($\left|z_0 \sin(\theta)\right|<3$~mm) reduce the pile-up contributions and improve the efficiency in selecting tracks from the hard-scatter vertex. Track-jets are required to have $\pt > 5$~GeV, more than one track within the jet radius, $|\eta|<2.5$, and not overlap with the leading non-$b$-tagged jet in the event ($\Delta R>0.4$). The standard $b$-tagging algorithm is employed for track-jets~\cite{ATLAS-CONF-2019-027} and the selection requirement is tighter than for regular jets, due to the larger amount of background at low \pt. The average identification efficiency for jets containing $b$-hadrons is $70\%$ as determined in simulated \ttbar\ events. Using the same simulated sample, a rejection factor of approximately 200 (10) is reached for jets initiated by light quarks and gluons (charm quarks).
 
Electron candidates are reconstructed from clusters of energy deposits in the
electromagnetic calorimeter that are matched to a track in the inner
detector. They are required to have $|\eta|<2.47$ and $\pT>4.5$~GeV, and
must pass a loose likelihood-based selection~\cite{PERF-2017-01,PERF-2017-03}. The impact parameter along the
beam direction is required to be less than $0.5$~mm. The electromagnetic shower of an electron can also be reconstructed as a jet such that a procedure is required to resolve this ambiguity. In the case where the separation\footnote{For the overlap removal, rapidity ($y$) is used instead of pseudorapidity: $y = \frac{1}{2} \ln \frac{E+p_z}{E-p_z}$, where $E$ is the energy and $p_z$ is the $z$-component of the momentum of the object. The separation is then defined as $\dRy \equiv \sqrt{(\Delta y)^2+(\Delta \phi)^2}$.} between an electron candidate and a non-$b$-tagged ($b$-tagged) jet is $\dRy< 0.2$, the candidate is considered to be an electron ($b$-tagged jet). This procedure uses a $b$-tagged jet definition that is looser than the one described earlier, to avoid selecting electrons from heavy-flavour hadron decays. If the
separation between an electron candidate and any jet satisfies $0.2<\dRy < 0.4$, the candidate is considered to be a jet, and the electron candidate is removed.
 
Muons are reconstructed by matching tracks in the inner detector to tracks
in the muon spectrometer and are required to have $|\eta|<2.7$ and
$\pT>4$~GeV~\cite{PERF-2015-10}. The impact parameter along the
beam direction is required to be less than $0.5$~mm. Events containing muons identified as originating from cosmic rays, $|d_0| > 0.2$~mm and $|z_0| > 1$~mm, or as poorly reconstructed, $\sigma(q/p)/|(q/p)| > 0.2$, are removed.
Here, $\sigma(q/p)/|(q/p)|$ is a measure of the momentum uncertainty for a particle with charge $q$. Muons are discarded if they are within $\dR = 0.4$ of jets that survive the electron--jet overlap removal, except when the number of tracks associated with the jet is less than three, where the muon is kept and the jet discarded.
 
The requirements on electrons and muons are tightened for the
selection of events in background control regions (described in
Section~\ref{sec:background}) containing at least one electron or muon.
The electrons and muons passing the tight selection are called `control' electrons or muons in the following, as opposed to `baseline' electrons and muons, which are only required to pass the requirements described above. Control electrons and muons are required to satisfy the `FCLoose' $\pt$-dependent track-based and calorimeter-based isolation criteria~\cite{EGAM-2018-01}. The calorimeter-based isolation is determined by taking the ratio of the sum of energy deposits in a cone of $\Delta R=0.2$ around the electron or muon candidate to the sum of energy deposits associated with the electron or muon. The track-based isolation is estimated in a similar way but using a variable cone size with a maximum value of $\Delta R=0.2$ for electrons and $\Delta R=0.3$ for muons. Electron candidates are required to pass a `tight' likelihood-based selection.
The impact parameter of the electron in the transverse plane is required to be less than five times the transverse impact parameter uncertainty ($\sigma_{d_0}$). Further selection criteria are also imposed on reconstructed muons: muon candidates are required to pass a `medium' quality selection and meet the $|d_0| < 3\sigma_{d_0}$ requirement.
 
The $\ptmiss$ vector is the negative vector sum
of the \pT\ of all selected and calibrated electrons, muons, and jets in the
event, plus an extra term (`soft' term) added to account for energy depositions in the event that are not associated with any of the objects. The `soft' term is calculated from inner-detector tracks ($\pT > 500~\MeV$ and matched to the primary vertex, to make it resilient to pile-up contamination) not associated with selected objects~\cite{PERF-2016-07,ATLAS-CONF-2018-023}.
The missing transverse momentum calculated using only the tracking system (denoted by \ptmisstrk, with magnitude \mettrk) is computed from the vector sum of the inner-detector tracks with $\pt>500~\MeV$ and $|\eta|<2.5$ that are associated with the event's primary vertex.
 
Hadronically decaying $\tau$-lepton candidates are identified as non-$b$-tagged jets with $|\eta|<2.5$ and a maximum of four inner-detector tracks matched to them. They are only used in some regions to veto events with $\tau$-lepton candidates most likely originating from $W \rightarrow \tau \nu$ decays, which are identified with the additional requirement that the $\Delta \phi$ between the $\tau$-lepton candidate and the $\ptmiss$ is less than $\pi / 5$.
% End of text imported from the .//tex/Reconstruction.tex input file
\FloatBarrier
 
% The next lines are included from the .//tex/SignalRegions.tex input file
\section{Signal region definitions}
 
The experimental signature of this search, for all signal topologies, consists of multiple jets, one or two of which are $b$-tagged, no electrons and muons (following the baseline definition described in Section~\ref{sec:reconstruction}), and large missing transverse momentum. The \met\ trigger is used to collect the data in all signal regions.
 
\begin{figure}[htpb]
\begin{center}
\includegraphics[width=0.7\textwidth]{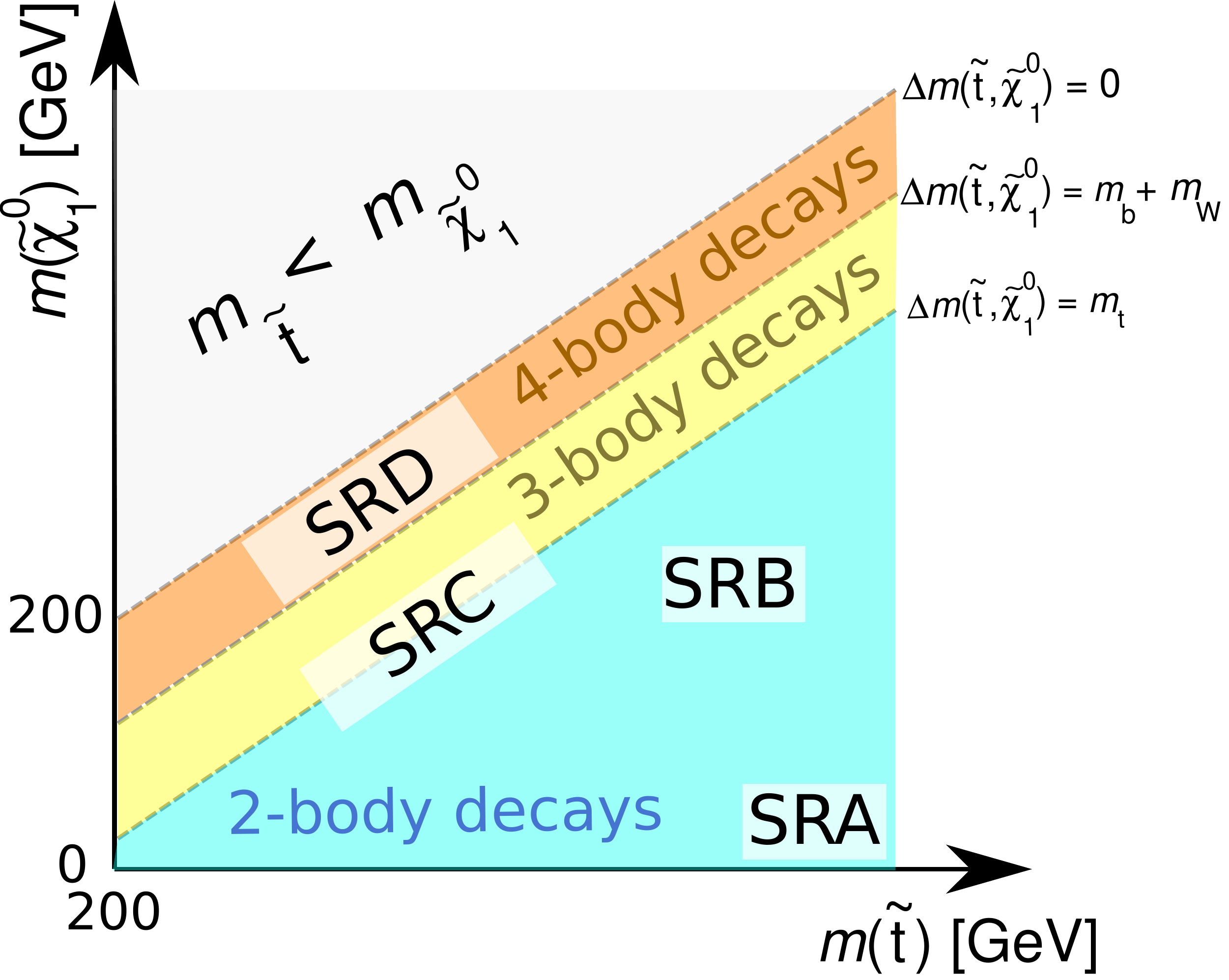}
 
\caption{Schematic representation of the various topologies targeted by the different signal regions defined in the analysis (SRA, SRB, SRC, SRD). SRA and SRB are orthogonal and the exact requirements made in the signal regions are detailed in the text and Table~\ref{tab:SignalRegionAB}. }
\label{fig:srSketch}
\end{center}
\end{figure}

Beyond these common requirements, four sets of signal regions (SRA--D) are defined to target each decay topology and kinematic regime, as shown in Figure~\ref{fig:srSketch}. SRA (SRB) is sensitive to the production of high-mass \stop\ pairs that each undergo a two-body decay with large (medium) $\Delta m(\stop,\ninoone)$, or the production of high-mass leptoquark pairs. Both SRA and SRB employ top-quark mass-reconstruction techniques to reject background, of which the dominant source is associated production of a $Z$ boson with heavy-flavour jets, with the $\Zboson$ decaying into neutrinos (\Zjets). SRC targets compressed two/three-body top squark decays with $\Delta m(\stop,\ninoone)\sim m_{t}$ and has \ttbar\ production as the dominant background contribution. A common preselection is defined for SRA--C: at least four jets are required ($\njets\ge4$), at least two of which must be $b$-tagged ($\nbjets\ge2$), and the leading four jets must satisfy $\pt>80,80,40,40$~GeV. SRD targets highly compressed four-body top squark decays and uses track-jets to identify $b$-hadrons with low \pt. As in SRA and SRB, the dominant source of background in SRD is \Zjets. In both SRC and SRD, a high-\pt\ jet originating from ISR is used to improve sensitivity to the targeted decays.
 
\subsection{Signal regions A and B}
 
SRA is optimised for exclusion at 95\% confidence level (CL) of the scenario where $\mstop=1300$~GeV\ and $\mLSP=1$~GeV, while SRB is optimised for $\mstop=700$~GeV\ and $\mLSP=400$~GeV.
SRA and SRB have the best sensitivity to up-type, third-generation scalar leptoquarks, when leptoquarks decay via $\lqThreeUp\rightarrow t\nu$.
 
To avoid a loss of efficiency when the top quark has $\pt>200$~GeV\ and its daughters are close to each other, the two hadronic top candidates are reconstructed by using the \antikt\ algorithm to cluster $R=0.4$ jets, using radius parameters of $R=0.8$ and $R=1.2$, similar to the technique used in the previous ATLAS search~\cite{SUSY-2016-15}. Each reclustered jet is assigned a mass which is computed from the four-momenta of its jet constituents.
Two $R=1.2$ reclustered jets, representing top candidates, are required, and the leading reclustered $R=1.2$ jet must have a mass (\mantikttwelvezero) greater than 120~GeV. To optimise signal efficiency regardless of the subleading top candidate reconstruction success (measured by how close the candidate mass is to the top quark mass), the events are divided into three categories based on the subleading $R=1.2$ reclustered jet mass (\mantikttwelveone): the `TT' category includes events with $\mantikttwelveone>120$~GeV, corresponding to successfully reconstructing a subleading top candidate; the `TW' category contains events with $60<\mantikttwelveone<120$~GeV, corresponding to successfully reconstructing a subleading \Wboson\ candidate; and the `T0' category represents events with $\mantikttwelveone<60$~GeV, corresponding to not reconstructing a top nor a \Wboson\ candidate.
 
In SRA, in addition to using the mass of the reclustered jets, information about the flavour content of the reclustered jet is used to improve background rejection. For all SRA categories, a \bjet\ is required to be within $\Delta R=1.2$ of the leading reclustered $R=1.2$ jet, \bantikttwelvezero, while in the SRA-TT category, the same selection is made for the subleading $R=1.2$ jet, \bantikttwelveone. A requirement is also made on the leading $R=0.8$ reclustered jet mass ($\mantikteightzero>60$~GeV) in SRA.
 
In order to reject events with mismeasured \met\ originating from multijet and hadronic \ttbar\ decays, the minimum difference in azimuthal angle between the \ptmiss\ and the leading four jets (\dphijetfourmet) is required to be greater than 0.4.
 
The most powerful rejection of background comes from requiring that the object-based \met\ significance (\objmetsig)~\cite{ATLAS-CONF-2018-038} is greater than 25 (14) in SRA (SRB). This variable characterises the \met\ according to the \pT, \pT\ resolution, and $\phi$ resolution of all objects in the event, and is defined as:
 
\begin{equation}
\textup{\objmetsig} = \frac{\met}{\sqrt{\sigma_{\textup{L}}^2(1- \rho_{\textup{LT}}^2)}}, \notag
\end{equation}
 
where $\sigma_{\textup{L}}$ is the expected resolution of the total longitudinal momentum (relative to the direction of \ptmiss) of all objects in the event as a function of the \pt\ of each object. Likewise, $ \rho_{\textup{LT}}$ is the correlation factor between the longitudinal and transverse momentum resolutions for all objects.
 
Substantial $\ttbar$ background rejection is provided by additional requirements to reject events in which one $\Wboson$ boson decays via a lepton plus neutrino. The first requirement is that the transverse mass (\mt) calculated from the \met\ and
the \bjet\ closest in $\phi$ to the \ptmiss\ direction and defined as:
 
\begin{equation}
\mtbmin\ = \sqrt{2\,\ptb\,\met \left[1-\cos{\Delta\phi\left(\vecptb,\ptmiss\right)}\right]}, \notag
\end{equation}
 
must be above $200$~GeV.
The second requirement consists of vetoing events containing hadronic $\tau$-lepton candidates likely to have originated from a $W \rightarrow \tau \nu$ decay ($\tau$-veto).
 
To reject events that contain \bjet s from gluon splitting, requirements are made on the angular distance between the two leading \bjet s, \drbjetbjet. In SRB, an additional requirement of $\mtbmax>200$~GeV\ is made, where \mtbmax\ is analogous to \mtbmin\ except that the transverse mass is computed with the \bjet\ that has the largest $\Delta\phi$ relative to the $\ptmiss$ direction. This requirement is a more stringent version of \mtbmin, requiring that the leading two \bjet s are not near the \ptmiss.
 
Finally, to allow the statistical combination of SRA and SRB, SRA is required to have the \mttwochisquare\ variable greater than 450~GeV, while SRB is required to have $\mttwochisquare<450$~GeV. The \mttwochisquare\ variable is based on \mttwo~\cite{Lester:1999tx,Barr:2003rg} and is constructed from the direction and magnitude of \ptmiss\ and the direction of each of the top candidates, reconstructed using a $\chi^2$-like method with $R=0.4$ jets as inputs.
The minimisation for finding the top candidates used in \mttwochisquare\ is performed in terms of a $\chi^2$-like penalty function, $\chi^2 = (m_{\mathrm{cand}}-m_{\mathrm{true}})^2/m_{\mathrm{true}}$, where $m_{\mathrm{cand}}$ is the top quark or \Wboson\ boson candidate mass and $m_{\mathrm{true}}$ is set to 80.4~GeV\ for \Wboson\ boson candidates and 173.2~GeV\ for top quark candidates.\footnote{These mass values were the world averages of the \Wboson\ boson and top quark masses at the time of the development of this method which was for the last iteration of this search~\cite{SUSY-2016-15}. Updated measurements of the masses of the \Wboson\ boson and top quark have a negligible effect on this method and thus were not included.} Initially, single or pairs of $R=0.4$ jets, whichever configuration results in a mass closest to $m_W$, form \Wboson\ boson candidates, which are then combined with additional \bjets\ in the event to construct top quark candidates. When calculating \mttwochisquare\ the momenta of top quark candidates selected by the $\chi^2$ method are used, while the masses of the top quarks are set to 173.2~GeV\ and the invisible particles are assumed to be massless. Table~\ref{tab:SignalRegionAB} summarises all the selection criteria used in SRA and SRB.
 
In addition to SRA and SRB, which are optimised for high $\mstop$ via a statistical combination, a signal region is optimised for discovery.
This region, SRA-TT-Disc, has the same requirements as SRA-TT, with the exception of a less stringent requirement of $\objmetsig > 11$. When setting exclusion limits on specific signal models, SRA-TT-Disc is not considered.
 
% The next lines are included from the .//tex/srabDefTable.tex input file
\begin{table}[htb]
\caption{Selection criteria for SRA and SRB. Each signal region is separated into three categories based on reconstructed top candidate masses. A dash indicates that no selection is applied.}
\begin{center}
\def\arraystretch{1.4}
\resizebox{\textwidth}{!}{
\begin{tabular}{l||c|c|c|c|c|c} \hline\hline
\textbf{ Variable/SR} & \textbf{ SRA-TT} & \textbf{ SRA-TW} & \textbf{ SRA-T0} & \textbf{ SRB-TT} & \textbf{ SRB-TW} & \textbf{ SRB-T0} \\ \hline \hline
Trigger               & \multicolumn{6}{c}{\met}                                                                                        \\ \hline
\met               & \multicolumn{6}{c}{$>250$~GeV}                                                                                        \\ \hline
\nlep                 & \multicolumn{6}{c}{exactly 0}                                                                                           \\ \hline
\njets         & \multicolumn{6}{c}{$\ge4$}                                                                                      \\ \hline
\pttwo                     & \multicolumn{6}{c}{$>80$~GeV}         \\ \hline
\ptfour                    & \multicolumn{6}{c}{$>40$~GeV}         \\ \hline
\dphijetfourmet                                                      & \multicolumn{6}{c}{$> 0.4$}                                                         \\ \hline
\nbjets                                                              & \multicolumn{6}{c}{$\ge2$}                                                          \\ \hline
\mtbmin & \multicolumn{6}{c}{$>200$~GeV}                                                     \\ \hline
$\tau$-veto & \multicolumn{6}{c}{\checkmark} \\ \hline
\mantikttwelvezero                                                   & \multicolumn{6}{c}{$>120$~GeV}                                                     \\ \hline
\mantikttwelveone                                                    & $>120$~GeV{} & $60$--$120$~GeV{} & $<60$~GeV{} & $>120$~GeV{} & $60$--$120$~GeV{} & $<60$~GeV{} \\ \hline
 
\mantikteightzero & \multicolumn{3}{c|}{$>60$~GeV} & \multicolumn{3}{c}{--}                           \\ \hline
\bantikttwelvezero      & \multicolumn{3}{c|}{\checkmark} & \multicolumn{3}{c}{--}                           \\ \hline
\bantikttwelveone   & \checkmark                      & \multicolumn{5}{c}{--}                \\ \hline
\drbjetbjet       & $>1.0$                          & \multicolumn{2}{c|}{--} & \multicolumn{3}{c}{$>1.4$}              \\ \hline
\mtbmax                 & \multicolumn{3}{c|}{--}         & \multicolumn{3}{c}{$>200$~GeV}                  \\ \hline
 
\objmetsig & \multicolumn{3}{c|}{$>25$ } & \multicolumn{3}{c}{$>14$ } \\ \hline
\mttwochisquare                          & \multicolumn{3}{c|}{$>450$~GeV} & \multicolumn{3}{c}{$<450$~GeV}          \\
\hline\hline
\end{tabular}
}
\end{center}
\label{tab:SignalRegionAB}
\end{table}
% End of text imported from the .//tex/srabDefTable.tex input file
\FloatBarrier
 
\subsection{Signal regions C}
 
SRC is optimised for the case where $\Delta m(\stop,\ninoone)\sim m_t$, a regime in which the signal topology is similar to SM \ttbar\ production. In the presence of high-momentum ISR jets, the di-top-squark system is boosted in the transverse plane and better discrimination can be achieved. A recursive jigsaw reconstruction technique, as described
in Ref.~\cite{RJR_ISR}, is used to divide each event into an ISR hemisphere (denoted by `ISR')
and a sparticle hemisphere (denoted by `S'), where the latter consists of the pair of
candidate top squarks.
 
Objects are grouped together according to their proximity in
the laboratory frame's transverse plane by minimising the reconstructed
transverse masses of the ISR system and sparticle system, simultaneously over all choices of object assignment. Kinematic variables are then defined, based on this assignment of objects to either the ISR system or the sparticle system.
 
The ratio of the \met\ to the \pt\ of the ISR system (\pTISR), defined as \rISR, is proportional to the ratio of the $\ninoone$ and $\stop$ masses~\cite{An,Macaluso}:
\begin{equation}
\rISR \equiv \frac{\met}{\pTISR} \sim \frac{m_{\ninoone}}{m_{\stop}}. \notag
\end{equation}
 
Due to the scaling of \rISR\ with the ratio of $m_{\ninoone}$ to $m_{\stop}$, signals with $\Delta m(\stop,\ninoone)\sim m_t$ are expected to form a peak in the \rISR\ distribution, with the location of the peak depending on $m_{\ninoone}$ divided by $m_{\stop}$. In order to maximise the sensitivity for a wide range of $m_{\ninoone}$ to $m_{\stop}$ ratio values, the events are divided into five
categories, defined by non-overlapping ranges of $\rISR$ and targeting
different top squark and $\ninoone$ masses. For instance, SRC1 is optimised
for $\mstop = 225$~GeV\ and $\mLSP=52$~GeV, and SRC5 is optimised for $\mstop
= 600$~GeV\ and $\mLSP=427$~GeV.
 
In addition, at least four jets must be
assigned to the sparticle hemisphere of the event (\nJetS), and at least two of those jets must be $b$-tagged. Requirements on \pTISR, the highest-\pt\ \bjet\ in the sparticle hemisphere (\pTSBOne), and the fourth-highest-\pt\ jet in the sparticle hemisphere (\pTSFour) are applied. To reject events with poorly measured \met, the difference in $\phi$ between the \ptmisstrk\ and \ptmiss, \dphimettrk, is required to be less than $\pi/3$ and the leading two jets are required to be separated in azimuthal angle from the \ptmiss: $\dphijettwomet>0.4$. The transverse mass of the sparticle system and \ptmiss, defined as $\mS$, is required to be greater than 400~GeV. The ISR system is also required to be separated in azimuthal angle from \ptmiss: $\dPhiISRMET>3.0$. The selection criteria for SRC are summarised in Table~\ref{tab:SignalRegionC}.
 
In addition to SRC1--5, a region optimised for
discovery, SRC-Disc, is defined. In SRC-Disc, the same requirements as in the other SRCs are applied, with the exception of requiring $\rISR>0.5$ and $\objmetsig > 11$. As with SRA-TT-Disc, when setting exclusion limits on specific signal models, this region is not considered.

% The next lines are included from the .//tex/srcDefTable.tex input file
 
\begin{table}[htpb]
\caption{Selection criteria for SRC. The signal regions are separated into five categories based on ranges of $\rISR$.}
\begin{center}
\def\arraystretch{1.4}
\begin{tabular}{l||c|c|c|c|c} \hline\hline
\textbf{Variable/SR} & \textbf{ SRC1} & \textbf{ SRC2} & \textbf{ SRC3} & \textbf{ SRC4} & \textbf{ SRC5} \\
\hline \hline
Trigger               & \multicolumn{5}{c}{\met}                                                                                        \\ \hline
\met               & \multicolumn{5}{c}{$>250$~GeV}                                                                                        \\ \hline
\nlep            & \multicolumn{5}{c}{exactly 0}                                     \\ \hline
\njets           & \multicolumn{5}{c}{$\ge4$}                                        \\ \hline
\pttwo           & \multicolumn{5}{c}{$>80$~GeV}                                    \\ \hline
\ptfour          & \multicolumn{5}{c}{$>40$~GeV}                                    \\ \hline
\nbjets          & \multicolumn{5}{c}{$\ge2$}                                        \\
\hline
$\mettrk$       & \multicolumn{5}{c}{$>30$~GeV}                                    \\
\hline
\dphimettrk        & \multicolumn{5}{c}{ $< \pi/3$}                                    \\
\hline
$\dphijettwomet$ & \multicolumn{5}{c}{$> 0.4$}                                       \\
\hline
\nJetS           & \multicolumn{5}{c}{$\ge 4$}                                       \\
\hline
\nBJetS            & \multicolumn{5}{c}{$\ge2$}                                        \\
\hline
\pTISR           & \multicolumn{5}{c}{$>400$~GeV}                                   \\
\hline
\pTSBOne         & \multicolumn{5}{c}{$>50$~GeV}                                    \\
\hline
\pTSFour         & \multicolumn{5}{c}{$>50$~GeV}                                    \\
\hline
\mS              & \multicolumn{5}{c}{$>400$~GeV}                                   \\
\hline
\dPhiISRMET      & \multicolumn{5} {c}{$>3.0$}                                       \\
\hline
\rISR            & $0.30$--$0.40$ & $0.40$--$0.50$ & $0.50$--$0.60$ & $0.60$--$0.70$ & $>0.70$ \\
\hline \hline
\end{tabular}
\end{center}
\label{tab:SignalRegionC}
\end{table}
% End of text imported from the .//tex/srcDefTable.tex input file
\FloatBarrier
 
\subsection{Signal regions D}
 
SRD aims to select four-body top squark decays, for which the kinematic properties depend mainly on $\Delta m(\stop,\ninoone)$. Four-body top squark decays result in final state particles with low \pt, which are particularly challenging to reconstruct. For instance, low-\pt\ $b$-hadrons originating from such decays are usually not contained in jets passing the minimum $\pt>20$~GeV\ requirement when $\Delta m(\stop,\ninoone)<50$~GeV, and therefore cannot be tagged the same way as in SRA--C. To circumvent this problem and identify the low-\pt\ $b$-hadrons produced in a larger part of the four-body decay phase space, $b$-tagging using track-jets with $\pt>5$~GeV\ is employed. Three signal region categories, \srfourbzero, \srfourbone, and \srfourbtwo, are defined according to the $b$-tagged jet multiplicity (zero, one, and two, respectively), and are optimised for ${\Delta m(\stop,\ninoone)=}$~20,~50,~80~GeV, respectively. In SRD0 and SRD1 the presence of at least one $b$-tagged track-jet is required to recover undetected jets that contain $b$-hadrons.
 
An event including a pair of four-body top squark decays with $\met>250$~GeV\ is likely to be caused by the presence of significant ISR emission. Thus the leading non-\bjet, identified as the ISR jet (\nonb), is required to have large \pt\ (\ptnonbjetone), as well as a large azimuthal separation \dphinonbjetmet\ from \ptmiss.
 
In order to reject events with mismeasured \met\ originating from multijet and hadronic \ttbar\ decays, requirements are made on \mettrk\ and \dphimettrk. Further background reduction is required in \srfourbzero\ and attained by selecting large \dphijetfourmet.
 
Only low-\pt\ jets and track-jets (\ptonevr, \ptbonevr, \ptbone) are considered in all three categories. Requirements are also made on $b$-tagged jet and track-jet pseudorapidities (\etabonevr, \etabone, \etabtwo) to ensure they are in the central region of the detector, which make them more likely to originate from a top squark decay and maximise the $b$-tagging performance. Only events with high $\htsig$ are kept, where \HT\ is the scalar sum of the transverse momenta of all jets. This kinematic variable was found to provide better signal versus background discrimination than the object-based \met\ significance when the final state is composed of low-\pt\ objects.
 
Given the absence of on-shell top quarks and \Wboson\ bosons, no top nor $W$ reconstruction methods are used, such that additional discrimination of the signal from the background relies on differences in angular separation between jets and track-jets. In \srfourbone\ (\srfourbtwo), requirements are made on the angular separation between the ISR jet and the \bjet (s), \dphinonbjetbjetone\ (\dphinonbjetbjetone\ and \dphinonbjetbjettwo), to ensure the \bjet (s) is (are) well-separated from the ISR jet. In \srfourbone, the minimum $\Delta\phi$ between the leading four track-jets and the ISR jet ($\mindphijetvrfournonb$) is also required to be large, to separate the low-\pt\ top squark decay products from the ISR jet. Further background rejection is required in \srfourbzero\ and attained by requiring significant azimuthal separation between the leading $b$-tagged track-jet and the ISR jet (\maxdphinonbjetbjetvr), and between the leading $b$-tagged track-jet and the next track-jet most likely to contain a $b$-hadron ($\dphibbvr$). Table~\ref{tab:SignalRegionD} summarises the full signal region selections for \srfourbzero-2.
 
% The next lines are included from the .//tex/srdDefTable.tex input file
\begin{table}[htpb]
\caption{Signal region selections for SRD. Variables involving track-jets are denoted with the label `track'. A dash indicates that no selection is applied.}
\begin{center}
\def\arraystretch{1.4}
\begin{tabular}{l||c|c|c} \hline\hline
\textbf{Variable/SR}                        & \textbf{\srfourbzero}       & \textbf{\srfourbone} & \textbf{\srfourbtwo} \\
\hline \hline
Trigger               & \multicolumn{3}{c}{\met}                                                                                        \\ \hline
\met               & \multicolumn{3}{c}{$>250$~GeV}                                                                                        \\ \hline
\nlep                 & \multicolumn{3}{c}{exactly 0}   \\ \hline
\nbjets                 & exactly 0                   & exactly 1            & $\ge2$               \\
\hline
\ptnonbjetone           & \multicolumn{3}{c}{$>250$~GeV}                                            \\
\hline
\dphinonbjetmet         & \multicolumn{3}{c}{$>2.4$}                                                \\
\hline
$\mettrk$               & \multicolumn{3}{c}{$>30$~GeV}                                                    \\
\hline
\dphimettrk             & \multicolumn{3}{c}{ $< \pi/3$}                                            \\
\hline
\nbjetsvr               & \multicolumn{2}{c|}{$\ge1$} & --                                           \\
\hline
$\dphijetfourmet$       & $> 0.4$                     & \multicolumn{2}{c}{--}                       \\
\hline
\etabonevr              & $< 1.2$                     & \multicolumn{2}{c}{--}                       \\
\hline
$\maxdphinonbjetbjetvr$ & $>2.2$                      & \multicolumn{2}{c}{--}                       \\
\hline
$\dphibbvr$             & $< 2.5$                     & \multicolumn{2}{c}{--}                       \\
\hline
\ptbonevr               & $< 50$~GeV                  & $>10$~GeV            & --                    \\
\hline
\ptonevr                & --                           & $<40$~GeV            & --                    \\
\hline
$\mindphijetvrfournonb$ & --                           & $> 1.2$              & --                    \\
\hline
\etabone                & --                           & $< 1.6$              & --                    \\
\hline
$\dphinonbjetbjetone$   & --                           & \multicolumn{2}{c}{$>2.2$}                  \\
\hline
\etabtwo                & \multicolumn{2}{c|}{--}      & $<1.2$                                      \\
\hline
\ptbone                 & \multicolumn{2}{c|}{--}      & $< 175$~GeV                                 \\
\hline
$\dphinonbjetbjettwo$   & \multicolumn{2}{c|}{--}      & $>1.6$                                      \\
\hline
$\htsig$                & $> 26\sqrt{\mathrm{GeV}}$           & \multicolumn{2}{c}{$> 22 \sqrt{\mathrm{GeV}}$}      \\
\hline\hline
\end{tabular}
\end{center}
\label{tab:SignalRegionD}
\end{table}
% End of text imported from the .//tex/srdDefTable.tex input file
% End of text imported from the .//tex/SignalRegions.tex input file
\FloatBarrier
 
% The next lines are included from the .//tex/Background.tex input file
\section{Background estimation}
\label{sec:background}
 
The main SM background process in SRA, SRB, and SRD is $\Zboson\to \nu\bar{\nu}$ production in association with heavy-flavour jets. In SRC, \ttbar\ production dominates, including mostly events where one \Wboson\ boson decays hadronically and the other \Wboson\ boson decays via a $\tau$-lepton and its corresponding neutrino (\Wjets). Other important background processes include leptonic $\Wboson$ decays produced in association with heavy-flavour jets, a single top quark produced with a \Wboson\ boson, and the irreducible background from $\ttZ$, where the $\Zboson$ boson decays into two neutrinos.
 
Significant background contributions are estimated primarily from comparisons between data and simulation in specially designed `control regions' (CRs), which have a selection orthogonal to all SRs and aim to enhance a particular background process, while probing a similar event topology. Sufficient data are needed to minimise the statistical uncertainties in the background estimates in the CRs, while the extrapolation from the CR to the SR, evaluated with simulated events, should be as small as possible to reduce the associated systematic uncertainties. Furthermore, CR selection criteria are chosen to minimise potential  contamination by signal. The signal contamination is below 10\% in all CRs for top squark and neutralino mass combinations that have not yet been excluded at 95\% confidence level by previous ATLAS searches~\cite{SUSY-2014-07,SUSY-2016-15,SUSY-2016-16,SUSY-2016-17,EXOT-2016-27,TOPQ-2016-10}.
 
Separate CRs are defined for SRA--B, SRC and SRD, with the observed number of events in each region included in one of the three dedicated binned profile likelihood fits~\cite{Baak:2014wma} of the analysis (SRA--B fit, SRC fit, SRD fit). The CRs are defined so that all CRs associated with a given signal region are orthogonal to the other CRs for that specified region. Partial overlaps remain possible between regions included in different fits. Each likelihood function is built as the product of Poisson probability density functions, describing the observed and expected numbers of events in the control regions. Additional terms, constrained by Gaussian probability density functions accounting for MC statistics and common systematic uncertainties (discussed in Section~\ref{sec:Systematics}) between the control and signal regions and their correlations, are included and treated as nuisance parameters in the fitting procedure.
 
Control regions targeting the \Zjets, \ttbar, \Wjets, single-top and \ttZ\ backgrounds are included in the SRA--B fit, while for the SRC fit only a \ttbar\ control region is defined. For the SRD fit, control regions are defined for \Zjets, \ttbar, and \Wjets\ backgrounds. For each fit (SRA--B, SRC, and SRD), the normalisations of these backgrounds are determined simultaneously in order to best match the observed data in each control region, including contributions from all backgrounds (background-only fit). No observed or expected number of events in the signal regions is considered at this stage. In cases where there are multiple control regions for one background in one fit, the fit yields one normalisation which best fits all regions.
 
Contributions from all-hadronic \ttbar\ and multijet production are found to be negligible in all signal regions except for SRC, where they are subdominant. These backgrounds are estimated from data collected by single-jet triggers using a jet smearing procedure described in Ref.~\cite{jetSmearing} and are fixed in the fit, with an uncertainty assigned to them (discussed in Section~\ref{sec:Systematics}). The contributions from all other background processes (diboson, $tZ$, $\ttbar H$, $\ttbar W$, $tWZ$) are less than 15\% of the total SM background expectations and are fixed at the value expected from the simulation, using the most accurate theoretical cross sections available, while their uncertainties are included as additional nuisance parameters in the fit.
In the following, the multijet, diboson, $tZ$, $\ttbar H$, $\ttbar W$, and $tWZ$ backgrounds are grouped together and referred to as `other'.
 
Validation regions (VRs) are defined for the major sources of background in each signal region such that they are orthogonal to the control regions and the signal regions. They usually suffer from a higher signal contamination (up to 20\%) than the CRs, but probe a kinematic region which is closer to that of the SRs. The background normalisation factors from the simultaneous fit are applied to their respective backgrounds and compared with data in each VR to verify good agreement and that the simultaneous fit is well-behaved.
 
\begin{figure}[htpb]
\begin{center}
\subfloat[]{\includegraphics[width=0.45\textwidth]{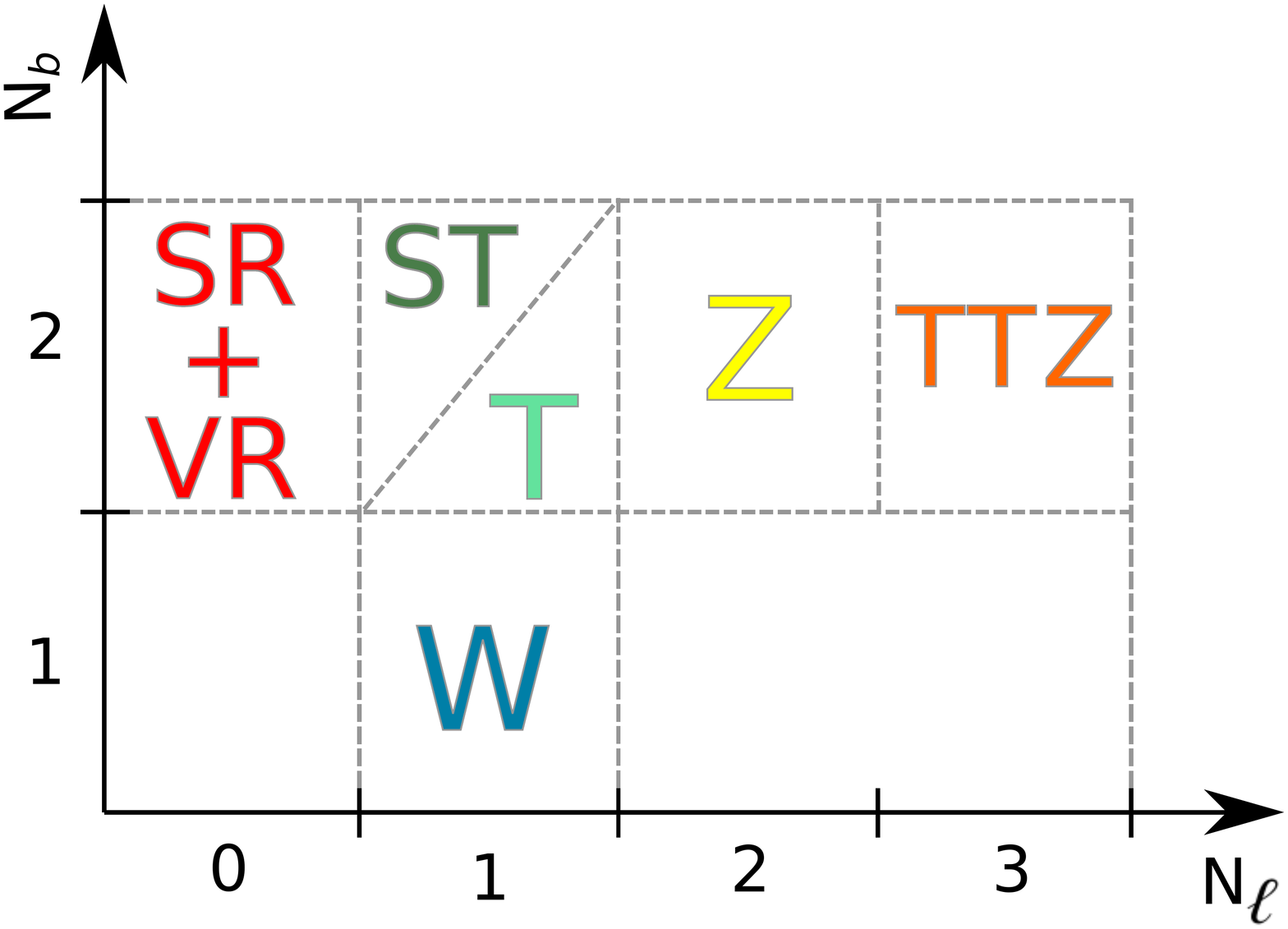}}
\subfloat[]{\includegraphics[width=0.45\textwidth]{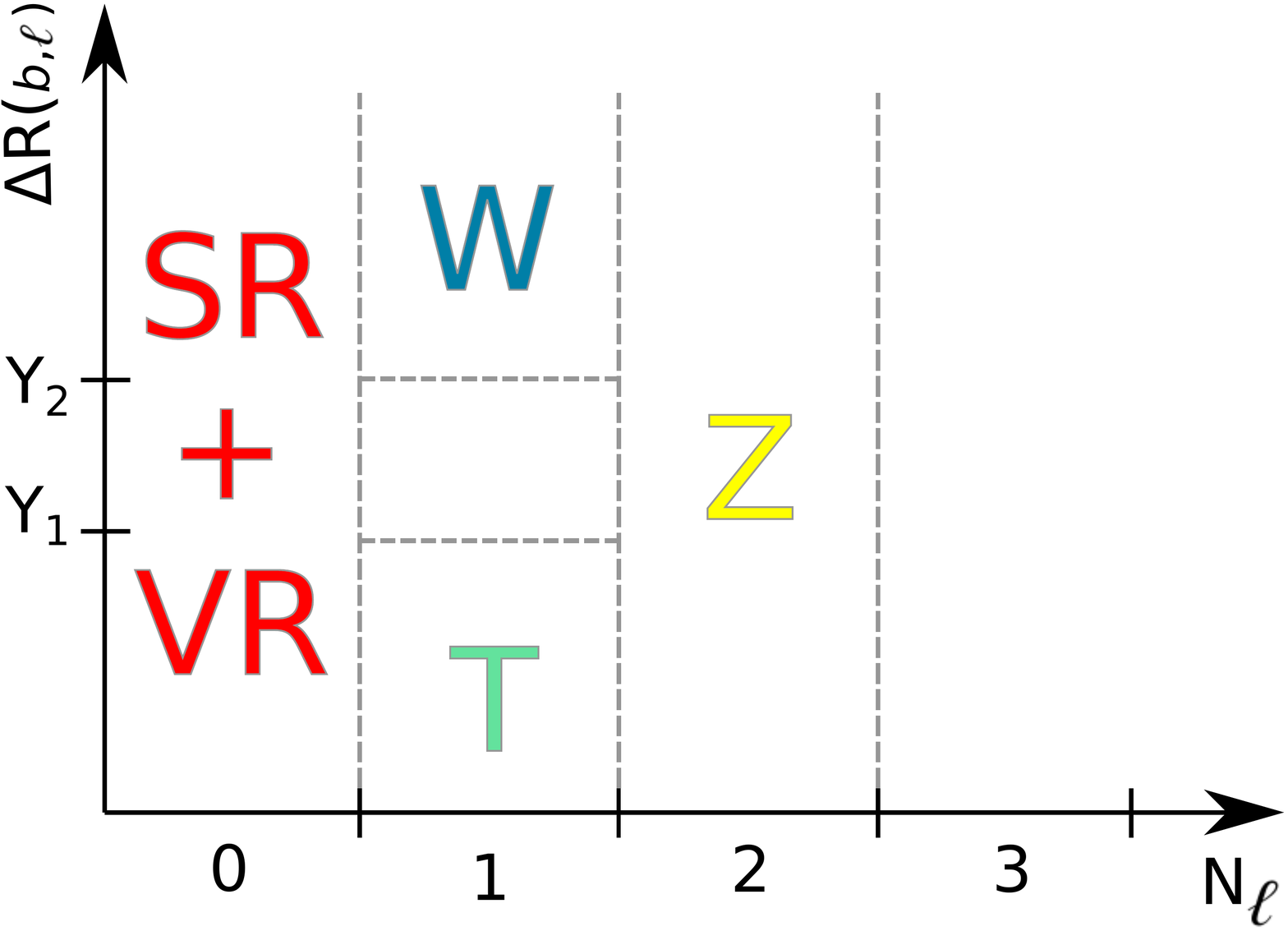}}
\caption{A summary of the background control region strategy used in the (a) SRA--B and (b) SRD fits. The orthogonality between the \Zjets\ (Z), \ttZ\ (TTZ), \ttbar\ (T), \Wjets\ (W), and single-top (ST) backgrounds' control regions and the signal and validation regions (SR+VR) included in the SRA--B fit rely on the number of leptons, $N_{\ell}$, and the number of $b$-tagged jets, $N_{b}$. T and ST are made orthogonal by selecting either low-\pt\ ($<20$~GeV) or high-\pt\ ($>27$~GeV) leptons, respectively. The orthogonality between the \Zjets\ (Z), \ttbar\ (T), and \Wjets\ (W) backgrounds' control regions and the signal and validation regions (SR+VR) included in the SRD fit relies on $N_{\ell}$ and, for $N_{\ell} =1$, the angular distance between the lepton and the closest $b$-tagged jet ($b$-tagged track-jet in CRWD0), \drblep. Placeholders for the values of \drblep\ are shown as $Y_1$ and $Y_2$ and vary in different SRD regions depending on $N_{b}$. Additional selections not appearing on the sketches ensure orthogonality between the SR and the VR. Additional extrapolations from CRs to SRs in other kinematic quantities not necessarily shown in this sketch are region-specific and detailed in the text.}
\label{fig:CRSketch}
\end{center}
\end{figure}

Detailed CR definitions for the estimation of \Zjets\ (CRZ), \ttZ\ (CRTTZ), \ttbar\ (CRT), \Wjets\ (CRW), and single-top (CRST) backgrounds are described in the following subsections, while a summary of the control region strategy in the SRA--B and SRD fits is shown in Figure~\ref{fig:CRSketch}. The strategy for SRC only involves one control region (extrapolating from an electron or muon multiplicity of zero in the SR to an electron or muon multiplicity of one in the CR) and one validation region (extrapolating over \dPhiISRMET) for the dominant \ttbar\ background.
 
\FloatBarrier
 
\subsection{\Zjets\ background estimation}
 
The normalisation of the simulation of $Z\to\nu\bar{\nu}$ produced in association with heavy-flavour jets is estimated from $Z\to e^{+}e^{-}$ and $Z\to\mu^{+}\mu^{-}$ events produced in association with heavy-flavour jets, which is the strategy adopted for SRA--B (CRZAB) and SRD (CRZD). Data events passing a single-electron or single-muon trigger are considered, and events with two control electrons or two control muons with opposite charge are selected. In CRZAB (CRZD), $\pT^\ell > 27, 20$~GeV\ ($\pT^\ell > 30, 20$~GeV) is required for the leading and subleading leptons, respectively, which must also have an invariant mass within 10~GeV\ of the $\Zboson$ boson mass, $m_{\Zboson}=91$~GeV. Events with $\MET\ > 50$~GeV\ ($\MET\ > 70$~GeV) in CRZAB (CRZD) are discarded in order to reject $t\bar{t}$ events. The transverse momenta of the selected electrons or muons are vectorially added to the \ptmiss\ to mimic the $\Znunujets$ decays in the SRs, forming the quantity \metprime. High-\pT\ $\Zboson$ bosons are then effectively selected by requiring large \metprime.
 
\begin{figure}[htpb]
\begin{center}
\subfloat[]{\includegraphics[width=0.5\textwidth]{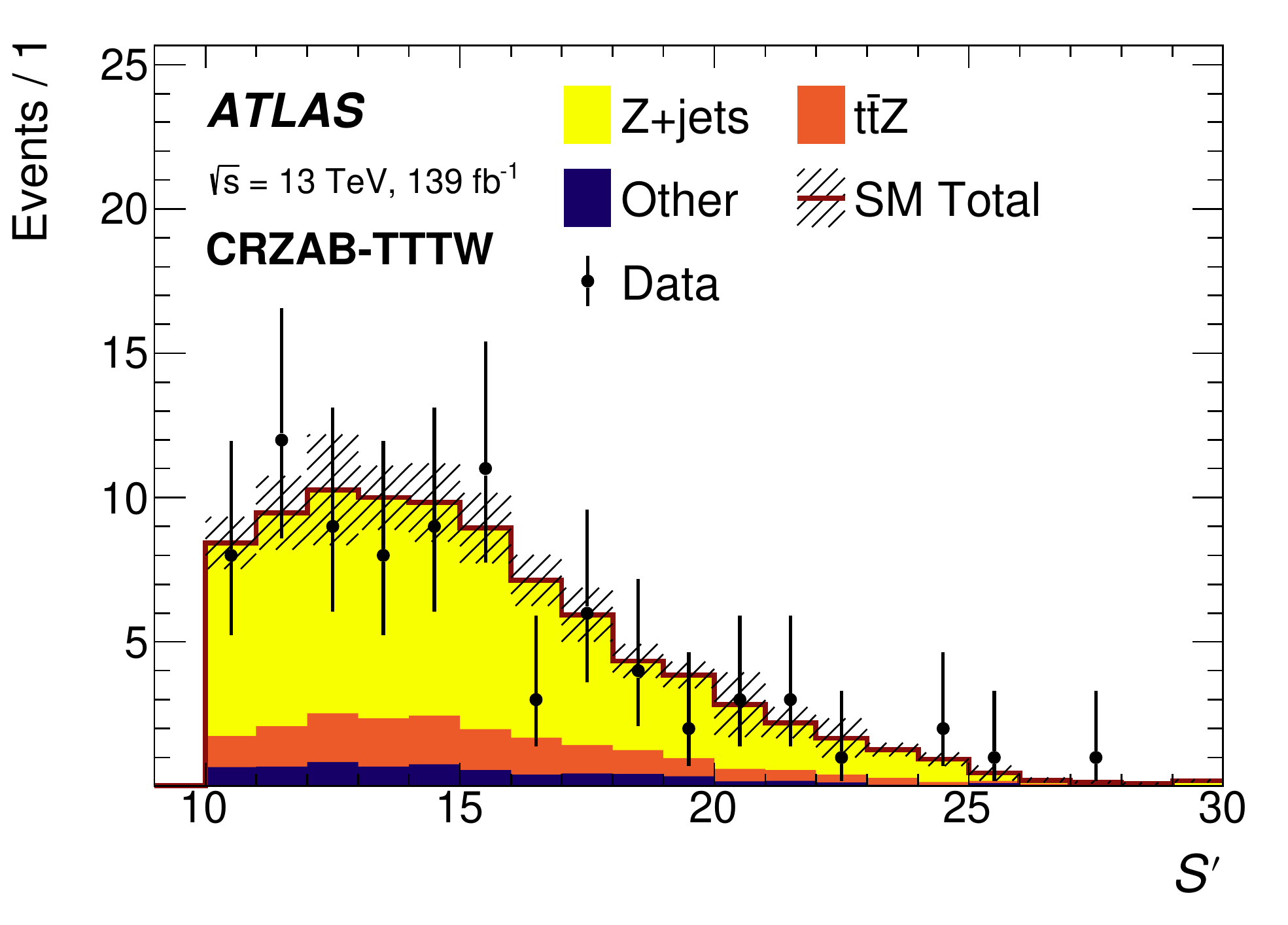}}
\subfloat[]{\includegraphics[width=0.5\textwidth]{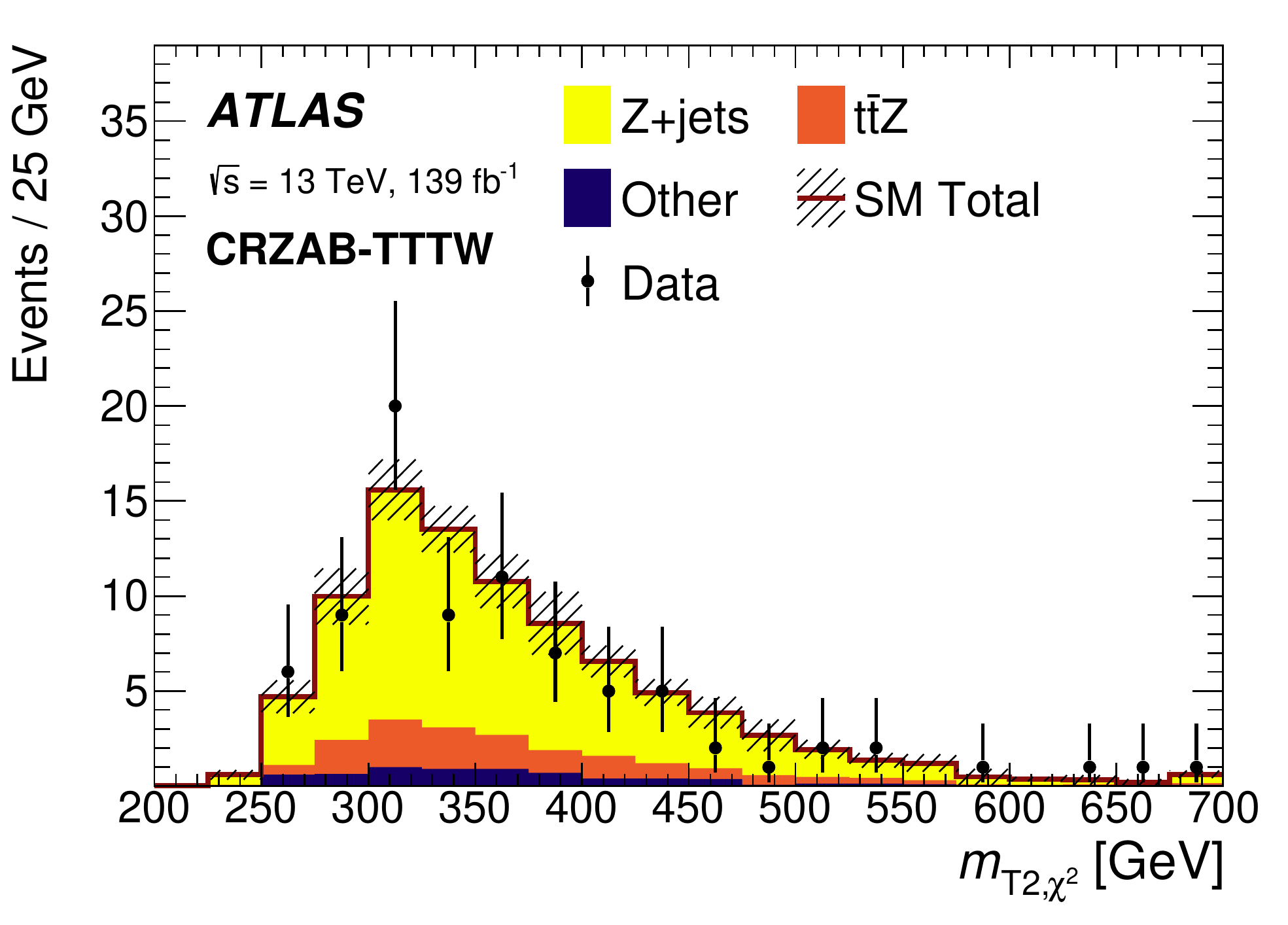}}\\
\subfloat[]{\includegraphics[width=0.5\textwidth]{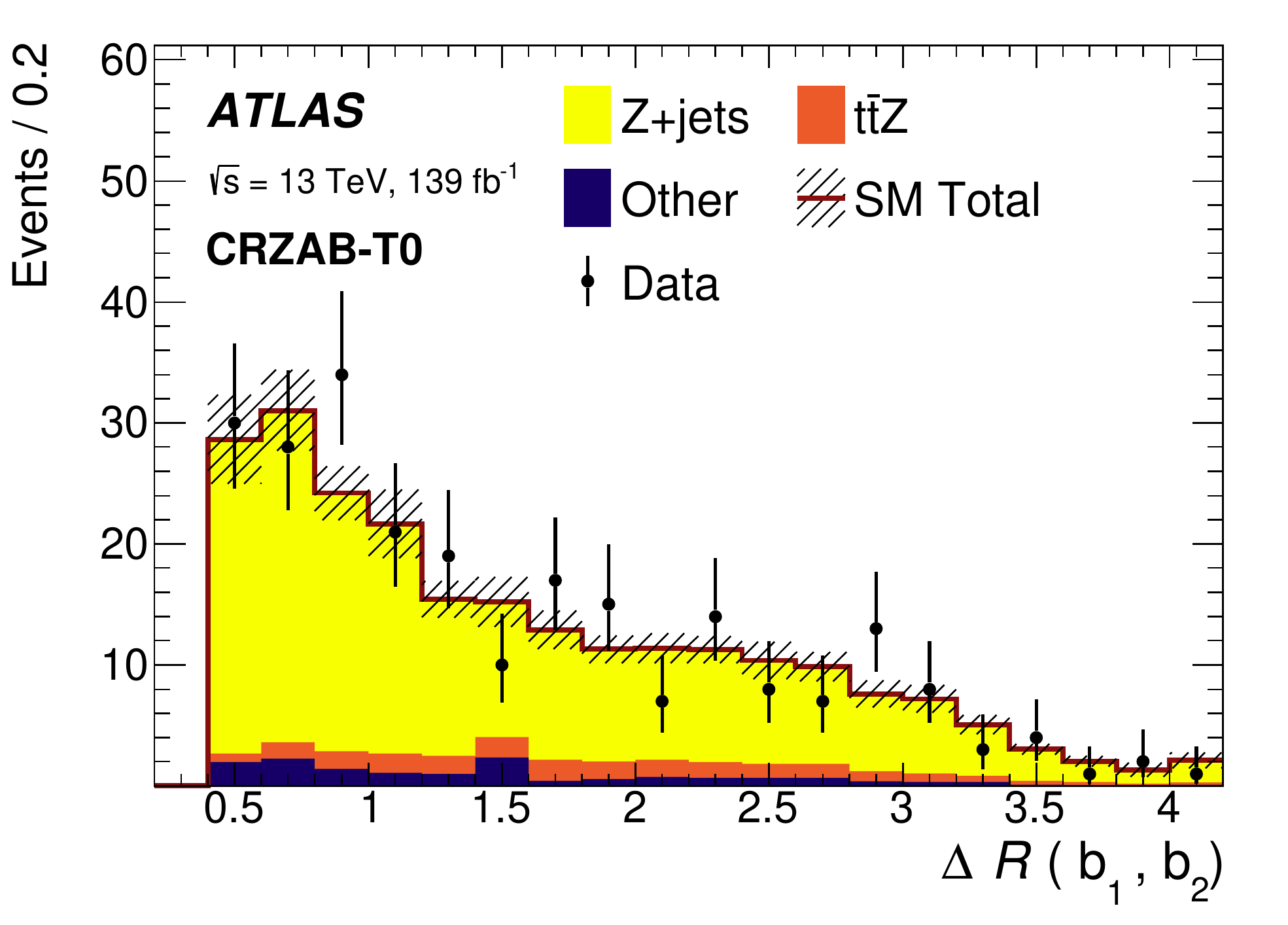}}
\subfloat[]{\includegraphics[width=0.5\textwidth]{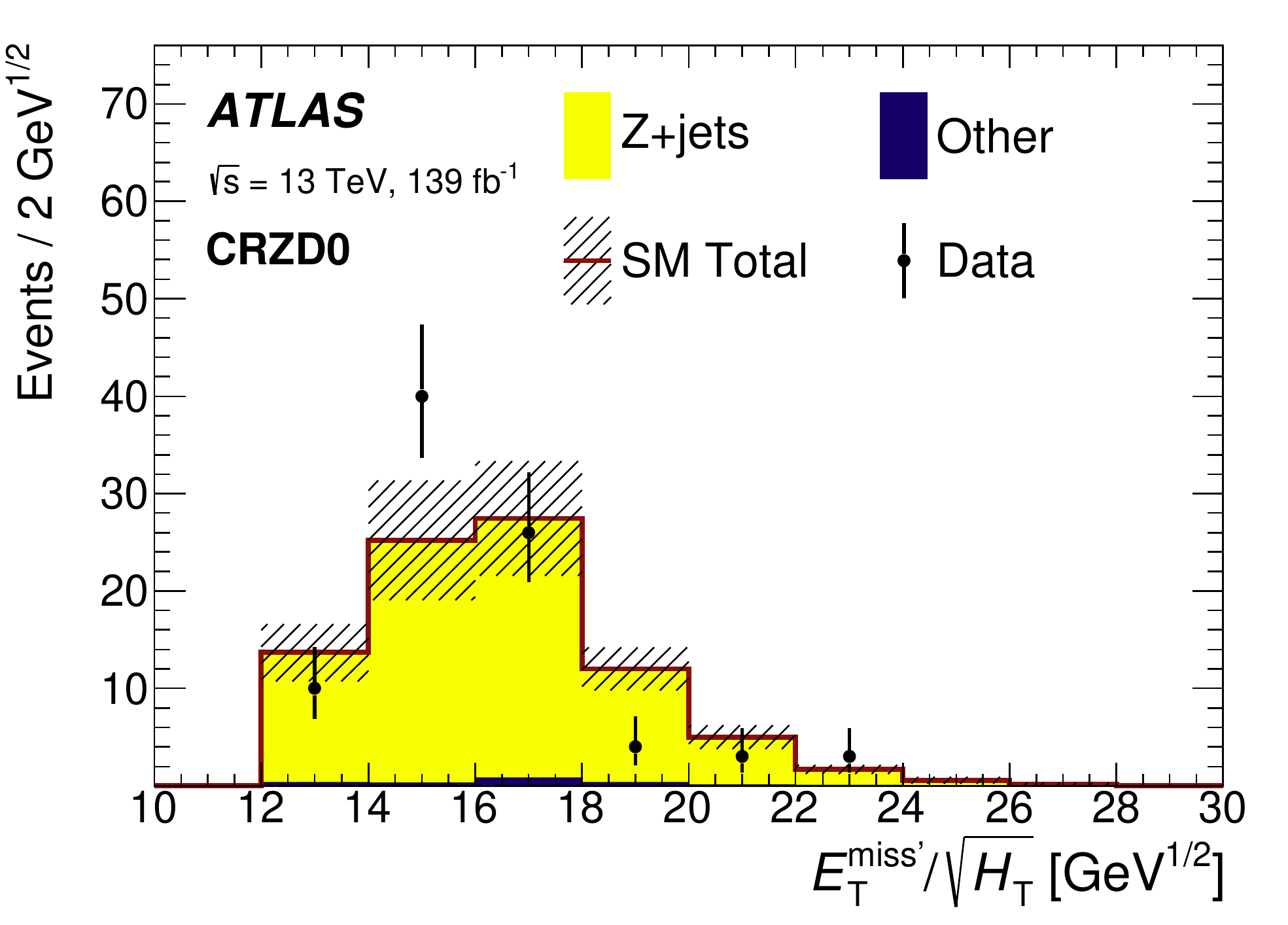}}
\caption{Distributions illustrating the level of agreement between data (points) and the SM expectation (stacked histograms, after simultaneously fitting to all backgrounds) in several \Zjets\  control regions: (a) \objmetsigprime\ and (b) \mttwochisquare\ for CRZAB-TTTW, (c) \drbjetbjet\ for CRZAB-T0, and (d) \htsigprime\ for CRZD0.
The hatched uncertainty band around the SM expectation includes the combination of MC statistical, theory-related and detector-related systematic uncertainties.
The rightmost bin in each plot includes all overflows.}
\label{fig:CRZ}
\end{center}
\end{figure}

Recalculated quantities that use \metprime\ instead of \met\ are identified by the addition of a prime (e.g.\ \mtbminprime). Where possible, the CR selection criteria are identical to the criteria used in the signal region; however, the criteria for key variables such as \mtbminprime\ and \objmetsigprime\ for CRZAB, or \htprimesig\ for CRZD, are loosened to enhance the number of data events in the CR. The $\Zjets$ CR included in the SRA--B (SRD) fit is split into two (three) categories depending on \mantikttwelveone\ (\nbjets), to minimise the extrapolation across the various SR categories. There are only two categories in CRZAB, CRZAB-TTTW (representing the background in the TT and TW signal categories) and CRZAB-T0, due to the limited number of data events. The detailed set of selection criteria for the \Zjets\ CRs are presented in Table~\ref{tab:CRZ}; representative distributions for CRZ variables that have looser requirements than in the SRs are shown in Figure~\ref{fig:CRZ}.
 
% The next lines are included from the .//tex/CRZTable.tex input file
 
\begin{table}[htpb]
\caption{Selection criteria for the $\Zjets$ control regions. The defining extrapolation for these control regions is over the number of leptons; two electrons or muons ($\ell$) from $\Zboson$ decays are required, compared with zero leptons in the signal regions. A dash indicates that no selection is applied. Variables for which the signal and control region requirements differ are highlighted by a thick border around the cell that contains the requirement. Requirements are made on the following variables in the signal regions but have no equivalent requirement in the control regions: $\tau$-veto, \mantikteightzero, \bantikttwelvezero, \bantikttwelveone, \drbjetbjet, \mtbmax, \mttwochisquare, \mettrk, and \dphimettrk.}
\begin{center}
\def\arraystretch{1.4}
\scriptsize
\begin{tabular}{c||c|c|c|c|c}
\hline \hline
\textbf{ Variable/CR }               & \textbf{ CRZAB-TTTW}                             & \textbf{ CRZAB-T0 }                       & \textbf{ CRZD0}      & \textbf{ CRZD1} & \textbf{ CRZD2}            \\
\hline\hline
Trigger                    & \multicolumn{5}{c}{single electron or muon}                                                                       \\
\hline
Control $\ell$              & \multicolumn{5}{c}{exactly $2$, same flavour / opposite sign}                                                     \\
\hline
Additional baseline $\ell$ & \multicolumn{5}{c}{0}                                                                                             \\
\hline
\minvll                    & \multicolumn{5}{c}{$81$--$101$~GeV}                                                                                  \\
\hline
\met                       & \multicolumn{2}{c|}{$<50$~GeV}          & \multicolumn{3}{c}{$<70$~GeV}                                           \\
\hline
$\pt^{\ell}$               & \multicolumn{2}{c|}{$>27$, $>20$~GeV }  & \multicolumn{3}{c}{$>30$, $>20$~GeV }                                   \\
\hline
\Cline{2-3}
\Cline{5-6}
 
\metprime                  & \multicolumn{2}{!{\vrule width 1pt}c!{\vrule width 1pt}}{$>200$~GeV}         & $>250$~GeV                      & \multicolumn{1}{!{\vrule width 1pt}c!{\vrule width 1pt}}{$>150$~GeV} & \multicolumn{1}{!{\vrule width 1pt}c!{\vrule width 1pt}}{$>200$~GeV}               \\
 
\Cline{2-3}
\Cline{5-6}
\hline
\njets                     & \multicolumn{2}{c|}{$\ge 4$}            & \multicolumn{3}{c}{--}                                                   \\
\hline
\pttwo                     & \multicolumn{2}{c|}{$>80$~GeV}         & \multicolumn{3}{c}{--}                                                   \\
\hline
\ptfour                    & \multicolumn{2}{c|}{$>40$~GeV}         & \multicolumn{3}{c}{--}                                                   \\
\hline
\nbjets                    & \multicolumn{2}{c|}{$\ge 2$}            & exactly 0                       & exactly 1  & $\ge 2$                  \\
\hline
\Cline{2-3}
\mantikttwelvezero         & \multicolumn{2}{!{\vrule width 1pt}c!{\vrule width 1pt}}{$> 80$~GeV}         & \multicolumn{3}{c}{--}                                                   \\
\Cline{2-3}
\hline
\Cline{2-2}
\mantikttwelveone          & \multicolumn{1}{!{\vrule width 1pt}c!{\vrule width 1pt}}{$ > 60$~GeV}                             & $<60$~GeV                       & \multicolumn{3}{c}{--}                 \\
\Cline{2-3}
\hline
\mtbminprime               & \multicolumn{2}{!{\vrule width 1pt}c!{\vrule width 1pt}}{$> 150$~GeV}        & \multicolumn{3}{c}{--}                                                   \\
\hline
\Cline{2-3}
\objmetsigprime            & \multicolumn{2}{!{\vrule width 1pt}c!{\vrule width 1pt}}{$> 10$} & \multicolumn{3}{c}{--}                                                   \\
\hline
\Cline{2-3}
\Cline{5-5}
\ptnonbjetone              & \multicolumn{2}{c|}{--}                  & $>250$~GeV & \multicolumn{1}{!{\vrule width 1pt}c!{\vrule width 1pt}}{$>200$~GeV} & $>250$~GeV                            \\
\Cline{5-5}
\hline
\dphinonbjetmet            & \multicolumn{2}{c|}{--}                  & \multicolumn{3}{c}{$>2.4$}                                              \\
\hline
\nbjetsvr                      & \multicolumn{2}{c|}{--}                  & \multicolumn{2}{c|}{$\ge1$}     & --                                     \\
\hline
\Cline{2-3}
$\dphijetfourmet$              & \multicolumn{2}{!{\vrule width 1pt}c!{\vrule width 1pt}}{--}                  & $> 0.4$                         & \multicolumn{2}{c}{--}                 \\
\Cline{2-3}
\hline
\etabonevr                     & \multicolumn{2}{c|}{--}                  & $< 1.2$                         & \multicolumn{2}{c}{--}                 \\
\hline
$\maxdphinonbjetbjetvr$        & \multicolumn{2}{c|}{--}                  & $>2.2$                          & \multicolumn{2}{c}{--}                 \\
\hline
$\dphibbvr$                    & \multicolumn{2}{c|}{--}                  & $< 2.5$                         & \multicolumn{2}{c}{--}                 \\
\hline
\ptbonevr                  & \multicolumn{2}{c|}{--}                  & $< 50$~GeV                      & $>10$~GeV  & --                        \\
\hline
\ptonevr                   & \multicolumn{3}{c|}{--}                  & $<40$~GeV                       & --                                     \\
\hline
$\mindphijetvrfournonb$        & \multicolumn{3}{c|}{--}                  & $> 1.2$                         & --                                     \\
\hline
\etabone                       & \multicolumn{3}{c|}{--}                  & $< 1.6$                         & --                                     \\
\hline
\Cline{5-5}
$\dphinonbjetbjetone$          & \multicolumn{3}{c|}{--}                  & \multicolumn{1}{!{\vrule width 1pt}c!{\vrule width 1pt}}{$>1.8$} & $>2.2$                                              \\
\Cline{5-5}
 
\hline
\etabtwo                       & \multicolumn{4}{c|}{--}                  & $<1.2$                                                                  \\
\hline
\ptbone                        & \multicolumn{4}{c|}{--}                  & $< 175$~GeV                                                             \\
\hline
$\dphinonbjetbjettwo$          & \multicolumn{4}{c|}{--}                  & $>1.6$                                                                  \\
\hline
\Cline{4-6}
$\htsigprime$                  & \multicolumn{2}{c|}{--}                  & \multicolumn{1}{!{\vrule width 1pt}c!{\vrule width 1pt}}{$> 12\sqrt{\mathrm{GeV}}$}               & \multicolumn{2}{!{\vrule width 0pt}c!{\vrule width 1pt}}{$> 8 \sqrt{\mathrm{GeV}}$} \\
\Cline{4-6}
\hline \hline
\end{tabular}
\end{center}
\label{tab:CRZ}
\end{table}
% End of text imported from the .//tex/CRZTable.tex input file
 
\FloatBarrier
 
\subsection{\ttZ\ background estimation}
 
The SM production of \ttZ, where $\Zboson \to \nu\bar{\nu}$, is a significant source of background in SRA and SRB and is largely irreducible. To estimate this background, a three-lepton (electrons and muons) region is defined, to maximise the purity of $\ttZ$.
 
Events that pass a single-electron or single-muon trigger are selected. The trigger electron or muon must pass the requirements for a control electron or muon and have offline $\pT>27$~GeV. Exactly two additional control leptons (electrons or muons) with $\pT > 20$~GeV\ are required. The sum of the charges of the three leptons is required to equal $1$ or $-1$, while two of the leptons are required to have the same flavour and opposite charge. The pair of same-flavour, opposite-sign leptons that is most consistent with the \Zboson\ boson mass forms the \Zboson\ boson candidate and is required to have an invariant mass satisfying $81\GeV<\minvll<101$~GeV. The \Zboson\ boson candidate is required to have $\pt>200$~GeV. The remaining lepton and the \ptmiss\ are treated as non-$b$-tagged jets in the computation of all jet-related variables (such as \pt), to mimic hadronic $\Wboson$ decays.
 
\begin{figure}[htpb]
\begin{center}
\subfloat[]{\includegraphics[width=0.5\textwidth]{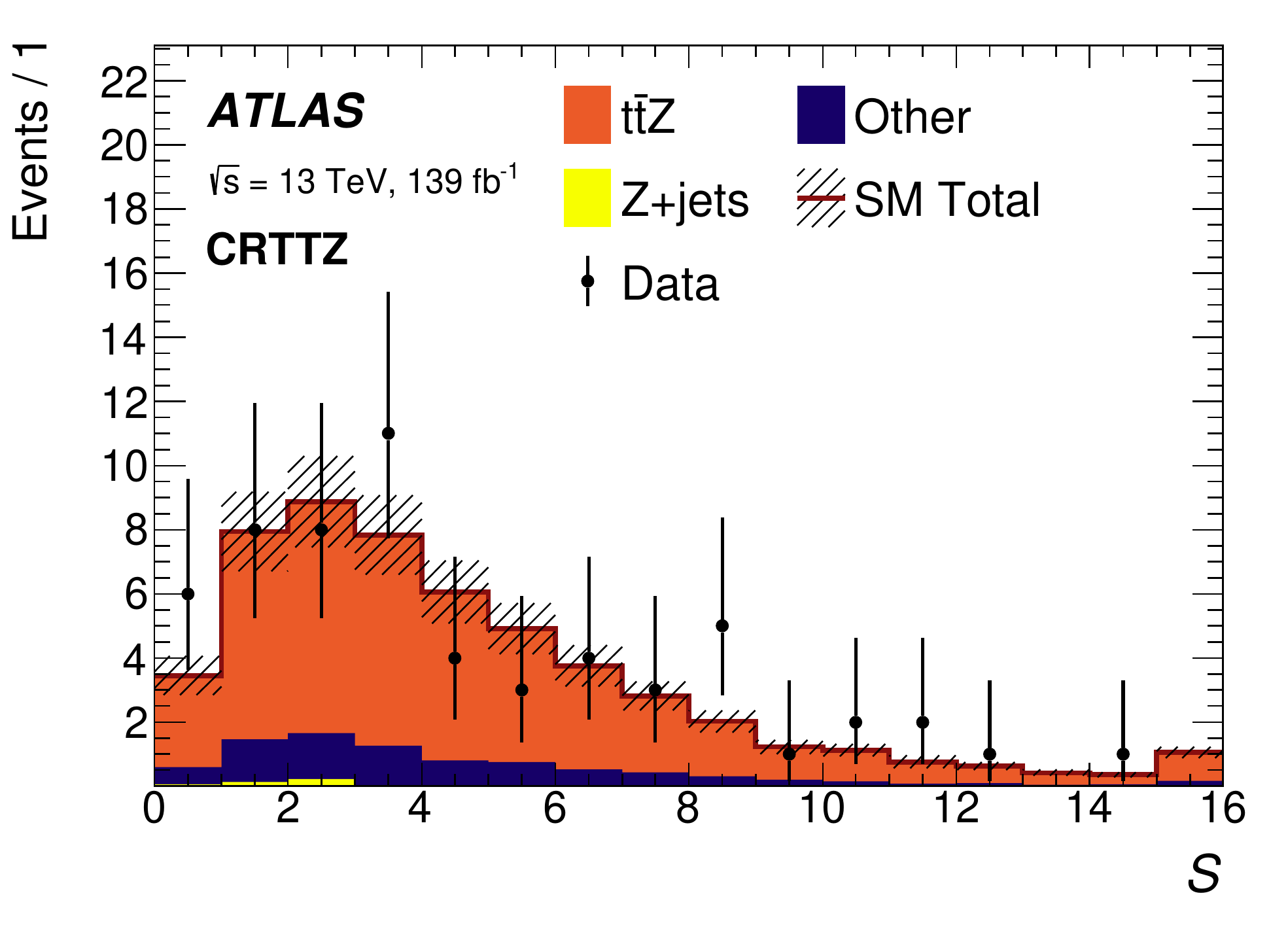}}
\subfloat[]{\includegraphics[width=0.5\textwidth]{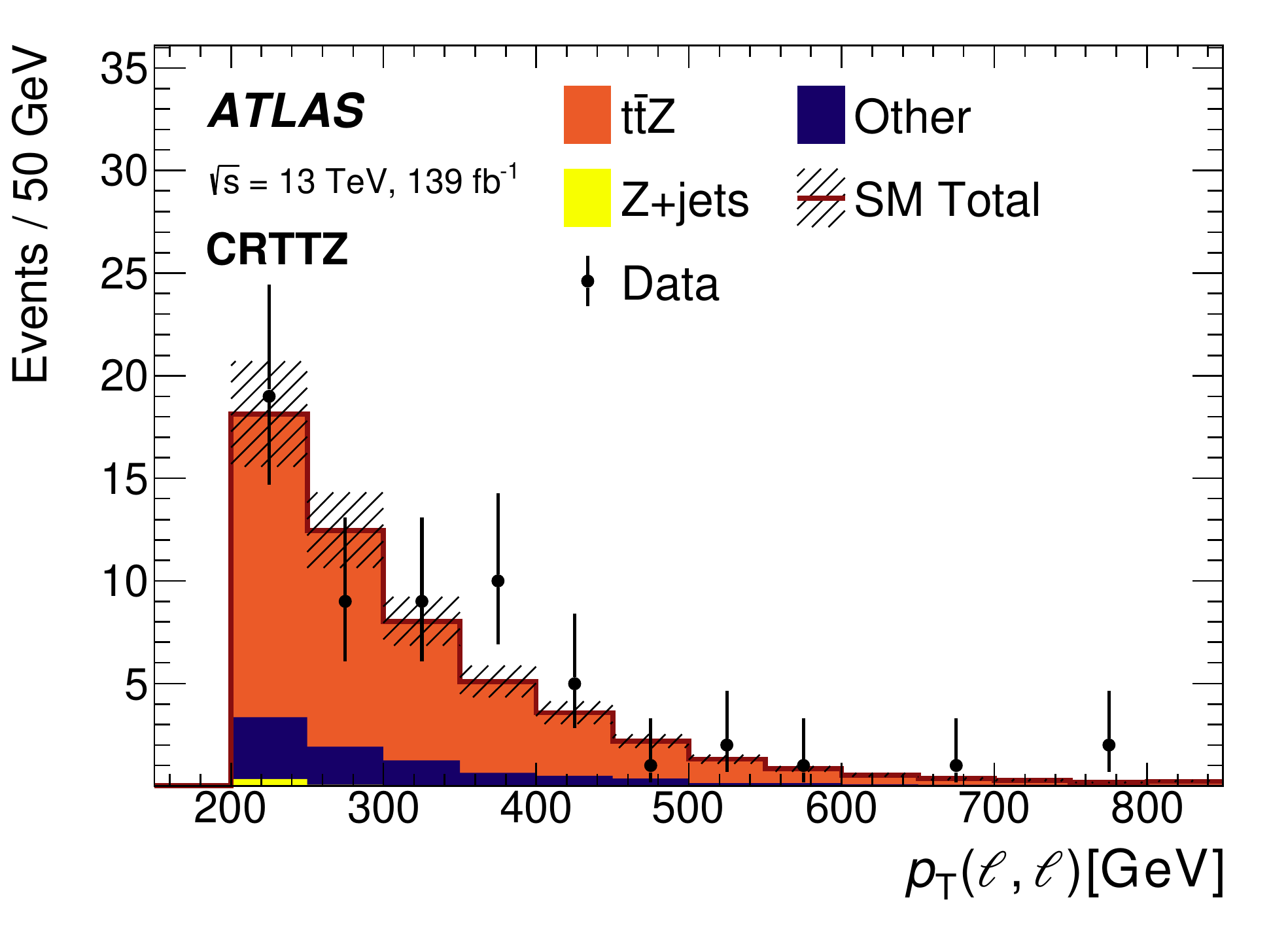}}
\caption{Distributions illustrating the level of agreement between data (points) and the SM expectation (stacked histograms, after simultaneously fitting to all backgrounds) in the \ttZ\  control region: (a) \objmetsig\ and (b) \ptll\ for CRTTZ.
The hatched uncertainty band around the SM expectation includes the combination of MC statistical, theory-related and detector-related systematic uncertainties.
The rightmost bin in each plot includes all overflows.}
\label{fig:CRttZ}
\end{center}
\end{figure}

Four jets are required to be in the event, in addition to the lepton not associated with the \Zboson\ boson candidate and the \ptmiss, and two of the jets are required to be \bjet s. The selection criteria are summarised in Table~\ref{tab:CRTTZ}. Representative distributions for CRTTZ variables that have looser requirements than in the SRs are shown in Figure~\ref{fig:CRttZ}.
 
% The next lines are included from the .//tex/CRTTZTable.tex input file
 
\begin{table}[htpb]
\caption{Selection criteria for the \ttZ\ control region. The defining extrapolation for these control regions is over the number of leptons; three leptons (a combination of electrons and muons) from $\Wboson$ and \Zboson\ decays is required, compared with zero leptons in the signal region. Variables for which the signal and control region requirements differ are highlighted by a thick border around the cell that contains the requirement. Requirements are made on the following variables in SRA and SRB but have no equivalent requirement in the control region: $\dphijetfourmet$, \mtbmin, $\tau$-veto, \mantikttwelvezero, \mantikttwelveone, \mantikteightzero, \bantikttwelvezero, \bantikttwelveone, \drbjetbjet, \mtbmax, \objmetsig, and \mttwochisquare.}
\begin{center}
\def\arraystretch{1.4}
\begin{tabular}{c||c}
\hline \hline
\textbf{ Variable/CR  }                                     & \textbf{ CRTTZ} \\
\hline\hline
Trigger                                           & single electron or muon  \\
\hline
Control $\ell$                                & exactly $3$              \\
\hline
Additional baseline $\ell$                      & 0                        \\
\hline
Sum of muon and electron charges                & $+1$ or $-1$                 \\
\hline
$\ell$ associated with $\Zboson$                & exactly $2$, same flavour / opposite sign             \\
\hline
\minvll                                       & $81$--$101$~GeV            \\
\hline
$\pt^{\ell}$                                  & $>27$, $>20$, $>20$~GeV \\
\hline
\Cline{2-2}
\ptll                                         & \multicolumn{1}{!{\vrule width 1pt}c!{\vrule width 1pt}}{$>200$~GeV}               \\
\Cline{2-2}
\hline
\njets                                          & $\ge 4$                  \\
\nbjets                                       & $\ge 2$                  \\
\hline
\hline
\pttwo\ (including \met\ and non-$Z$ $\ell$)  & $>80$~GeV               \\
\hline
\ptfour\ (including \met\ and non-$Z$ $\ell$) & $>40$~GeV               \\
\hline
\hline
\end{tabular}
\end{center}
\label{tab:CRTTZ}
\end{table}
% End of text imported from the .//tex/CRTTZTable.tex input file
 
\FloatBarrier
 
\subsection{\ttbar, \Wjets, and single-top background estimation}
 
The \ttbar\ background in SRB, SRC, and SRD originates from events where a $\Wboson$ boson decays into a hadronically decaying $\tau$-lepton, where the $\tau$-lepton is either not reconstructed (due to falling below the jet \pt\ threshold of $20$~GeV), or is reconstructed as a jet. In order to model this process in the CRs, events that pass the same \MET\ trigger as the signal region, but also have a control electron or muon, are selected. The electron or muon is used as a proxy for the $\tau$-lepton in the SRs.
 
In SRA and SRB, the hadronically decaying $\tau$-leptons are most likely to have fallen below the jet $\pT>20$~GeV\ requirement, such that for the $\ttbar$ and $\Wjets$ control regions (CRTAB and CRWAB, respectively), exactly one control electron in the range $4.5 < \pT^{e} < 20$~GeV\ or muon in the range $4.0 < \pT^{\mu} < 20$~GeV\ is required. In SRC and SRD, the hadronically decaying $\tau$-leptons have higher {\pT}, such that one control electron or muon with $\pT > 20$~GeV\ is required, and is treated as a non-$b$-tagged jet in the computation of all jet-related variables.
 
\begin{figure}[htpb]
\begin{center}
\subfloat[]{\includegraphics[width=0.5\textwidth]{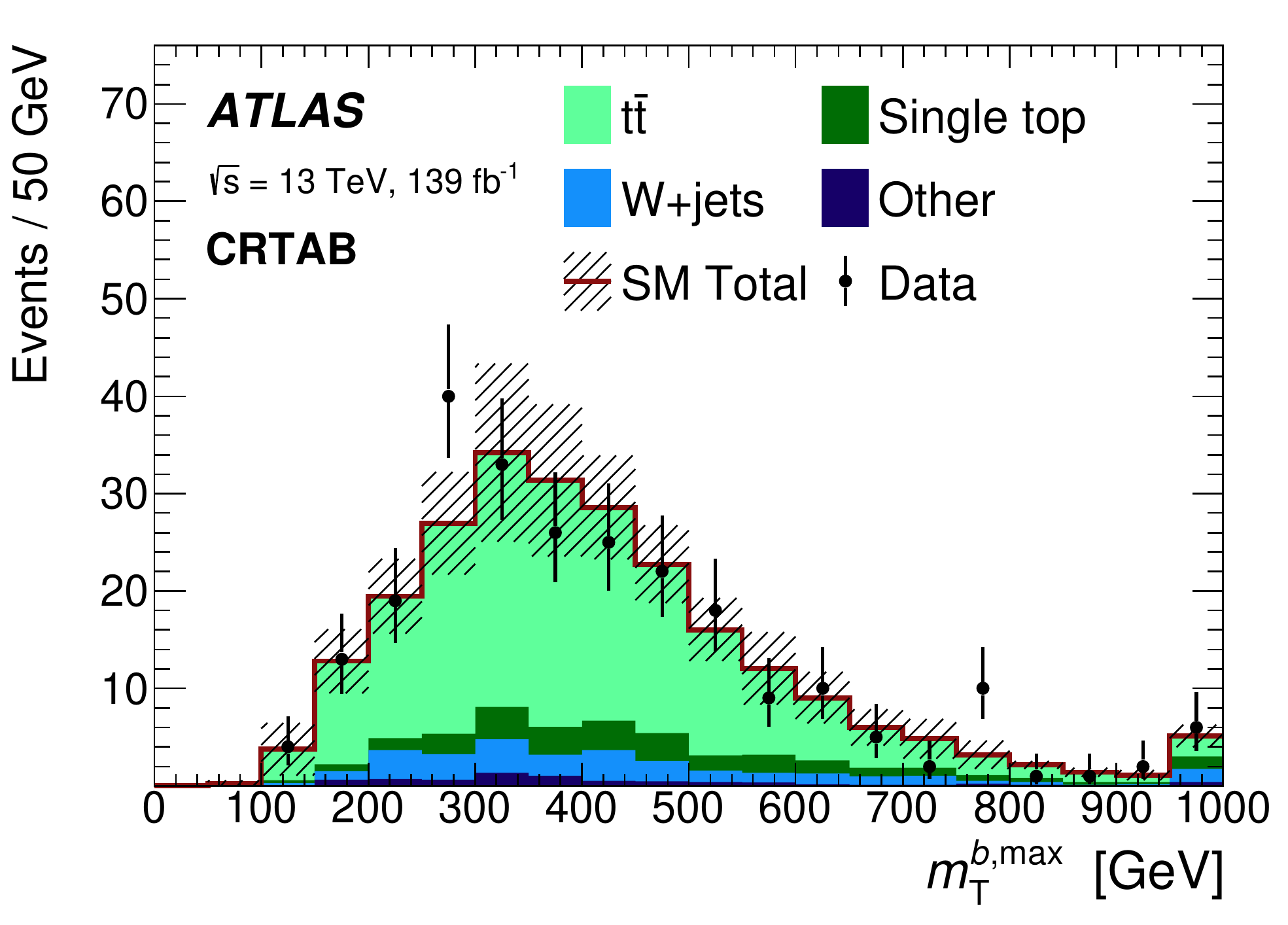}}
\subfloat[]{\includegraphics[width=0.5\textwidth]{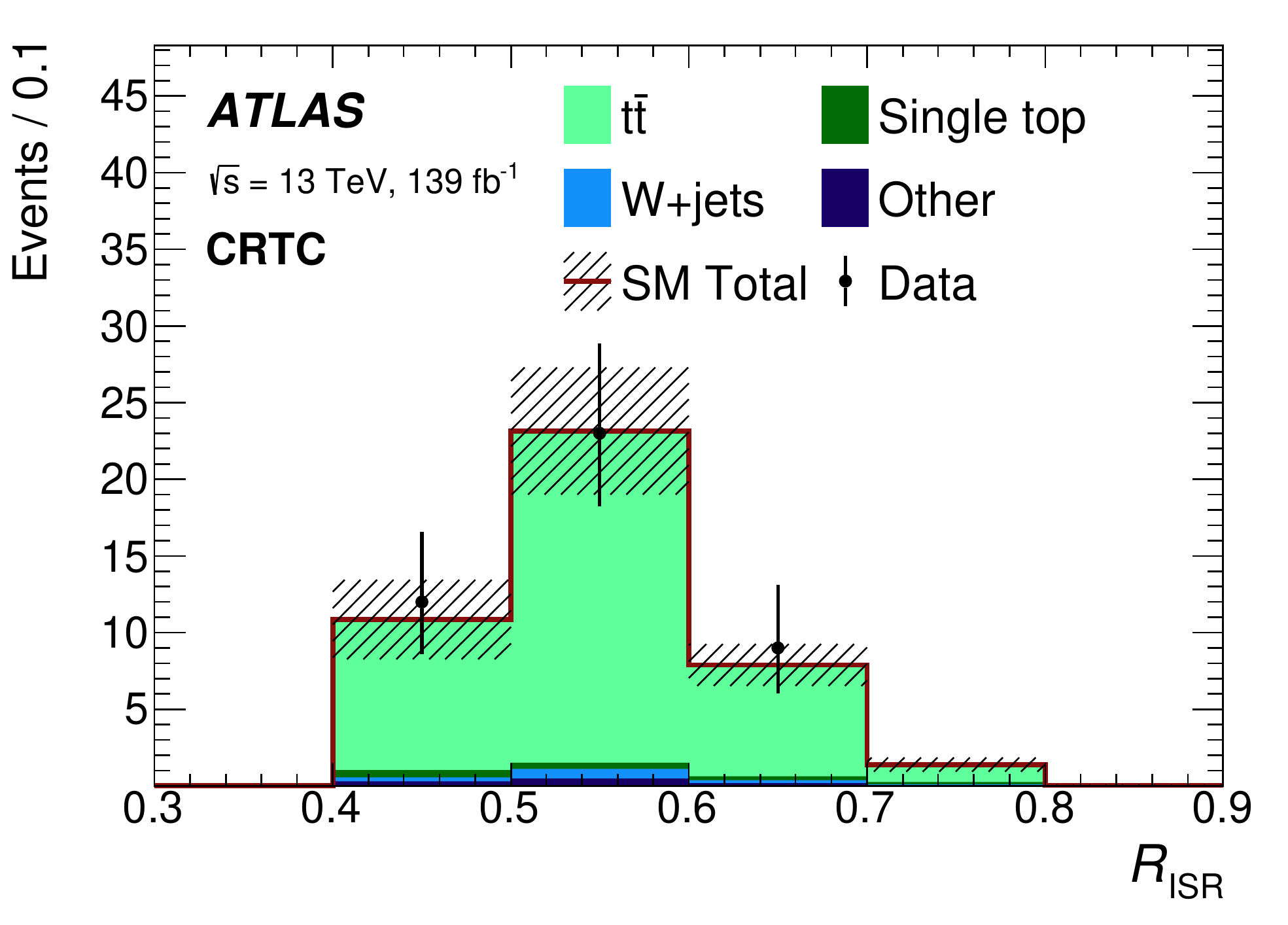}}\\
\subfloat[]{\includegraphics[width=0.5\textwidth]{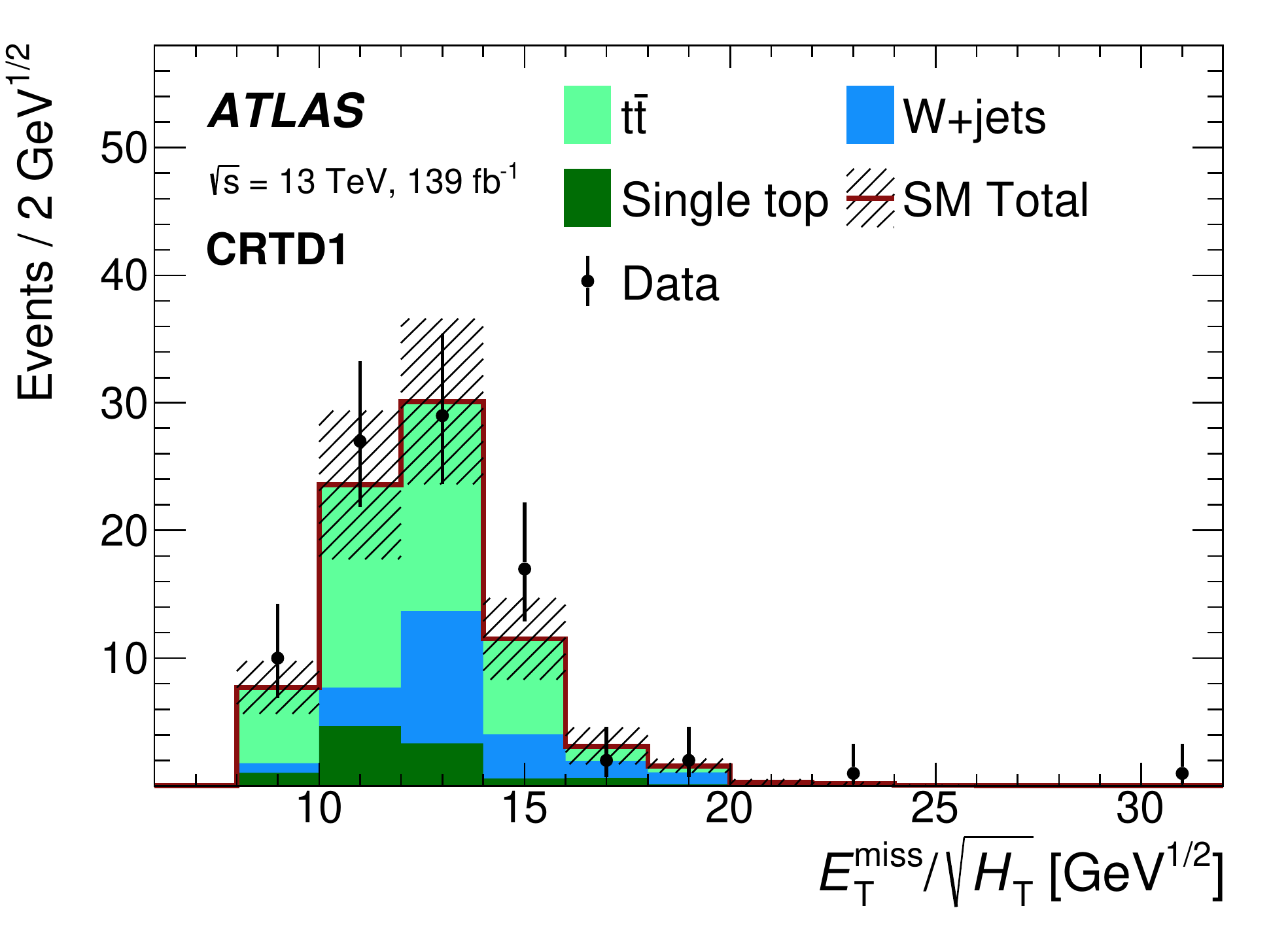}}
\subfloat[]{\includegraphics[width=0.5\textwidth]{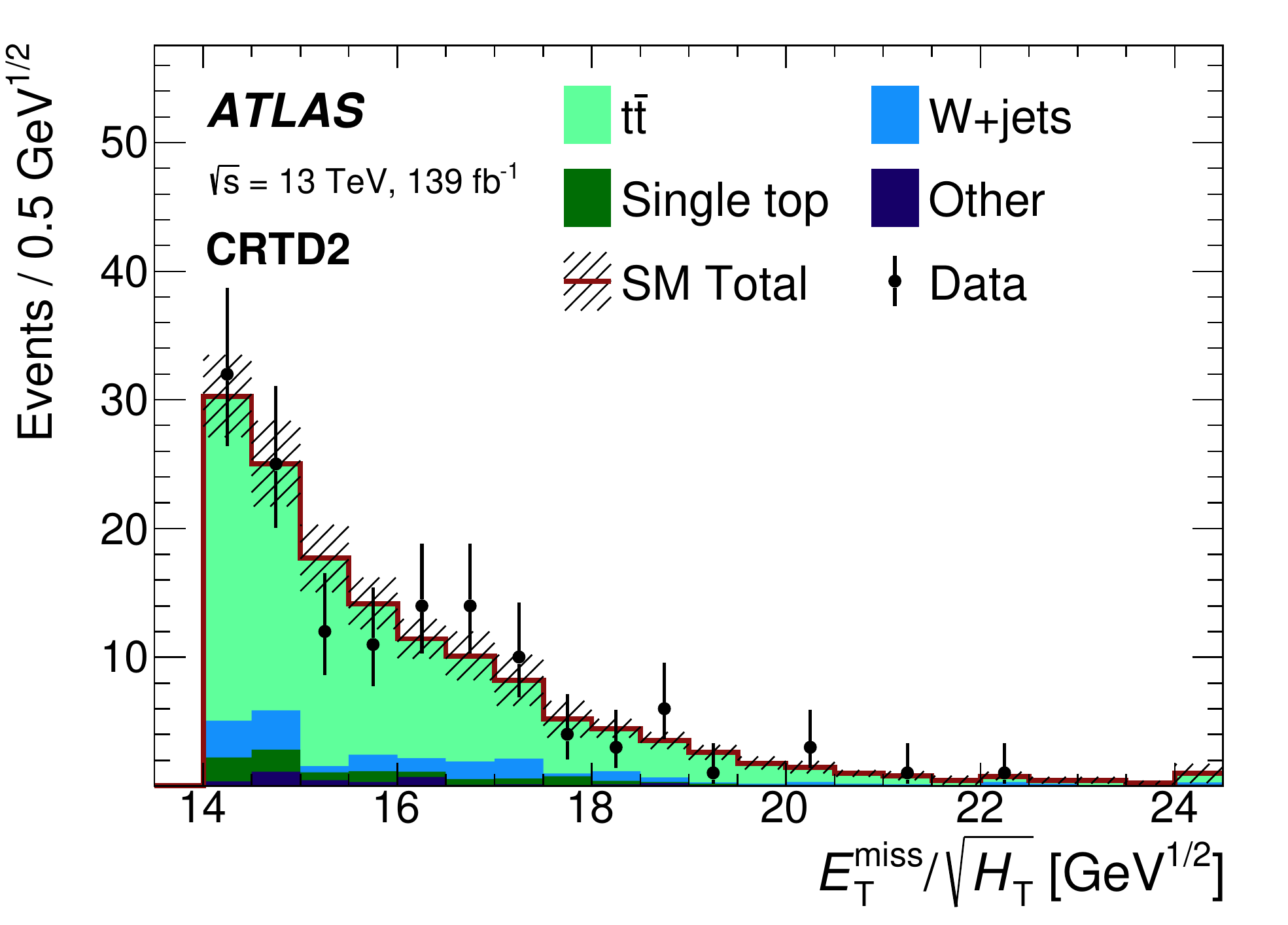}}
\caption{Distributions illustrating the level of agreement between data (points) and the SM expectation (stacked histograms, after simultaneously fitting to all backgrounds) in the \ttbar\ control regions: (a) \mtbmax\ for CRTAB, (b)~\rISR\ for CRTC, and ~\htsig\ for (c) CRTD1 and (d) CRTD2.
The hatched uncertainty band around the SM expectation includes the combination of MC statistical, theory-related and detector-related systematic uncertainties.
The rightmost bin in each plot includes all overflows.}
\label{fig:CRT}
\end{center}
\end{figure}

In the $\ttbar$ control regions (CRTC, CRTD), the angular separation between the electron or muon and the \bjet\ closest to the electron or muon, $\drblep$, is used to enhance the \ttbar\ purity. In CRTD, \drblep\ is also used to ensure orthogonality with the $\Wjets$ control region (CRWD). All $\ttbar$ control regions (CRTAB, CRTC, CRTD) have an upper bound on $\mtlepmet$ to preserve orthogonality between the CRs and the signal regions of other ATLAS ongoing studies in the one-lepton plus missing transverse momentum channel, as well as to reduce potential signal contamination. In addition to the variables used in SRC, CRTC has a $\mV/\mS < 0.75$ requirement, where \mS\ is the variable used in SRC and \mV\ is the invariant mass of all visible objects, which provides additional signal rejection. The $\ttbar$ CR included in the SRD fit is split into two categories (CRTD1 or CRTD2, which require exactly one or at least two $b$-tagged jets, respectively) to minimise the extrapolation across the various SR categories. The various \ttbar\ control regions designed for the analysis are defined in Table~\ref{tab:CRT}. Representative distributions are shown in Figure~\ref{fig:CRT}.
 
% The next lines are included from the .//tex/CRTTable.tex input file
 
\begin{table}[htpb]
\caption{Selection criteria for the $\ttbar$ control regions. The defining extrapolation for these control regions is over the number of leptons; one electron or muon ($\ell$) from $\Wboson$ decays is required, compared with zero leptons in the signal region. A dash indicates that no selection is applied. Variables for which the signal and control region requirements differ are highlighted by a thick border around the cell that contains the requirement. Requirements are made on the following variables in the signal regions but have no equivalent requirement in the control regions: \rISR, $\tau$-veto, \mantikteightzero, \bantikttwelvezero, \bantikttwelveone, \mtbmax, \mttwochisquare.}
\begin{center}
\def\arraystretch{1.9}
\tiny
\begin{tabular}{c||c|c|c|c}
\hline \hline
\textbf{ Variable/CR }                 & \textbf{ CRTAB }                                 & \textbf{ CRTC }            & \textbf{ CRTD1} & \textbf{ CRTD2}                 \\
\hline\hline
Trigger                      & \multicolumn{4}{c}{\met}                                                                   \\
\hline
\met                         & \multicolumn{4}{c}{$>250$ GeV}                                                             \\
\hline
Control $\ell$                & \multicolumn{4}{c}{exactly $1$}                                                            \\
\hline
Additional baseline $\ell$ & \multicolumn{4}{c}{0}                                                                      \\
\hline
$\pt^{\ell}$                 & $4.5\,(4.0) < \pt^{e\,(\mu)} < 20$ GeV & \multicolumn{3}{c}{$\pt>20$ GeV}                  \\
\hline
\mtlepmet                    & $<120$ GeV                             & $<100$ GeV       & \multicolumn{2}{c}{$<120$ GeV} \\
\hline
\njets                       & \multicolumn{2}{c|}{$\ge 4$}           & $\ge 3$          & --                             \\
\hline
\pttwo                       & \multicolumn{2}{c|}{$>80$~GeV}        & \multicolumn{2}{c}{--}                            \\
\hline
\ptfour                    & \multicolumn{2}{c|}{$>40$~GeV}        & \multicolumn{2}{c}{--}                            \\
\hline
\nbjets                      & \multicolumn{2}{c|}{$\ge 2$}           & exactly 1        & $\ge 2$                        \\
\hline
\dphijetfourmet            & $>0.4$                                 & \multicolumn{3}{c}{--}                            \\
\hline
\mantikttwelvezero           & $>120$ GeV                             & \multicolumn{3}{c}{--}                            \\
\hline
\Cline{2-2}
\mtbmin                    & \multicolumn{1}{!{\vrule width 1pt}c!{\vrule width 1pt}}{$>150$ GeV}                             & \multicolumn{3}{c}{--}                            \\
\Cline{2-2}
\hline
\Cline{2-2}
\drbjetbjet                & \multicolumn{1}{!{\vrule width 1pt}c!{\vrule width 1pt}}{$>1.4$}                                 & \multicolumn{3}{c}{--}                            \\
\Cline{2-2}
\hline
\Cline{2-3}
\objmetsig                   & \multicolumn{1}{!{\vrule width 1pt}c!{\vrule width 1pt}}{$>14$}                      & \multicolumn{1}{!{\vrule width 0pt}c!{\vrule width 1pt}}{$>5$} & \multicolumn{2}{c}{--}         \\
\Cline{2-3}
\hline
\Cline{3-3}
\dphijettwomet               & --                                     & \multicolumn{1}{!{\vrule width 1pt}c!{\vrule width 1pt}}{$>0.2$}           & \multicolumn{2}{c}{--}         \\
\Cline{3-3}
\hline
\nJetS                     & --                                     & $\ge 4$          & \multicolumn{2}{c}{--}         \\
\hline
\nBJetS                    & --                                     & $\ge 2$          & \multicolumn{2}{c}{--}         \\
\hline
\pTISR                     & --                                     & $>400$ \gev      & \multicolumn{2}{c}{--}         \\
\hline
\Cline{3-3}
\pTSBOne                   & --                                     & \multicolumn{1}{!{\vrule width 1pt}c!{\vrule width 1pt}}{$>40\gev$}        & \multicolumn{2}{c}{--}         \\
\Cline{3-3}
\hline
\pTSFour                   & --                                     & $>50$ \gev       & \multicolumn{2}{c}{--}         \\
\hline
\mS                        & --                                     & $>400\gev$       & \multicolumn{2}{c}{--}         \\
\hline
\dPhiISRMET                & --                                     & $>3.0$           & \multicolumn{2}{c}{--}         \\
\hline
\mV/\mS                    & --                                     & $<0.75$          & \multicolumn{2}{c}{--}         \\
\hline
\drblep                    & --                                     & $<2.0$           & \multicolumn{2}{c}{$<1.8$}     \\
\hline
$\mettrk$                 & --                                     & \multicolumn{2}{c}{$>30$ GeV}                     \\
\hline
\dphimettrk                & --                                     & \multicolumn{2}{c}{$< \pi/3$}                     \\
\hline
\ptnonbjetone      & \multicolumn{2}{c|}{--}                & \multicolumn{2}{c}{$>250$ GeV}                    \\
\hline
\dphinonbjetmet            & \multicolumn{2}{c|}{--}                & \multicolumn{2}{c}{$>2.4$}                        \\
\hline
$\dphinonbjetbjetone$   & \multicolumn{2}{c|}{--} & \multicolumn{2}{c}{$>2.2$}           \\
\hline
\nbjetsvr                    & \multicolumn{2}{c|}{--}                & $\ge1$           & --                             \\
\hline
\ptbonevr                    & \multicolumn{2}{c|}{--}                & $>10$ GeV        & --                             \\
\hline
\ptonevr                     & \multicolumn{2}{c|}{--}                & $<40$ GeV        & --                             \\
\hline
$\mindphijetvrfournonb$      & \multicolumn{2}{c|}{--}                & $> 1.2$          & --                             \\
\hline
\etabone                     & \multicolumn{2}{c|}{--}                & $< 1.6$          & --                             \\
\hline
\Cline{4-5}
$\htsig$                & \multicolumn{2}{c|}{--} & \multicolumn{1}{!{\vrule width 1pt}c!{\vrule width 1pt}}{$> 8\sqrt{\mathrm{GeV}}$} & \multicolumn{1}{!{\vrule width 0pt}c!{\vrule width 1pt}}{$> 14\sqrt{\mathrm{GeV}}$} \\
\Cline{4-5}
\hline
\etabtwo                & \multicolumn{3}{c|}{--} & $<1.2$                               \\
\hline
\ptbone               & \multicolumn{3}{c|}{--} & $< 175$ GeV                          \\
\hline
$\dphinonbjetbjettwo$ & \multicolumn{3}{c|}{--} & $>1.6$                               \\
\hline \hline
\end{tabular}
\end{center}
\label{tab:CRT}
\end{table}
% End of text imported from the .//tex/CRTTable.tex input file
 
\begin{figure}[htpb]
\begin{center}
\subfloat[]{\includegraphics[width=0.5\textwidth]{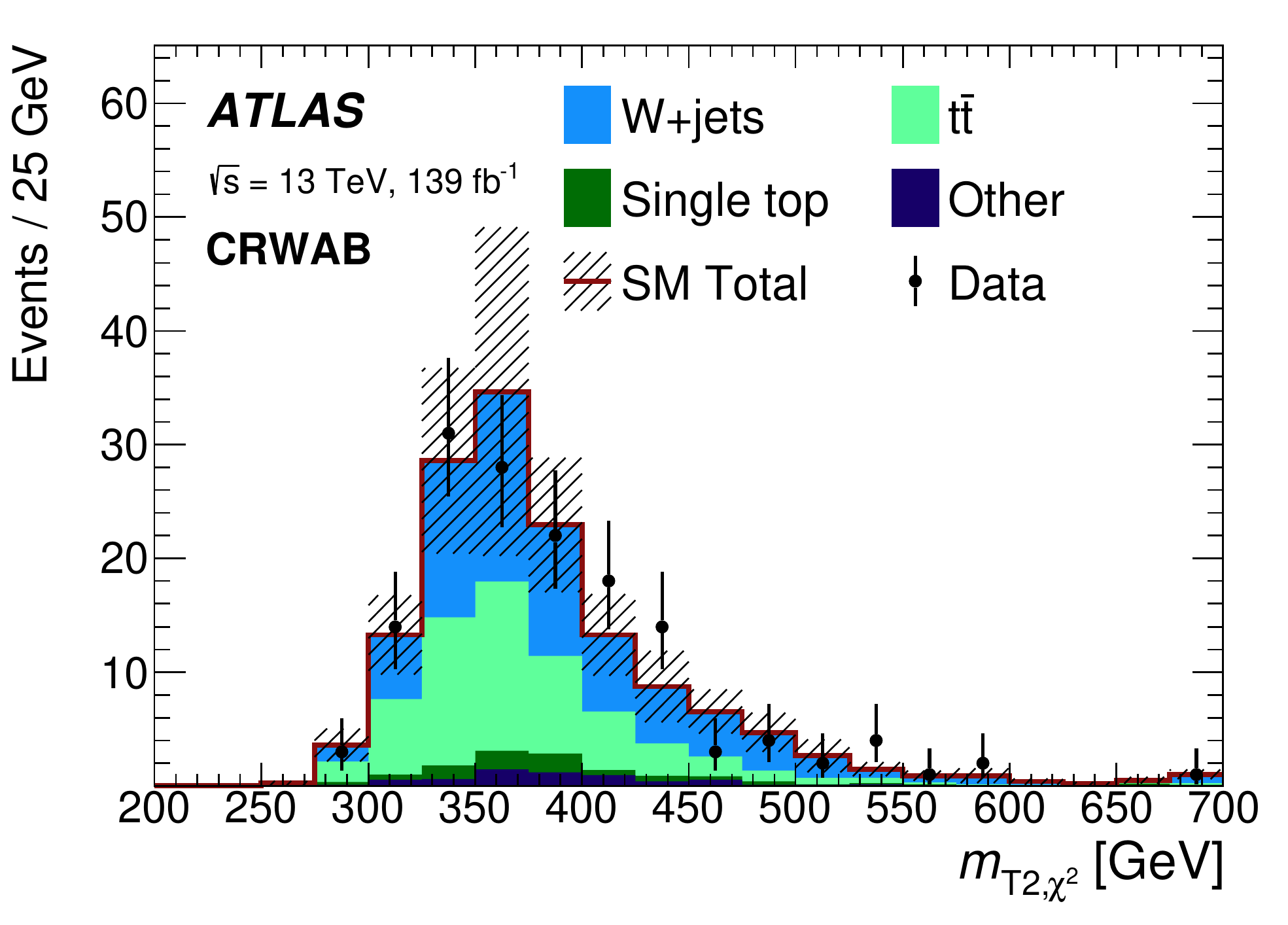}}
\subfloat[]{\includegraphics[width=0.5\textwidth]{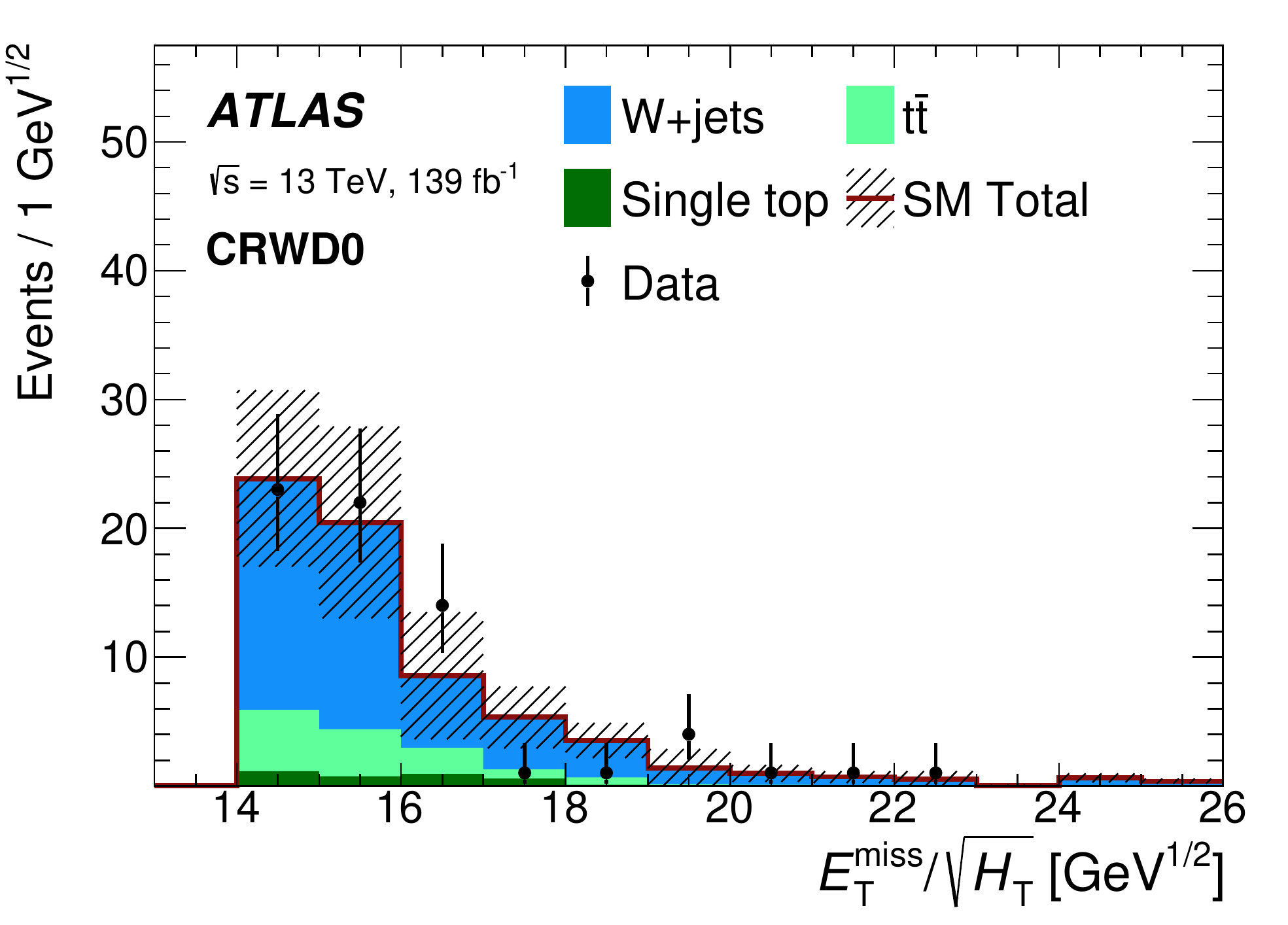}}\\
\subfloat[]{\includegraphics[width=0.5\textwidth]{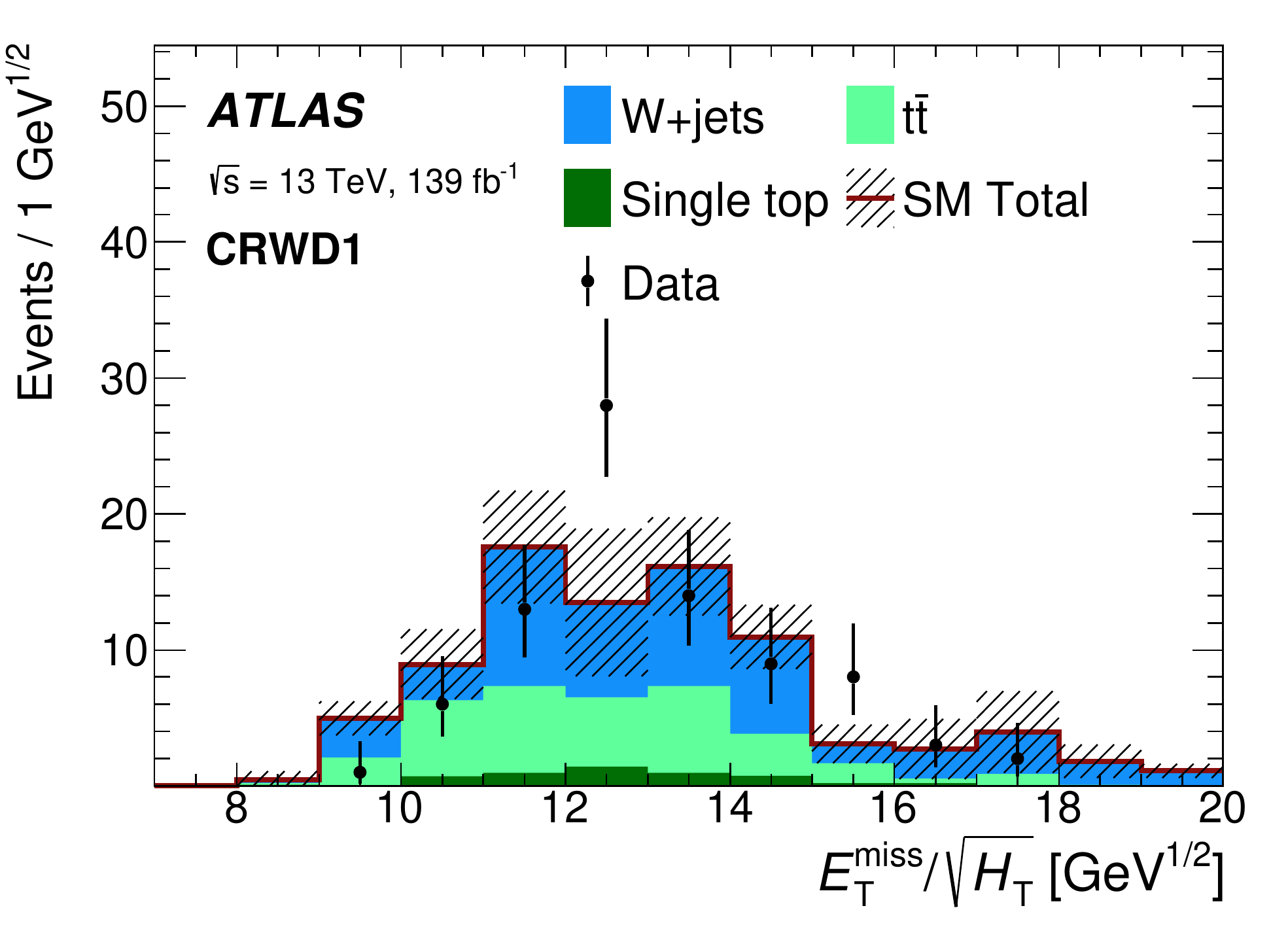}}
\subfloat[]{\includegraphics[width=0.5\textwidth]{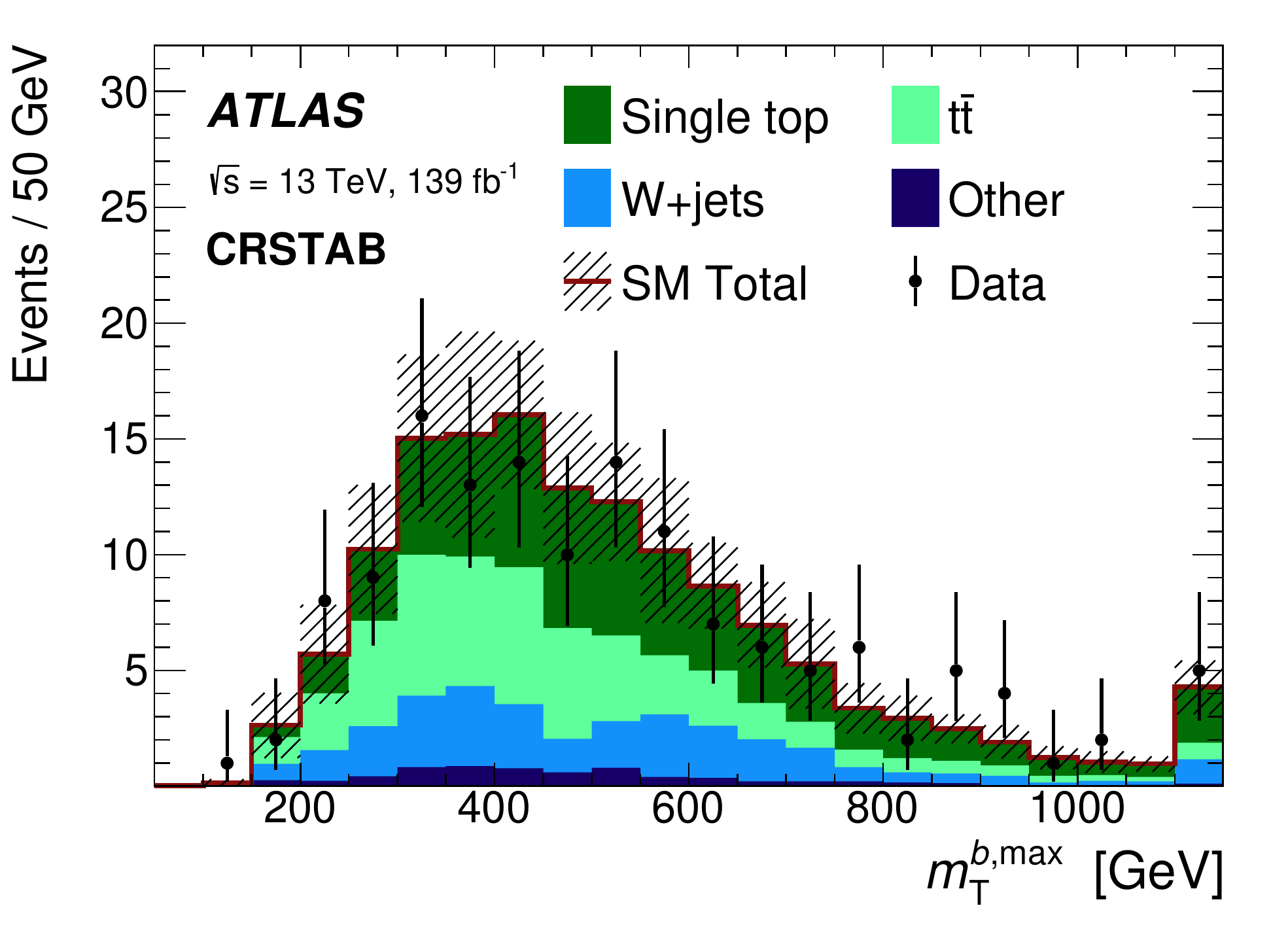}}
\caption{Distributions illustrating the level of agreement between data (points) and the SM expectation (stacked histograms, after simultaneously fitting to all backgrounds) in several \Wjets\ and single-top control regions: (a) \mttwochisquare\ for CRWAB, \htsig\ for (b) CRWD0 and (c) CRWD1, and (d)~\mtbmax\ for CRSTAB.
The hatched uncertainty band around the SM expectation includes the combination of MC statistical, theory-related and detector-related systematic uncertainties.
The rightmost bin in each plot includes all overflows.}
\label{fig:CRWST}
\end{center}
\end{figure}

The \Wjets\ background is important for SRA--B and SRD, while the single-top background is significant for SRA--B only; corresponding control regions (CRWAB, CRWD, and CRSTAB, respectively) are defined in Table~\ref{tab:CRWST}. The \Wjets\ background in SRA--B originates from $\Wboson$ boson decays into low-\pT\ $\tau$-leptons; thus, the strategy is similar to that described for CRTAB except that exactly one $b$-tagged jet is required, which makes CRWAB orthogonal to CRTAB. The single-top control region, CRSTAB, is defined as having exactly one control electron or muon with $\pT > 20$~GeV\ (making CRSTAB orthogonal to both CRWAB and CRTAB) and two or more $b$-tagged jets. A requirement of $\pT > 20$~GeV\ is used in CRWD because the \Wjets\ background in SRD is dominated by high-\pt\ electrons, muons, and $\tau$-leptons. To enhance the purity of the $\Wjets$ background in CRWD and ensure orthogonality with CRTD, lower bounds are put on \drblep, which is defined with respect to the $b$-tagged jet ($b$-tagged track-jet) closest to the lepton in CRWD1--2 (CRWD0). Representative distributions for the various \Wjets\ and single-top control regions defined in the analysis are shown in Figure~\ref{fig:CRWST}.
 
% The next lines are included from the .//tex/CRWSTTable.tex input file
 
\begin{table}[htpb]
\caption{Selection criteria for the \Wjets\ and single-top control regions. The defining extrapolation for these control regions is over the number of leptons; one electron or muon ($\ell$) from $\Wboson$ decays is required compared with zero leptons in the signal regions. A dash indicates that no selection is applied. Variables for which the signal and control region requirements differ are highlighted by a thick border around the cell that contains the requirement. Requirements are made on the following variables in the signal regions but have no equivalent requirement in the control regions: \mantikttwelveone, \mantikteightzero, \bantikttwelvezero, \bantikttwelveone, \mtbmax, \mttwochisquare.}
\begin{center}
\def\arraystretch{1.4}
\scriptsize
\begin{tabular}{c||c|c|c|c|c}
\hline \hline
\textbf{ Variable/CR}                  & \textbf{ CRSTAB}                                 & \textbf{ CRWAB}                                  & \textbf{ CRWD0}     & \textbf{ CRWD1}     & \textbf{ CRWD2}    \\
\hline\hline
Trigger                      & \multicolumn{5}{c}{\met}                                                                                           \\
\hline
\met                         & \multicolumn{5}{c}{$>250$ GeV}                                                                                     \\
\hline
Control $\ell$                & \multicolumn{5}{c}{exactly $1$}                                                                                    \\
\hline
Additional baseline $\ell$ & \multicolumn{5}{c}{0}                                                                                              \\
\hline
$\pt^{\ell}$                 & $\pt>20$ GeV                           & $4.5\,(4.0) < \pt^{e\,(\mu)} < 20$ GeV & \multicolumn{3}{c}{$\pt>20$ GeV} \\
\hline
\mtlepmet                    & \multicolumn{2}{c|}{$<100$ GeV}        & \multicolumn{2}{c|}{$<120$ GeV}        & $<100$ GeV                       \\
\hline
\njets                       & \multicolumn{2}{c|}{$\ge 4$}           & \multicolumn{3}{c}{--}                                                    \\
\hline
\pttwo                       & \multicolumn{2}{c|}{$>80$~GeV}        & \multicolumn{3}{c}{--}                                                    \\
\hline
\ptfour                    & \multicolumn{2}{c|}{$>40$~GeV}        & \multicolumn{3}{c}{--}                                                    \\
\hline
\Cline{3-3}
\nbjets                      & $\ge 2$                                & \multicolumn{1}{!{\vrule width 1pt}c!{\vrule width 1pt}}{exactly 1}                              & exactly 0 & exactly 1 & $\ge 2$  \\
\Cline{3-3}
\hline
\dphijetfourmet            & \multicolumn{3}{c|}{$>0.4$}            & \multicolumn{2}{c}{--}                                                    \\
\hline
\Cline{3-3}
\mantikttwelvezero           & $>120$ GeV                              & \multicolumn{1}{!{\vrule width 1pt}c!{\vrule width 1pt}}{$<60$ GeV}                             & \multicolumn{3}{c}{--}           \\
\Cline{3-3}\hline
\mtbmin                    & \multicolumn{2}{c|}{$>200$ GeV}        & \multicolumn{3}{c}{--}                                                    \\
\hline
\Cline{2-3}
\Cline{6-6}
\drbjetbjet                & \multicolumn{1}{!{\vrule width 1pt}c!{\vrule width 1pt}}{$>1.4$} & \multicolumn{1}{!{\vrule width 1pt}c!{\vrule width 1pt}}{--}                                 & \multicolumn{2}{c|}{--}                 &   \multicolumn{1}{!{\vrule width 1pt}c!{\vrule width 1pt}}{$<1.0$}                          \\
\Cline{2-3}
\Cline{6-6}
\hline
\mlbmin                     & $>100$ GeV                             & \multicolumn{4}{c}{--}                                                    \\
\hline
\Cline{3-3}
$\tau$-veto                   & yes                                     & \multicolumn{1}{!{\vrule width 1pt}c!{\vrule width 1pt}}{--} & \multicolumn{3}{c}{--}                                                    \\
\Cline{3-3}
\hline
\Cline{2-3}
\objmetsig                   & \multicolumn{2}{!{\vrule width 1pt}c!{\vrule width 1pt}}{$>14$} & \multicolumn{3}{c}{--}                                                    \\
\Cline{2-3}
\hline
\drblep                    & --                                & $>2.0$                                      & $>1.6$   & $>1.8$    & $>2.2$   \\
\hline
\ptnonbjetone              & \multicolumn{2}{c|}{--}                  & \multicolumn{3}{c}{$>250$ GeV}                            \\
\hline
$\mettrk$                 & \multicolumn{2}{c|}{--}                & \multicolumn{3}{c}{$>30$ GeV}                                             \\
\hline
\dphimettrk                & \multicolumn{2}{c|}{--}                & \multicolumn{3}{c}{$< \pi/3$}                                             \\
\hline
\dphinonbjetmet            & \multicolumn{2}{c|}{--}                & \multicolumn{3}{c}{$>2.4$}                                                \\
\hline
\nbjetsvr                    & \multicolumn{2}{c|}{--}                & \multicolumn{2}{c|}{$\ge1$}            & --                               \\
\hline
\etabonevr                   & \multicolumn{2}{c|}{--}                & $< 1.2$                                & \multicolumn{2}{c}{--}           \\
\hline
$\maxdphinonbjetbjetvr$      & \multicolumn{2}{c|}{--}                & $>2.2$                                 & \multicolumn{2}{c}{--}           \\
\hline
$\dphibbvr$                  & \multicolumn{2}{c|}{--}                & $< 2.5$                                & \multicolumn{2}{c}{--}           \\
\hline
\ptbonevr                    & \multicolumn{2}{c|}{--}                & $<50$ GeV                              & $>10$ GeV & --                   \\
\hline
\ptonevr                     & \multicolumn{3}{c|}{--}                & $<40$ GeV                              & --                               \\
\hline
$\mindphijetvrfournonb$      & \multicolumn{3}{c|}{--}                & $> 1.2$                                & --                               \\
\hline
\etabone                     & \multicolumn{3}{c|}{--}                & $< 1.6$                                & --                               \\
\hline
\ptbone                      & \multicolumn{4}{c|}{--}                & $< 175$ GeV                                                               \\
\hline
\etabtwo                   & \multicolumn{4}{c|}{--}                & $<1.2$                                                                    \\
 
\hline
\Cline{4-6}
$\htsig$ & \multicolumn{2}{c|}{--} & \multicolumn{1}{!{\vrule width 1pt}c!{\vrule width 1pt}}{$>14 \sqrt{\mathrm{GeV}}$} & \multicolumn{1}{!{\vrule width 0pt}c!{\vrule width 0pt}}{$>8 \sqrt{\mathrm{GeV}}$} & \multicolumn{1}{!{\vrule width 1pt}c!{\vrule width 1pt}}{$>12 \sqrt{\mathrm{GeV}}$} \\
\Cline{4-6}
\hline \hline
\end{tabular}
\end{center}
\label{tab:CRWST}
\end{table}
% End of text imported from the .//tex/CRWSTTable.tex input file
 
\begin{figure}[htpb]
\begin{center}
\includegraphics[width=0.65\textwidth]{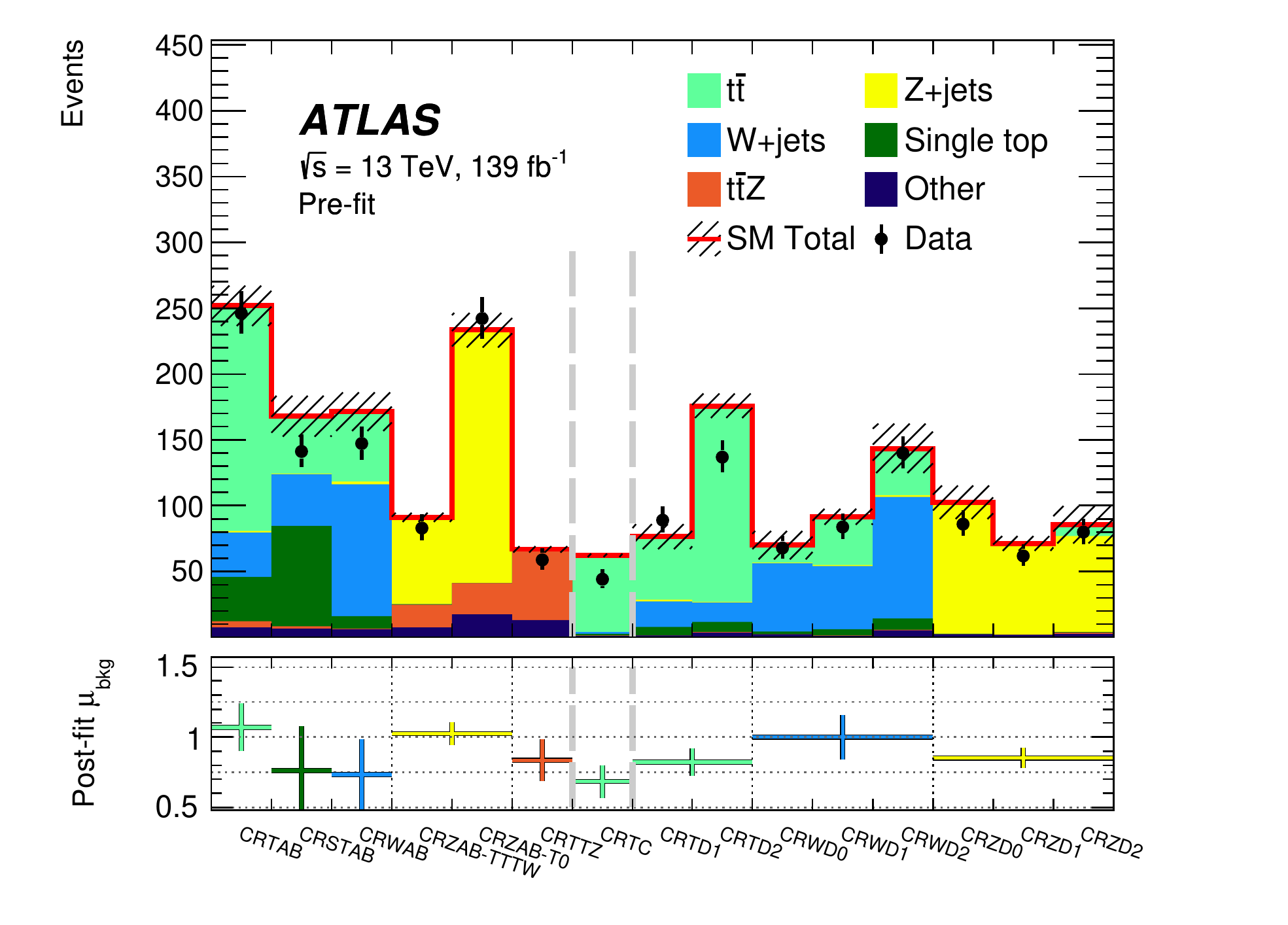}
\caption{A summary of the normalisation factors determined from the various background-only fits. The total number of data events (points) and the SM expectation (stacked histograms) are shown in each control region before the fit. The uncertainty associated with the SM expectation includes the combination of MC statistical uncertainties, theory-related and detector-related systematic uncertainties.  The normalisation factor applied to each background source ($\mu_{\textrm{bkg}}$) after the fit and respective uncertainty, including the combination of MC statistical uncertainties, theory-related and detector-related systematic uncertainties, is shown in the lower panel. The control regions included in the SRA--B, SRC and SRD fits are separated by vertical dashed lines.}
\label{fig:CRSummary}
\end{center}
\end{figure}
 
\FloatBarrier
 
\subsection{Validation of background estimates}
 
The background normalisation factors derived from the SRA--B, SRC and SRD background-only fits are summarised in Figure~\ref{fig:CRSummary}. Most normalisation factors are within 1$\sigma$ of unity, where $\sigma$ denotes the total uncertainty, including the data statistical uncertainty in the CRs and the theory-related and detector-related systematic uncertainties (described in Section~\ref{sec:Systematics}). However, the $\ttbar$ ($\ttbar$ and $\Zjets$) normalisation factors derived from the SRC (SRD) fit are lower than unity by one to two $\sigma$. Significant amounts of ISR radiation are required in SRC, SRD, and the associated control regions, unlike SRA--B and the associated control regions. The simulated event yields in $\ttbar$-enriched regions compare differently with data in SRA--B control regions and SRC--D control regions, overestimating the number of events in the latter, while fairly good agreement is observed in the former. A similar effect is observed in CRZAB and CRZD. These observations point to a mismodelling possibly related to the ISR system in $\ttbar$ and $\Zjets$ events. The fitting procedure corrects for this mismodelling and is validated in the VRs discussed below.
 
Validation regions are defined to check the validity of the normalisation factors in the signal regions and to check the ability of the MC to describe the shapes of the kinematic variables
over which extrapolations are made in propagating background estimates from the control regions to the signal regions. The defining extrapolation from control to signal regions is in the lepton multiplicity, whereas the validation regions include only events with zero leptons, as in the signal regions. Validation regions are designed for the \Zjets\ background in SRA (VRZA) and SRB (VRZB-TTTW, VRZB-T0) and SRD (VRZD0--2), as well as for the \ttbar\ background in SRA--B (VRTAB), SRC (VRTC), and SRD (VRTD1--2). Requirements applied in the SRs are modified in the VRs to ensure orthogonality with the SRs, to limit signal contamination, and to retain a sufficient number of events expected in data. Signal contamination in the VRs, for all signals considered in this search, is kept below 20\%.
 
VRZA is made to be orthogonal to SRA--B by vetoing events where the leading reclustered jet contains a \bjet\ while still requiring at least two \bjet s. Orthogonality between VRZB-TTTW and SRA--B is achieved by inverting the \drbjetbjet\ requirement made in SRB, $\drbjetbjet<1.4$, and selecting a lower \objmetsig\ window than in SRA, $15<\objmetsig<17$. For VRZD0 and VRZD1--2, the orthogonality with SRD is ensured by inverting the \maxdphinonbjetbjetvr\ and \dphinonbjetbjetone\ requirements, respectively.
 
VRTAB is orthogonal to SRA--B due to the inversion of the \mtbmin\ requirement, while VRTC is orthogonal to SRC by inverting the \dPhiISRMET\ requirement. In VRTD1--2, the \dphinonbjetbjetone\ requirement is inverted, as is done in VRZD1--2. The purity of \ttbar\ events in VRTD1 is enhanced by introducing a $\drbjetbjet>2.0$ requirement.
Only one \bjet\ is required in VRTD1 (as in SRD1) and therefore \drbjetbjet\ is defined as the angular distance between the one \bjet\ and the next jet most likely to contain a $b$-hadron, quantified by the $b$-tagging discriminator output value.
 
\begin{figure}[htpb]
\begin{center}
\subfloat[]{\includegraphics[width=0.5\textwidth]{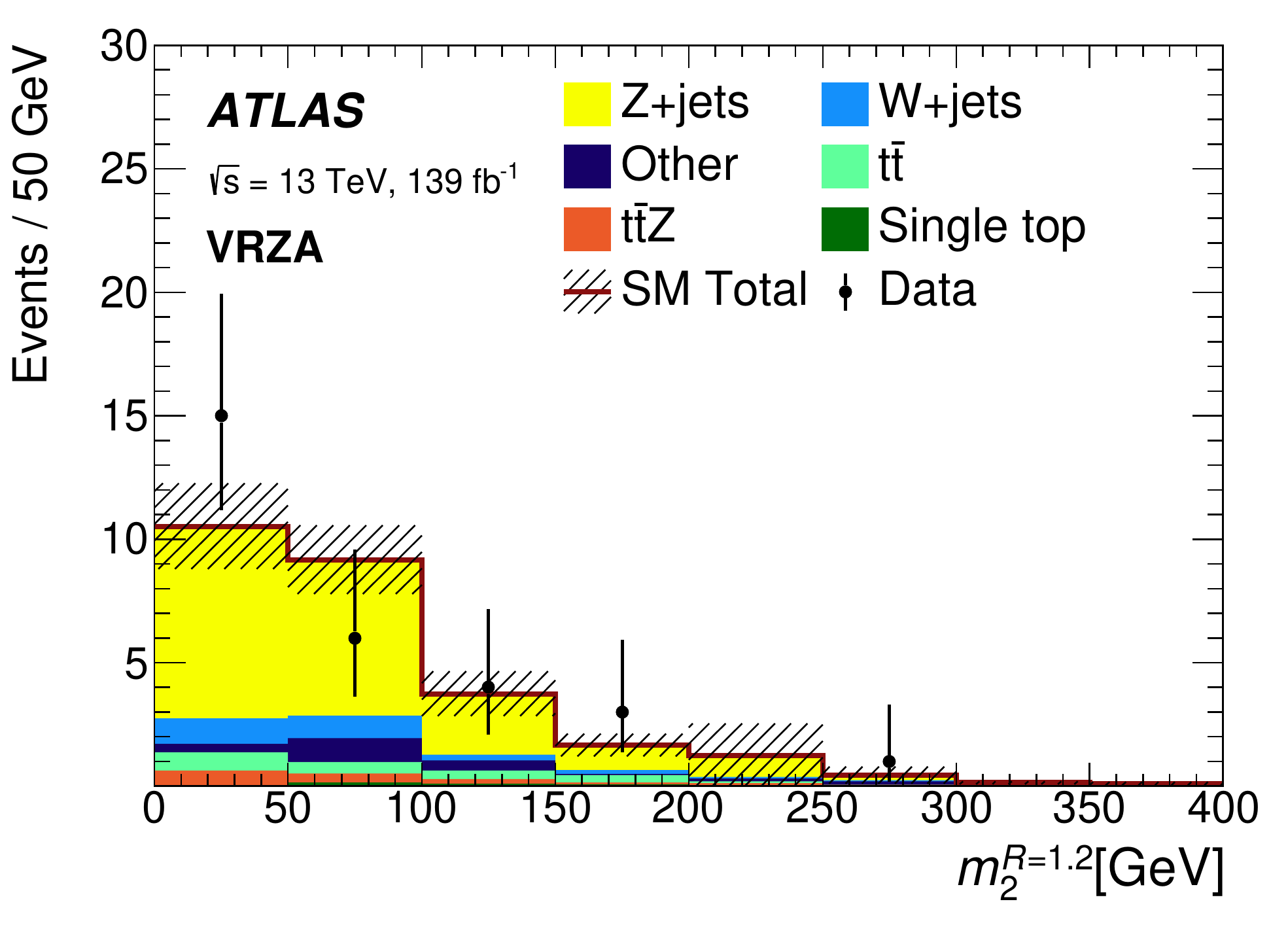}}
\subfloat[]{\includegraphics[width=0.5\textwidth]{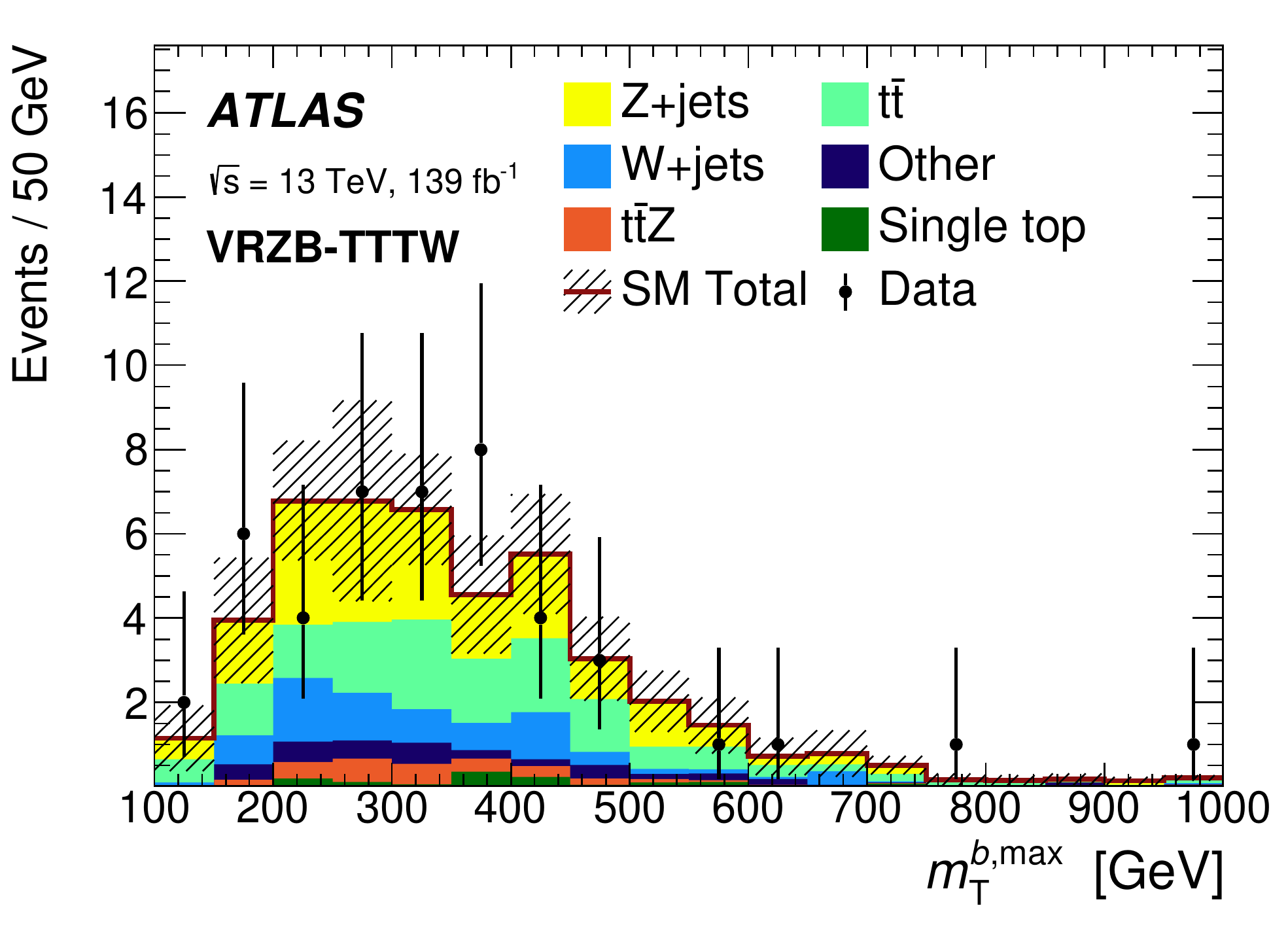}}\\
\subfloat[]{\includegraphics[width=0.5\textwidth]{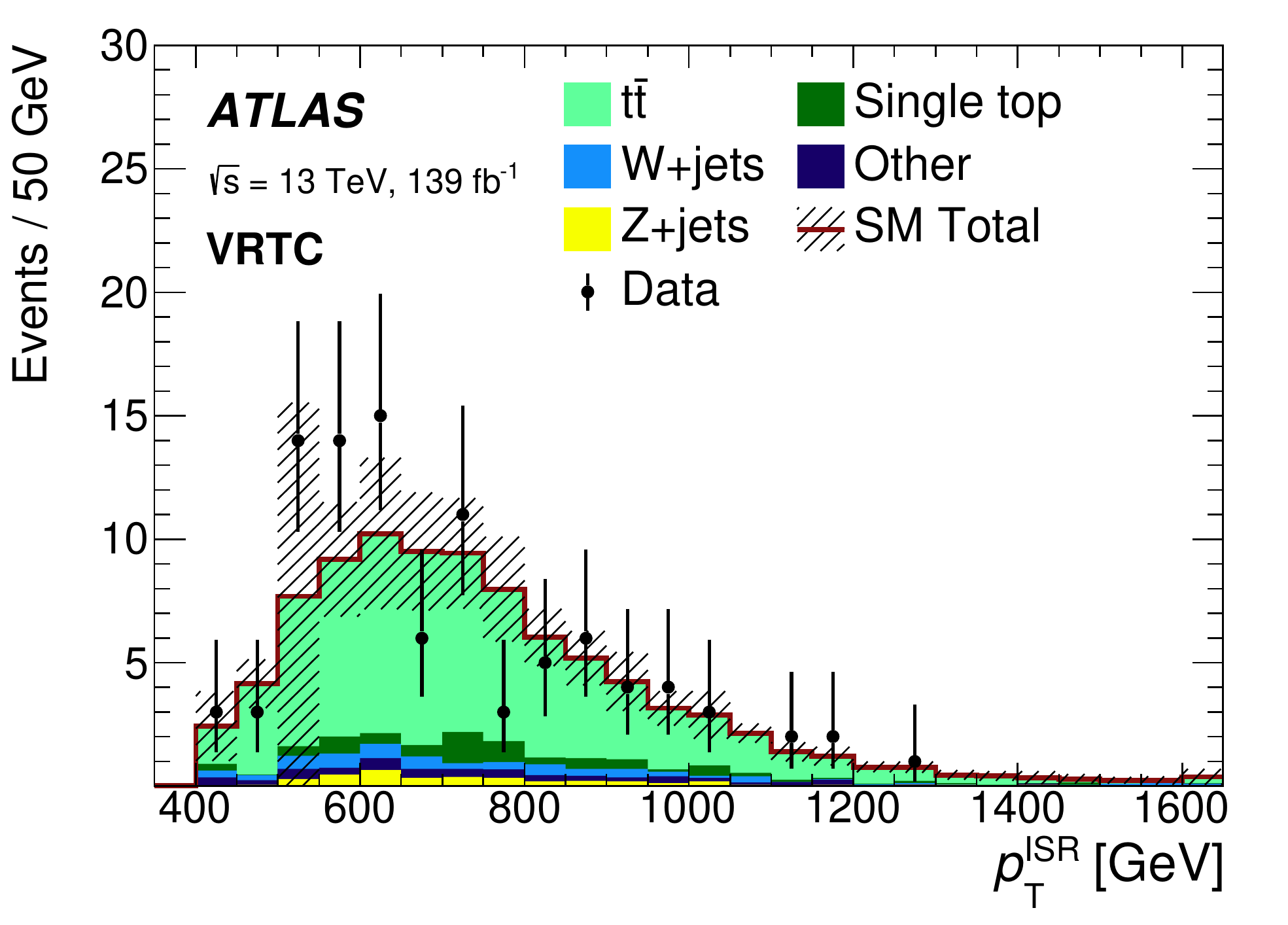}}
\subfloat[]{\includegraphics[width=0.5\textwidth]{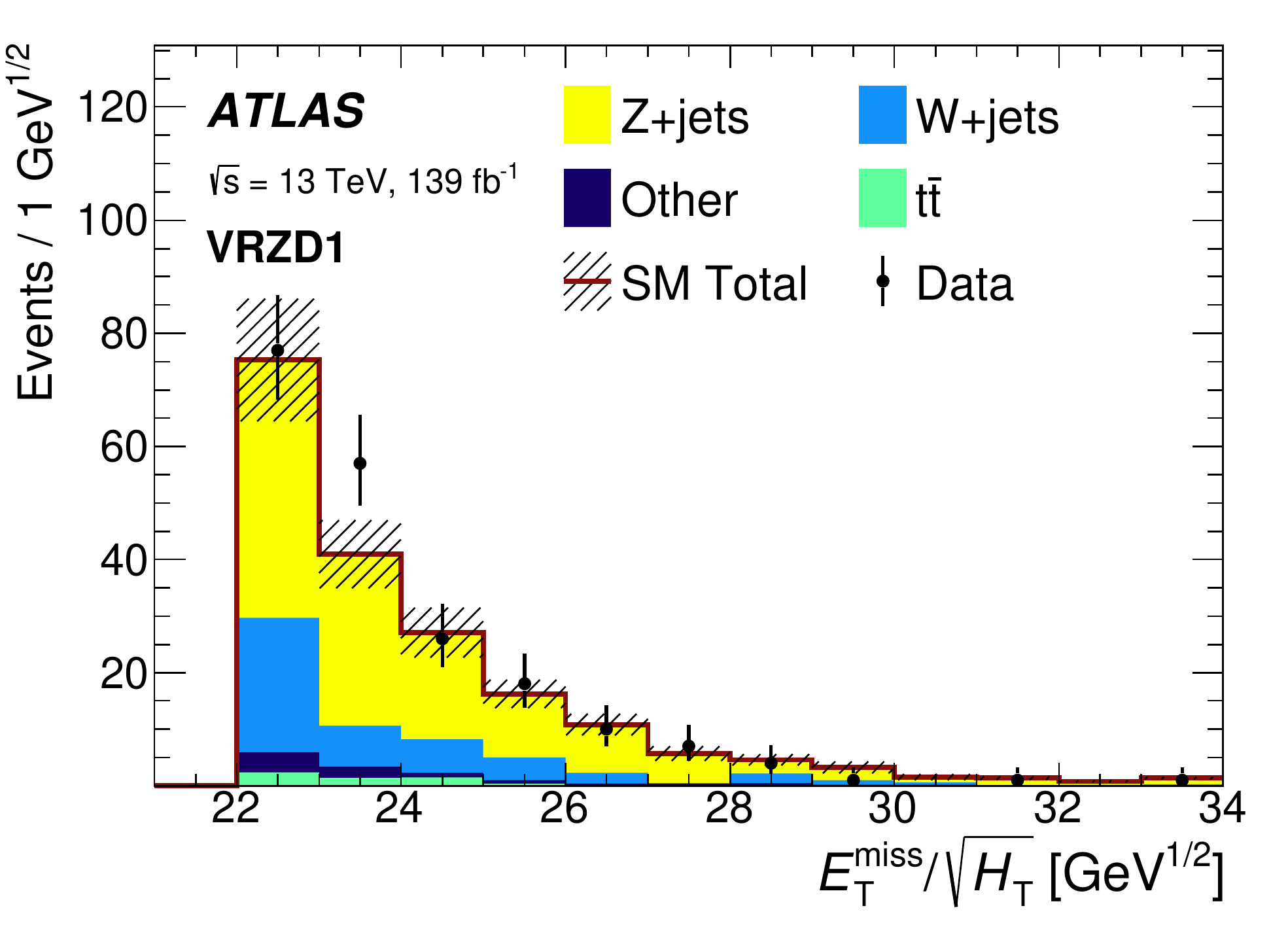}}
\caption{Distributions illustrating the level of agreement between data (points) and the SM expectation (stacked histograms, after simultaneously fitting to all backgrounds) in several validation regions: (a) \mantikttwelveone\ in VRZA, (b) \mtbmin\ in VRZB-TTTW, (c) \pTISR\ in VRTC, and (d) \htsig\ in VRZD1.
The hatched uncertainty band around the SM expectation includes the combination of MC statistical, theory-related and detector-related systematic uncertainties.
The rightmost bin in each plot includes all overflows.}
\label{fig:VR}
\end{center}
\end{figure}

\begin{figure}[htpb]
\begin{center}
\includegraphics[width=0.65\textwidth]{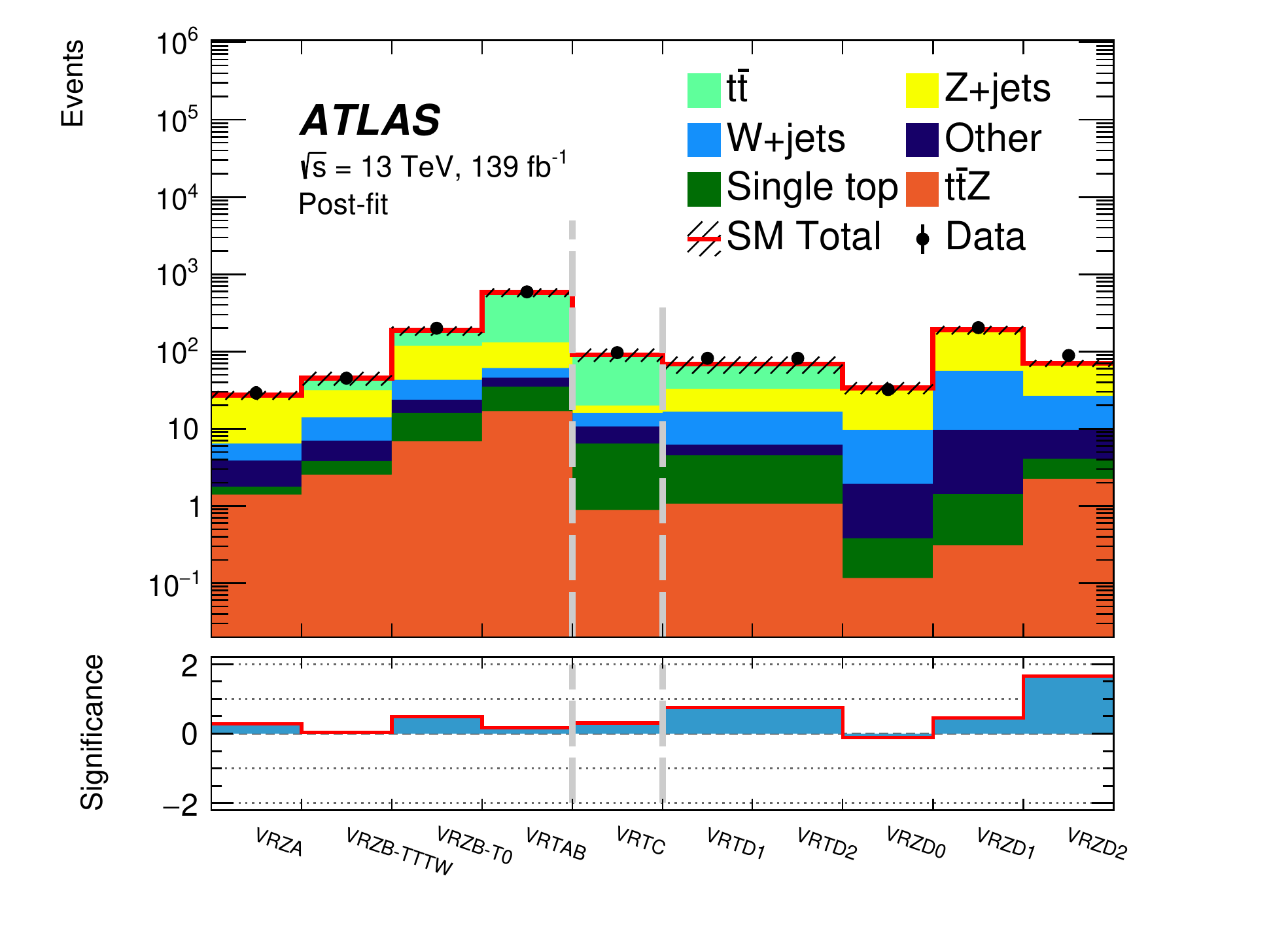}
\caption{The total number of data events (points) and the SM expectation (stacked histograms) in all validation regions after the SRA--B, SRC and SRD background-only fits. The stacked histograms show the SM prediction and the hatched uncertainty band around the SM prediction shows the total uncertainty, which includes the MC statistical uncertainties, theory-related and detector-related systematic uncertainties. The lower panel shows the significance of the difference between data and the background prediction calculated with the method described in Ref.~\cite{Choudalakis2012EPJP12725C}. The validation regions considered in the SRA--B, SRC and SRD fits are separated by vertical dashed lines.}
\label{fig:VRSummary}
\end{center}
\end{figure}

Representative distributions for the validation regions defined in the analysis are shown in Figure~\ref{fig:VR}. A summary of the expected and observed yields in the VRs after the SRA--B, SRC and SRD background-only fits is shown in Figure~\ref{fig:VRSummary}. All the background predictions in the VRs agree with the data within 1$\sigma$ except the predictions in VRZD2, which agree with the data within 2$\sigma$.
 
% End of text imported from the .//tex/Background.tex input file
\FloatBarrier

% The next lines are included from the .//tex/Systematics.tex input file
\section{Systematic uncertainties}
\label{sec:Systematics}
 
Uncertainties affecting the sensitivity of the analysis which originate from statistical sources are considered together with systematic uncertainties related to the detector calibration (detector-related uncertainties) and physics modelling of signal and background (theory-related uncertainties). The data statistical uncertainty in the number of events in the SRs dominates the total uncertainty in SRA and SRD, while uncertainties related to the physics modelling of the background play a significant role in SRB and SRC.
 
The impact of detector-related and theory-related systematic uncertainties in the background predictions are included in the profile likelihood fits (described in Section~\ref{sec:background}) as nuisance parameters constrained by Gaussian probability density functions. Their impact is reduced by scaling the dominant background components in the SRs using the data observed in the CRs via the introduction of free-floating normalisation parameters. After the SRA--B and SRD (SRC) background-only fit, none of the nuisance parameters are pulled significantly and most (all) of them are not constrained. The largest constraints are observed in the SRD fit and are on the $t\bar{t}$ modelling uncertainties and reach 30\%.
 
The dominant systematic uncertainties in the background estimates in SRA and SRB (SRC and SRD), expressed as percentages of the total background expectations, are shown in Table~\ref{tab:srABSysts} (Table~\ref{tab:srCDSysts}). By convention, the data statistical uncertainty in the numbers of events in the CRs is accounted for as a systematic uncertainty and included in table rows indicated by the normalisation factors for each background source ($\mu_{\textrm{bkg}}$).
 
\begin{table}[htpb]
\caption{Systematic uncertainties (in percent) greater than $1\%$ for at
least one category within SRA and SRB. Uncertainties are expressed relative to the total
background estimates. The uncertainties due to the scaling of background events based on data in control regions are indicated for each background component by $\mu_{\ttbar}$, $\mu_{\ttZ}$,  $\mu_{Z}$, $\mu_{W}$, and $\mu_{\mathrm{single~top}}$. The theory uncertainties quoted for each background include the different distribution shape uncertainties described in the text.}
\begin{center}
{\renewcommand{\arraystretch}{1.2}
% The next lines are included from the .//tex/systSummarySRAB.tex input file
\begin{tabular}{
S[table-alignment=left]||
S[table-number-alignment=right,table-alignment=right]
S[table-number-alignment=right,table-alignment=right]
S[table-number-alignment=right,table-alignment=right]
S[table-number-alignment=right,table-alignment=right]
S[table-number-alignment=right,table-alignment=right]
S[table-number-alignment=right,table-alignment=right]
}
\hline\hline
& {\textbf{SRA-TT}} & {\textbf{SRA-TW}} & {\textbf{SRA-T0}} & {\textbf{SRB-TT}} & {\textbf{SRB-TW}} & {\textbf{SRB-T0}}\\ \hline
\hline
{Total syst. unc.} & {15} & {12} & {10} & {14} & {9} & {9}\\
\hline
{\ttbar\ theory} & {2} & {2} & {1} & {11} & {6} & {4}\\
{Single-top theory} & {7} & {5} & {4} & {1} & {$<$1} & {1}\\
{$\ttbar Z$ theory} & {3} & {$<$1} & {$<$1} & {$<$1} & {$<$1} & {$<$1}\\
{$Z$ theory} & {$<$1} & {$<$1} & {1} & {$<$1} & {$<$1} & {$<$1}\\
\hline
{$\mu_{\ttbar}$} & {$<$1} & {$<$1} & {$<$1} & {4} & {4} & {4}\\
{$\mu_{\ttZ}$} & {6} & {2} & {2} & {4} & {3} & {1}\\
{$\mu_{Z}$} & {3} & {5} & {5} & {3} & {3} & {3}\\
{$\mu_{W}$} & {2} & {3} & {3} & {4} & {4} & {3}\\
{$\mu_{\mathrm{single~top}}$} & {6} & {4} & {5} & {3} & {4} & {5}\\
\hline
{JER} & {7} & {3} & {2} & {6} & {2} & {3}\\
{JES} & {4} & {4} & {2} & {2} & {$<$1} & {$<$1}\\
{$b$-tagging} & {5} & {3} & {3} & {2} & {1} & {2}\\
{\met\ soft term} & {2} & {1} & {1} & {$<$1} & {$<$1} & {$<$1}\\
{MC statistics} & {7} & {7} & {5} & {3} & {3} & {2}\\
\hline\hline
\end{tabular}
% End of text imported from the .//tex/systSummarySRAB.tex input file
}
\end{center}
\label{tab:srABSysts}
\end{table}
 
\begin{table}[htpb]
\caption{Systematic uncertainties (in percent) greater than $1\%$ for at
least one category within SRC and SRD. Uncertainties are expressed relative to the total
background estimates. The uncertainties due to the scaling of background events based on data in control regions are indicated for each background component by $\mu_{\ttbar}$, $\mu_{Z}$, and $\mu_{W}$. The theory uncertainties quoted for each background include the different distribution shape uncertainties described in the text.}
\begin{center}
{\renewcommand{\arraystretch}{1.2}
% The next lines are included from the .//tex/systSummarySRCD.tex input file
\begin{tabular}{
S[table-alignment=left]||
S[table-number-alignment=right,table-alignment=right]
S[table-number-alignment=right,table-alignment=right]
S[table-number-alignment=right,table-alignment=right]
S[table-number-alignment=right,table-alignment=right]
S[table-number-alignment=right,table-alignment=right]
S[table-number-alignment=right,table-alignment=right]
S[table-number-alignment=right,table-alignment=right]
S[table-number-alignment=right,table-alignment=right]
}
\hline\hline
& {\textbf{SRC1}} & {\textbf{SRC2}} & {\textbf{SRC3}} & {\textbf{SRC4}} & {\textbf{SRC5}} & {\textbf{SRD0}} & {\textbf{SRD1}} & {\textbf{SRD2}}\\ \hline
\hline
{Total syst. unc.} & {25} & {18} & {20} & {27} & {27} & {18} & {31} & {12}\\
\hline
{\ttbar\ theory} & {20} & {11} & {12} & {16} & {21} & {4} & {9} & {5}\\
{Single-top theory} & {$<$1} & {$<$1} & {$<$1} & {$<$1} & {$<$1} & {$<$1} & {4} & {2}\\
{$Z$ theory} & {$<$1} & {$<$1} & {1} & {2} & {4} & {7} & {3} & {2}\\
{$W$ theory} & {$<$1} & {$<$1} & {1} & {2} & {3} & {$<$1} & {$<$1} & {$<$1}\\
\hline
{$\mu_{\ttbar}$} & {12} & {13} & {14} & {14} & {11} & {$<$1} & {2} & {5}\\
{$\mu_{Z}$} & {$<$1} & {$<$1} & {$<$1} & {$<$1} & {$<$1} & {5} & {3} & {2}\\
{$\mu_{W}$} & {$<$1} & {$<$1} & {$<$1} & {$<$1} & {$<$1} & {4} & {5} & {3}\\
\hline
{JER} & {5} & {$<$1} & {8} & {15} & {7} & {8} & {18} & {4}\\
{JES} & {$<$1} & {1} & {$<$1} & {4} & {6} & {1} & {4} & {2}\\
{$b$-tagging} & {2} & {2} & {2} & {2} & {2} & {3} & {5} & {7}\\
{Track-jet flavour} & {$<$1} & {$<$1} & {$<$1} & {$<$1} & {$<$1} & {4} & {7} & {$<$1}\\
{Track-jet flavour (low \pt)} & {$<$1} & {$<$1} & {$<$1} & {$<$1} & {$<$1} & {7} & {4} & {1}\\
{\met\ soft term} & {$<$1} & {$<$1} & {$<$1} & {$<$1} & {3} & {$<$1} & {$<$1} & {$<$1}\\
{Pile-up} & {$<$1} & {$<$1} & {$<$1} & {1} & {$<$1} & {2} & {12} & {$<$1}\\
{MC statistics} & {3} & {2} & {3} & {4} & {6} & {11} & {17} & {5}\\
\hline\hline
\end{tabular}
% End of text imported from the .//tex/systSummarySRCD.tex input file
}
\end{center}
\label{tab:srCDSysts}
\end{table}
 
The dominant detector-related systematic uncertainty in the background estimates originates from sources related to the jet energy scale (JES) and resolution (JER)~\cite{PERF-2016-04}, which encompass both the modelling of the detector response and the analysis techniques used to derive the calibration, the $b$-tagging performance, which includes the uncertainty in the $b$-tagging efficiency of true $b$-jets~\cite{FTAG-2018-01} and in the $b$-tagging rate of light-flavour jets~\cite{ATLAS-CONF-2018-006} and $c$-jets~\cite{ATLAS-CONF-2018-001}, and the energy scale and resolution of the \met\ soft term~\cite{PERF-2016-07,ATLAS-CONF-2018-023}. The uncertainty in the modelling of pile-up events contributes significantly to the total uncertainty only in SRD2.
 
The JES uncertainty is derived as a function of the $\pT$ and $\eta$ of each jet, the pile-up conditions, and the jet flavour. It is determined using a combination of simulated samples and collision data, including measurements of dijet, multijet, $\Zboson+$jet, and $\gamma+$jet events. The JER uncertainty is derived as a function of the $\pT$ and $\eta$ of each jet, and is determined from a random cone technique applied to data recorded without selection bias, and studies of asymmetries in dijet events. The uncertainty in the JER is significant in many signal regions (maximally $18\%$ in SRD1), while the most significant impact of the JES uncertainty reaches $6\%$ in SRC5.
 
The uncertainty originating from the $b$-tagging performance of jets is estimated by varying the $\pT$- and flavour-dependent per-jet scale factors, applied to each jet, within predefined ranges determined from efficiency and mis-tag rate measurements in data. The $b$-tagging uncertainty is highest in SRA and SRD and does not exceed $7\%$ (reached in SRD2).
 
Uncertainties in the $b$-tagging performance of track-jets, which are only relevant in SRD, are estimated for track-jets with $\pt>10$~GeV\ in the same way as for jets. The largest contribution from this systematic uncertainty is in SRD1, where it is 7\%. For track-jets with $5<\pt<10$~GeV, which may be selected in SRD0, no evaluation of the $b$-tagging performance in data is available, so the uncertainty is evaluated by comparing the $b$-tagging performance observed in \Zll~events generated with \SHERPA\ and \MADGRAPH\ interfaced with \PYTHIAV{8.2}, resulting in an uncertainty of 7\% in SRD0.
The difference between the predictions of these two generators was found to always be larger than the difference between the predictions from extrapolation from the neighbouring bins, hence assigning the uncertainty based on the generator comparison is assumed to be conservative.
 
All jet-, electron-, and muon-related uncertainties are propagated to the calculation of the \met\ when evaluated, and additional uncertainties in the energy scale and resolution of the \met\ soft term are evaluated. The uncertainty in the soft term of the \met\ is derived using $\Zboson\to\mu^{+}\mu^{-}$ events and is less than $3\%$ in all SRs.
The uncertainty due to the reweighting of the simulated samples to match the distribution of pile-up in data is negligible in all SRs except SRD, where it is at most $12\%$ (reached in SRD1).
 
Uncertainties in electron and muon reconstruction and identification uncertainties are also considered but have a negligible impact on the final background estimates. The uncertainty in the combined 2015--2018 integrated luminosity is $1.7\%$~\cite{ATLAS-CONF-2019-021,LUCID2} and has a negligible impact on the analysis.
 
Theoretical uncertainties in the physics modelling of the background processes are also evaluated for each background component. For the \ttbar\ background, uncertainties are estimated from the comparison of different matrix-element calculations (\POWHEGBOX\ vs \AMCatNLO), the choice of parton-showering model (\PYTHIA\ vs \HERWIGV{7}), and the emission of ISR and final-state radiation (FSR) within \textsc{Pythia}~8 while leaving all other parameters for each comparison unchanged. The effects of ISR and FSR are explored by reweighting the nominal \ttbar events in a manner that reduces (reduces and increases) initial-state (final-state) parton-shower radiation~\cite{ATL-PHYS-PUB-2016-020} and by using an alternative \POWHEGBOXV{v2} + \PYTHIAV{8.2} sample with $h_{\textrm{damp}}$ set to $3m_t$ and parameter variation group \textsc{Var3} (described in Ref.~\cite{ATL-PHYS-PUB-2016-020}) increased, leading to increased ISR. SRC is most sensitive to \ttbar\ theory systematic uncertainties, ranging from 11\% to 21\%, followed by SRB, ranging from 4\% to 11\%.
 
Most of the single-top background events include a $\Wboson$ boson in the final state ($\Wboson t$). To account for the interference between $\Wboson t$ and \ttbar\ production, yields in the signal and control regions are compared between a $\Wboson t$ simulated sample that uses the diagram-subtraction scheme, and the nominal sample that uses the diagram-removal scheme~\cite{Frixione:2008yi}.
The final single-top uncertainty relative to the total background estimate is maximally 7\% (in SRA-TT).
 
For the \ttbar+$W/Z$ background, largely dominated by $\ttbar+Z$, the modelling uncertainty is estimated through variations of the renormalisation and factorisation scales simultaneously by factors of 2.0 and 0.5, and a comparison of parton-showering models (\PYTHIA\ vs \HERWIGV{7}), resulting in a maximum uncertainty of 4\% (in SRB-TT).

The modelling uncertainties for the \Vjets\ background processes due to missing higher orders are evaluated~\cite{Bothmann:2016nao} using both coherent and independent variations of the QCD factorisation and renormalisation scales in the matrix elements by factors of $0.5$ and $2$, avoiding variations in opposite directions. The matrix-element matching scale between jets from the matrix element and the parton shower, and the resummation scale for soft gluon emission within \SHERPA, are also varied by factors of $0.5$ and $2$. The resulting impact on the total background yields from the \Vjets\ modelling is at most $7\%$ (in SRD0).
 
Uncertainties in each background from scale variations are fully correlated across regions and categories, and uncorrelated between processes. In some cases this may result in uncertainties cancelling out, while the higher-order corrections may not cancel out. The sensitivity of the results to the correlation assumptions was tested by redoing the fit with scale variations uncorrelated across all regions and categories, which resulted in negligible changes in the excluded cross sections near the edge of the exclusion region.
 
Detector and theory-related systematic uncertainties in signal yields are also evaluated when setting exclusion limits on specific signals (see Section~\ref{sec:result}). Detector-related uncertainties consider the same sources as for the background and are usually smaller than the modelling uncertainties. Signal theory uncertainties include sources related to signal acceptance, which are included in the profile likelihood fits as a single nuisance parameter, and the uncertainty in the total cross section, which is accounted for by repeating the exclusion procedure for the central and $\pm 1\sigma$ values of the cross section. The uncertainty in the total cross section is 7--16\% for direct top squark production~\cite{Beenakker:2016lwe,Beenakker:2014sma,Beenakker:2013mva,Beenakker:2011sf,Beenakker:2009ha,Kulesza:2009kq,Kulesza:2008jb,Beenakker:1996ch}, depending on the top squark mass. The same uncertainty is used for leptoquark production, due to the similarities between the two types of signals.
 
The main detector-related uncertainties in the signal yields originate from the JER, ranging from 2\% to 15\%, the JES, ranging from 2\% to 20\%, and the $b$-tagging performance (including track-jet $b$-tagging in SRD), ranging from 2\% to 11\%. Uncertainties in the signal acceptance due to variations of the renormalisation and factorisation scales and the matching scale (each varied up and down by a factor of two), and the parton-shower tune variations, are also taken into consideration. In regions where a high-\pt\ ISR system is selected (SRC and SRD), the \pt\ scale of the ISR is large enough such that the leading ISR jet is included in the matrix elements. The uncertainty in the ISR is therefore accounted for when varying the renormalisation and factorisation scales. The total uncertainty in the signal acceptance, considering the full range of $\mstop$ and $\mLSP$ used in this search, is at most 12--13\% in SRA--C, and 25\% for SRD.
% End of text imported from the .//tex/Systematics.tex input file
 
\FloatBarrier
 
% The next lines are included from the .//tex/Results.tex input file
\section{Results and interpretation}
\label{sec:result}
 
\begin{figure}[htpb]
\begin{center}
\includegraphics[width=0.8\textwidth]{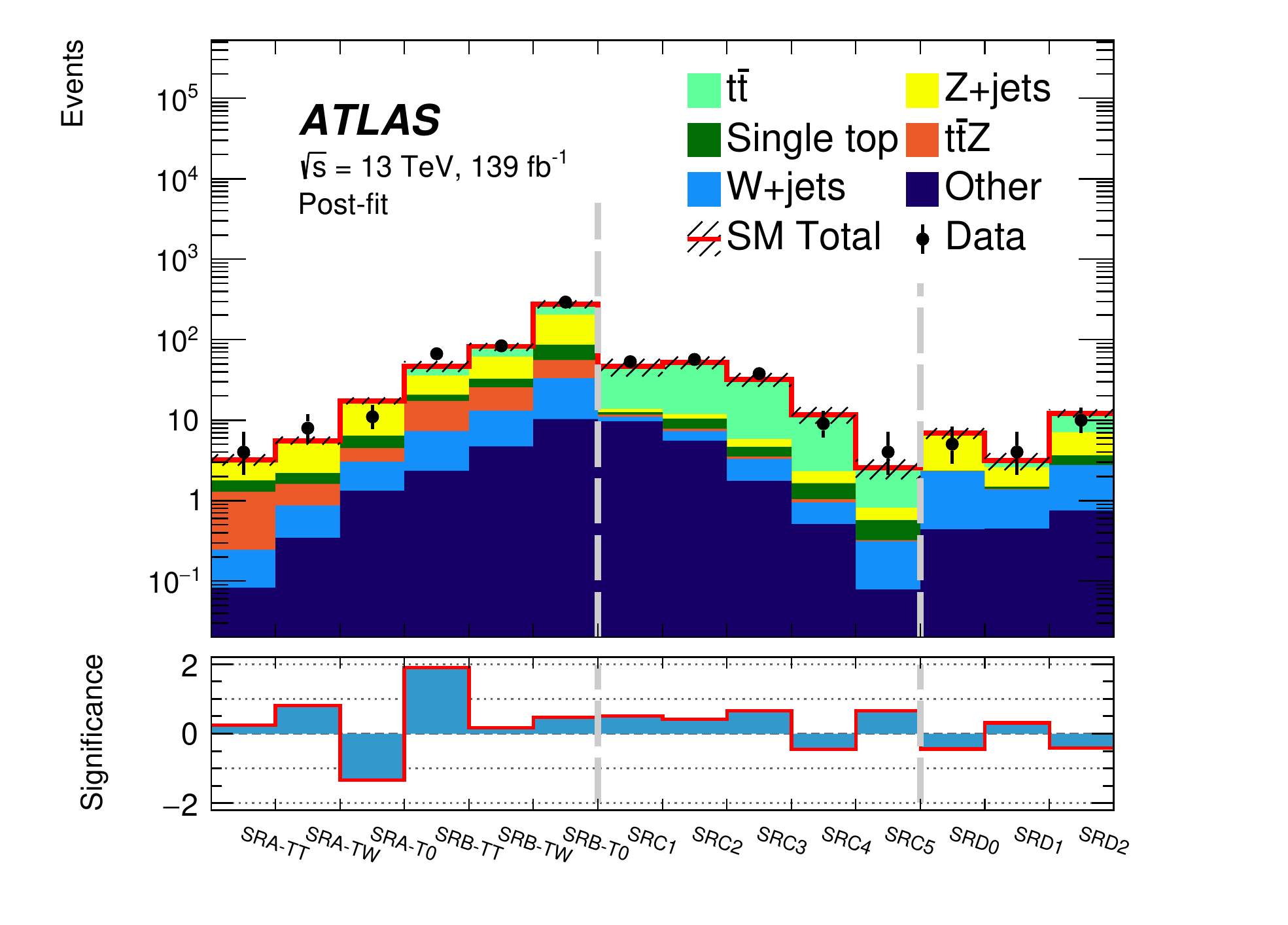}
\caption{Event yields comparing data (points) to the SM prediction (stacked histograms) in all signal regions after the SRA--B, SRC and SRD background-only fits. The hatched uncertainty band around the SM prediction shows the total uncertainty, which includes the MC statistical uncertainties, detector-related systematic uncertainties, and theoretical uncertainties. The signal regions included in the SRA--B, SRC and SRD fits are separated by vertical dashed lines.}
\label{fig:srSummary}
\end{center}
\end{figure}

\begin{figure}[htpb]
\begin{center}
\subfloat[]{\includegraphics[width=0.49\textwidth]{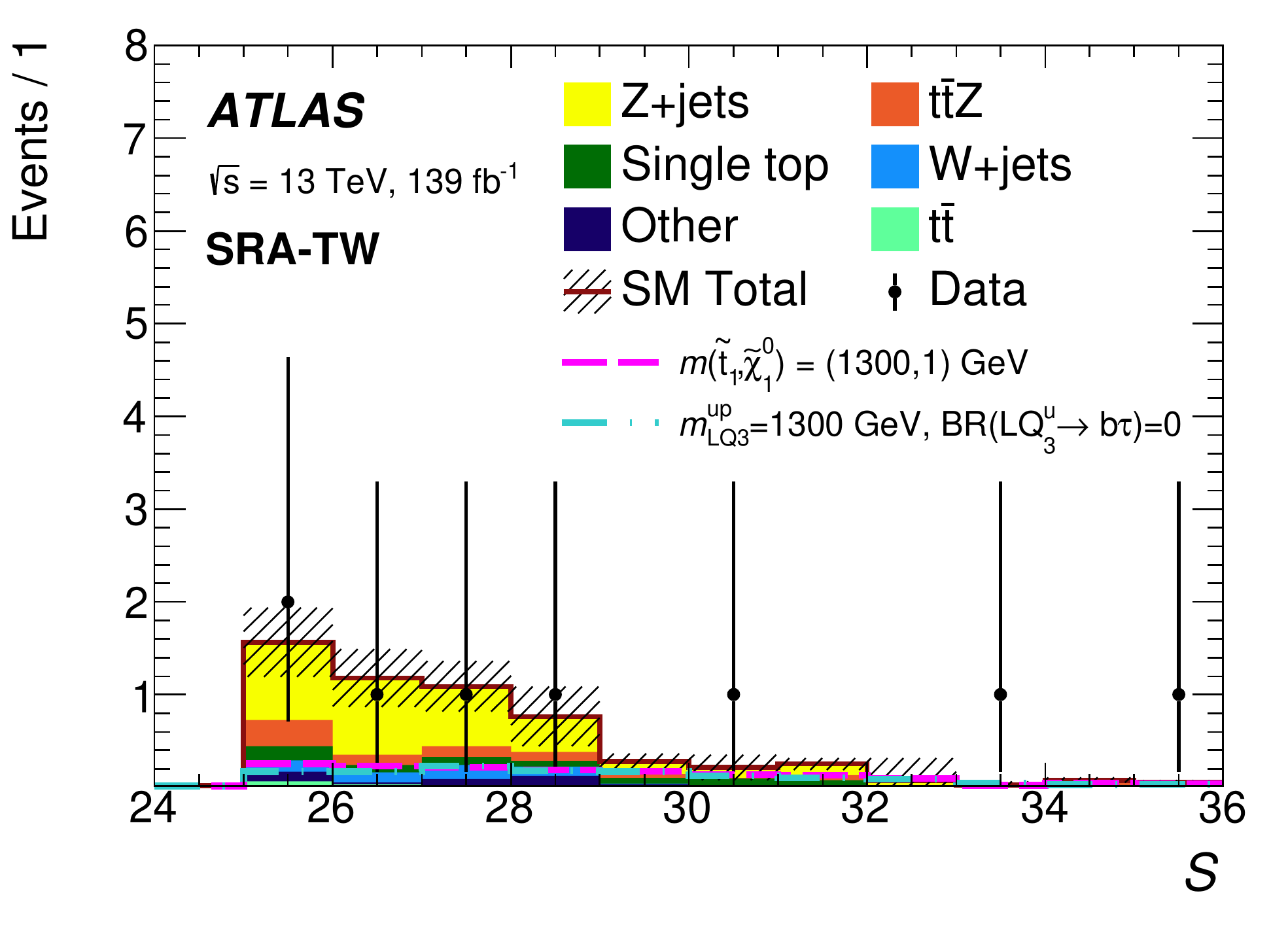}}
\subfloat[]{\includegraphics[width=0.49\textwidth]{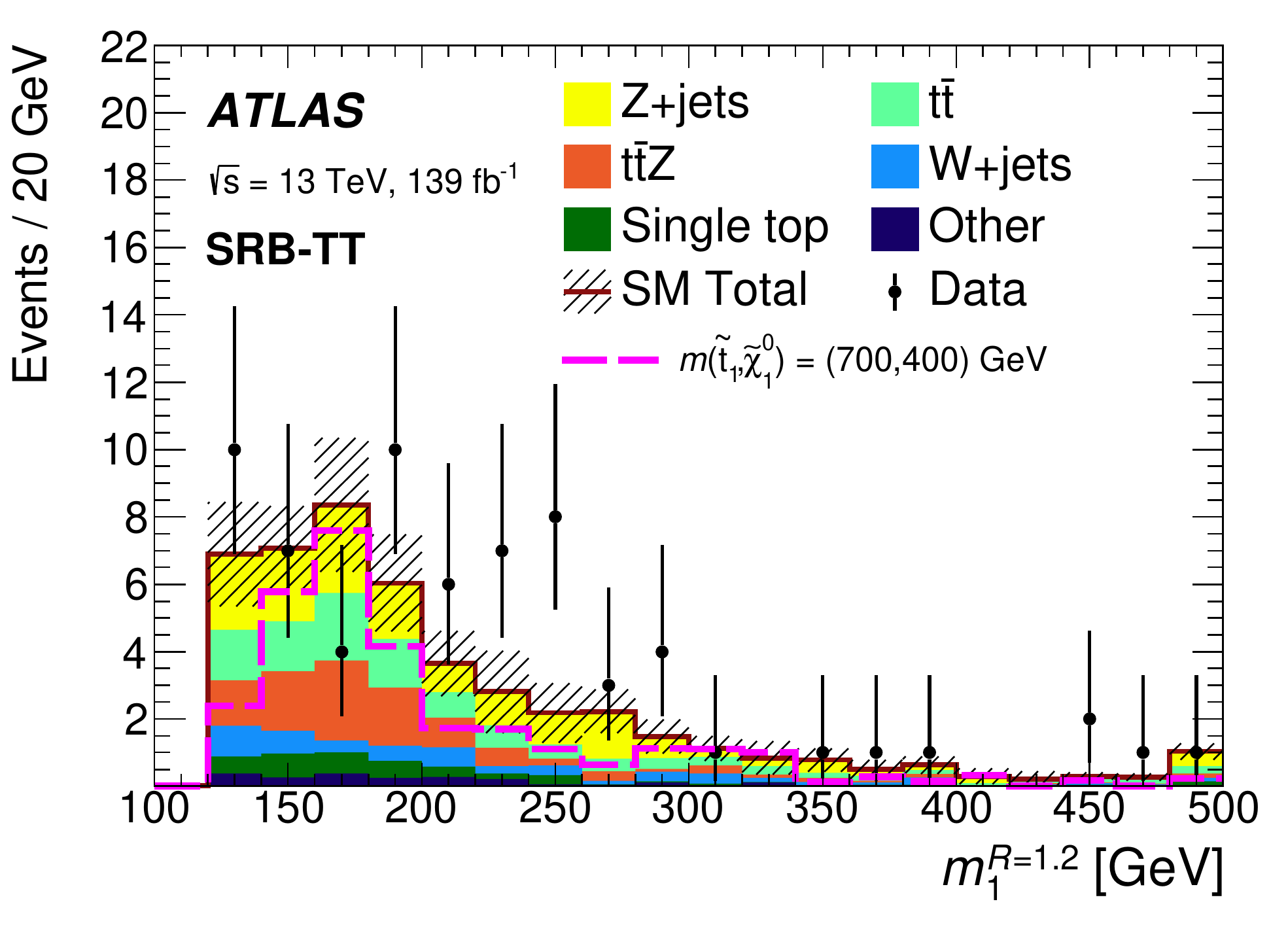}}\\
\subfloat[]{\includegraphics[width=0.49\textwidth]{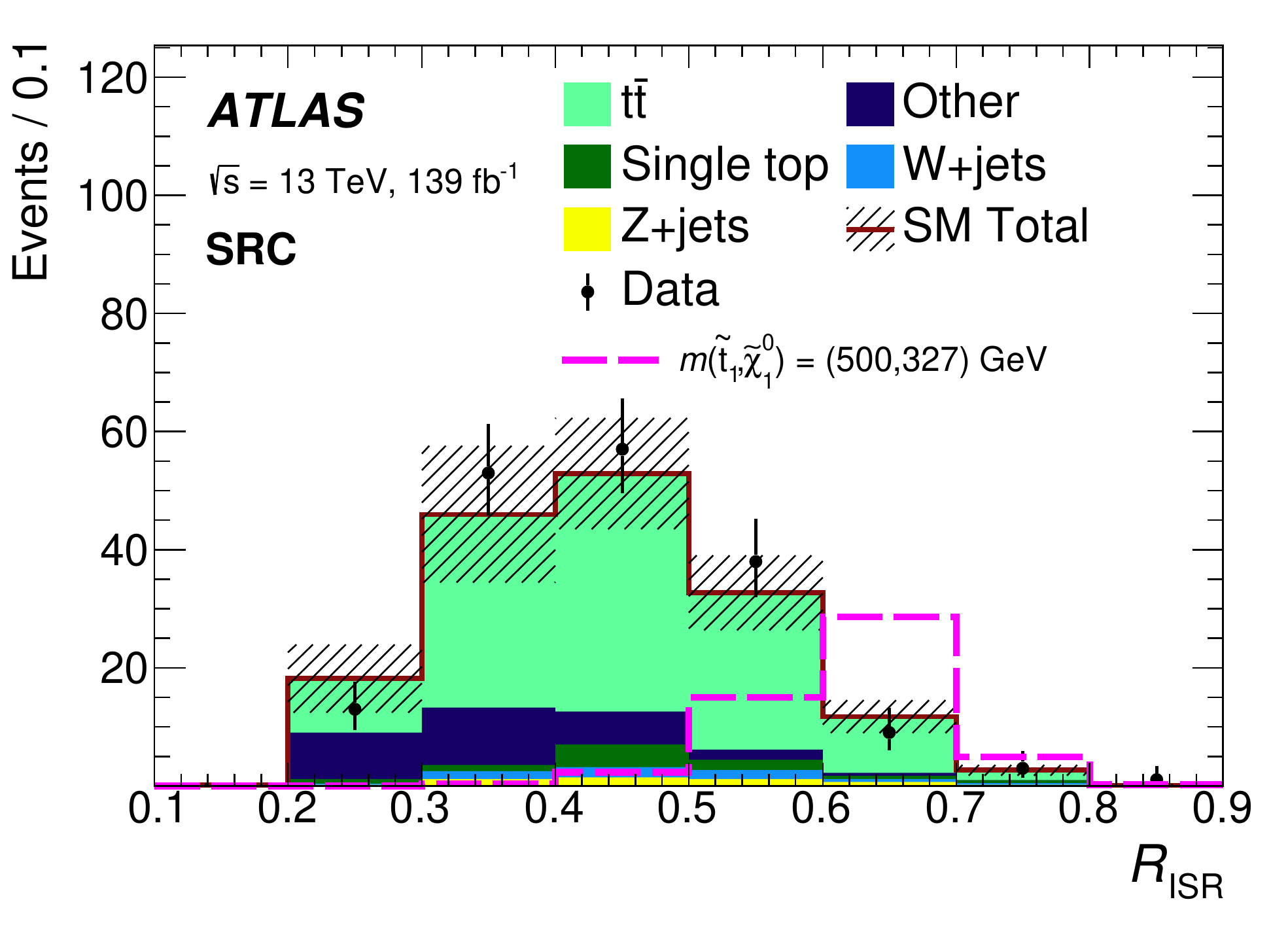}}
\subfloat[]{\includegraphics[width=0.49\textwidth]{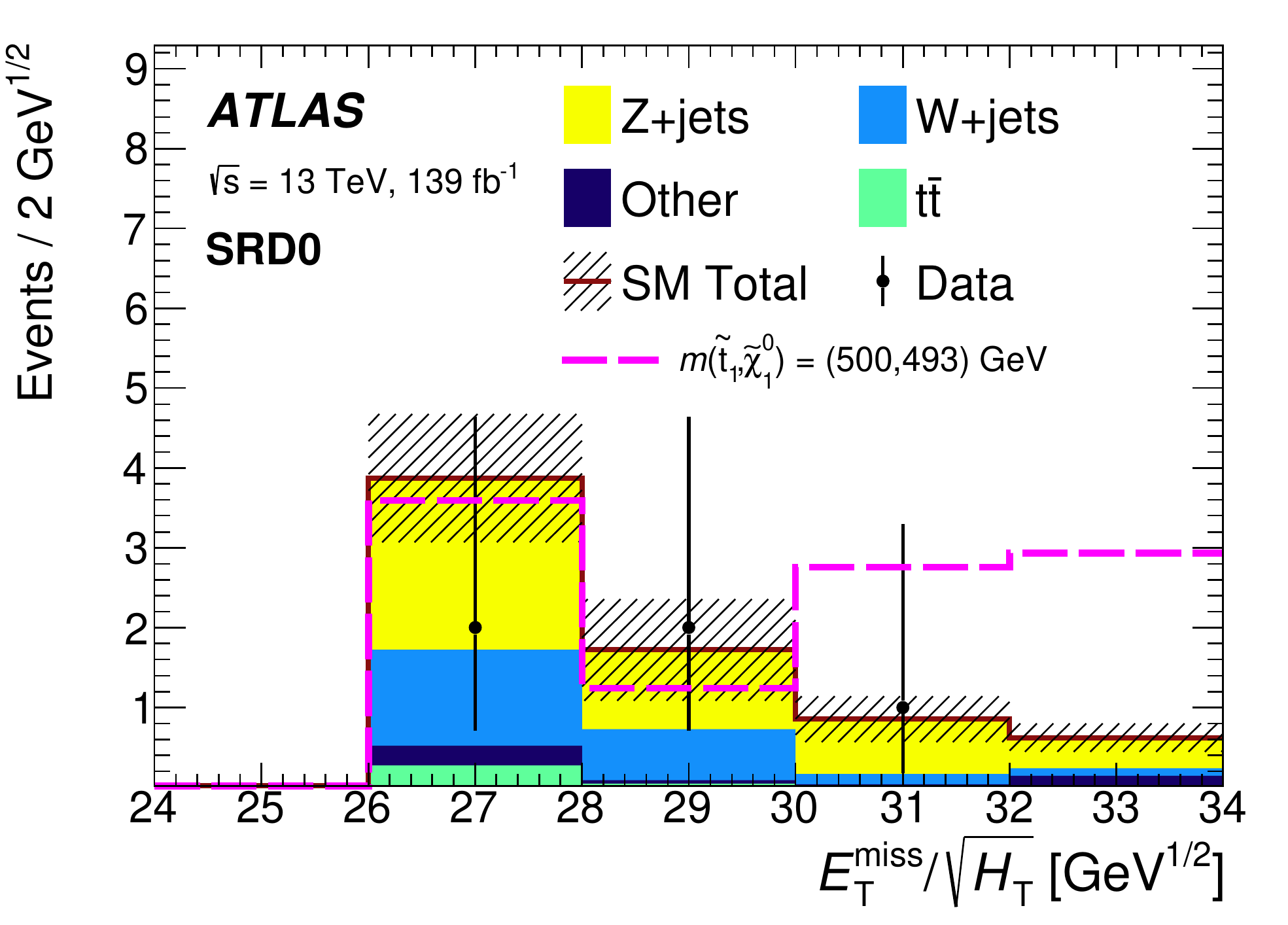}}\\
\subfloat[]{\includegraphics[width=0.49\textwidth]{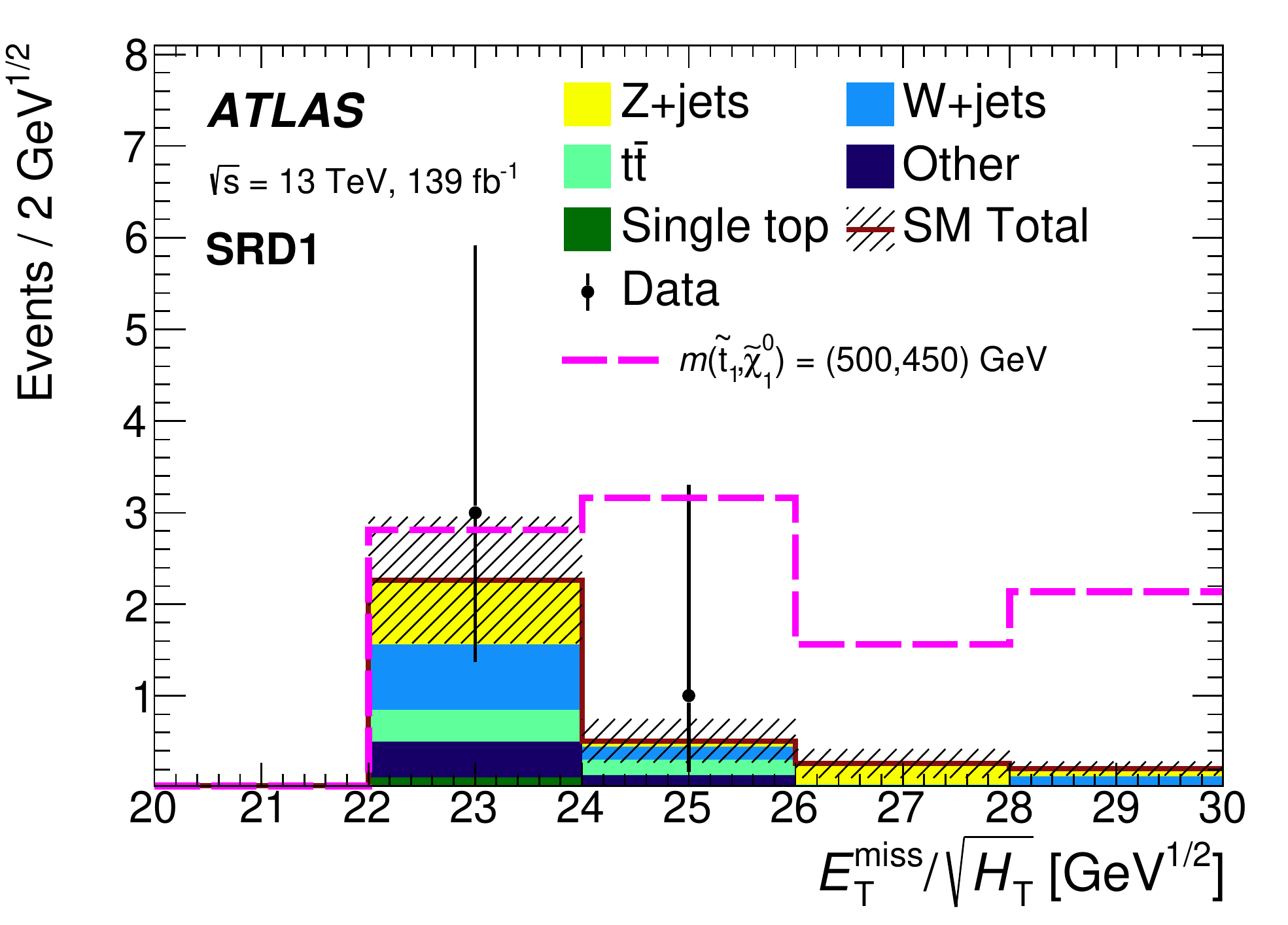}}
\subfloat[]{\includegraphics[width=0.49\textwidth]{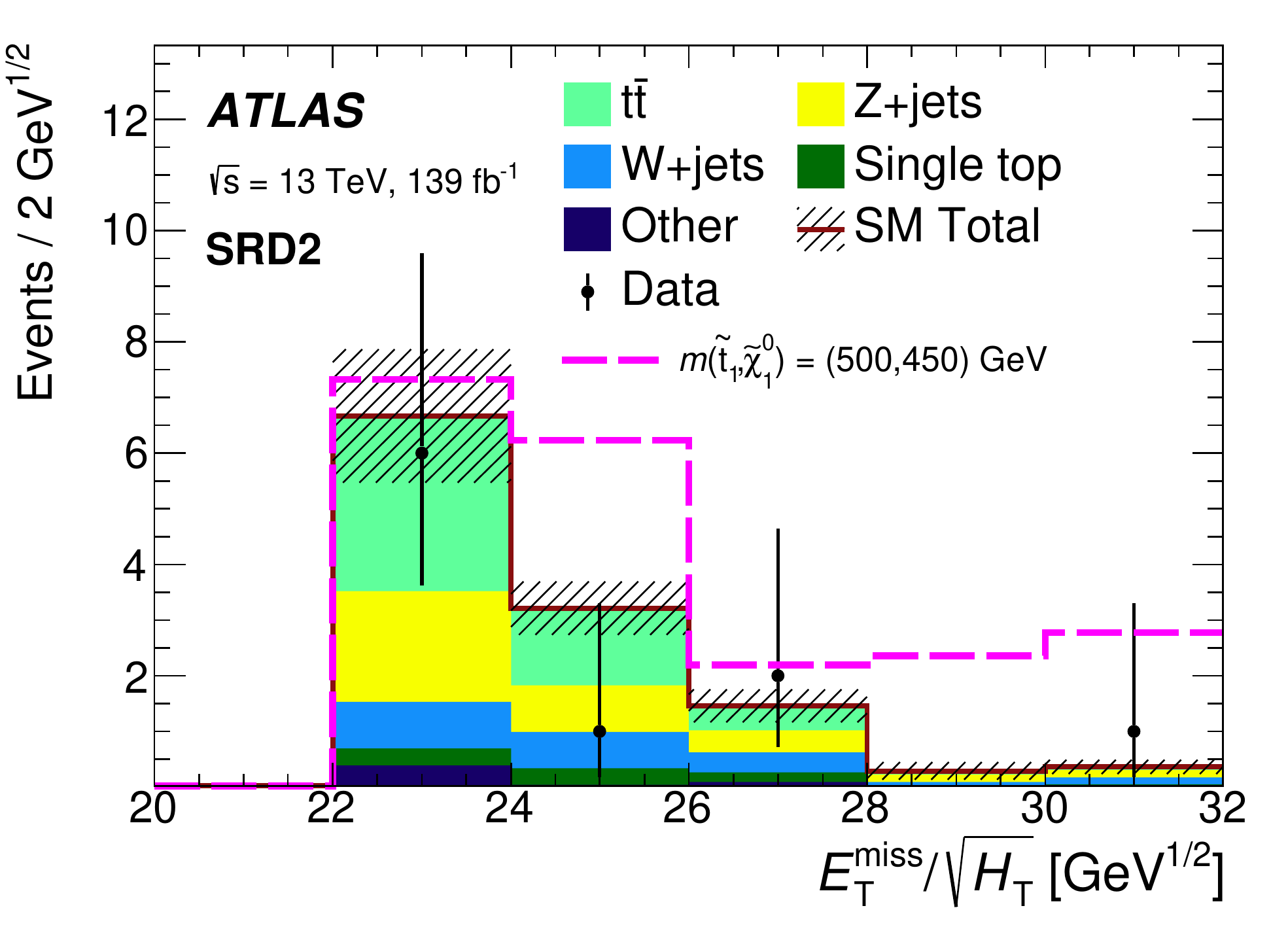}}
\caption{Distributions of (a) \objmetsig\ in SRA-TW, (b) \mantikttwelvezero\ in SRB-TT, (c) \rISR\ in SRC, and (d--f) \htsig\  in SRD0--2 after the SRA--B, SRC and SRD background-only likelihood fits. The stacked histograms show the SM prediction and the hatched uncertainty band around the SM prediction shows the total uncertainty, which includes the MC statistical uncertainties, detector-related systematic uncertainties, and theoretical uncertainties. The data (points) are overlaid. For each variable, the distribution for a representative signal hypothesis is overlaid as a dashed line. The rightmost bin includes overflow events.}
\label{fig:SRs}
\end{center}
\end{figure}

The background originating from SM processes is determined separately for each set of signal regions (SRA--B, SRC, SRD) from three profile likelihood `background-only' fits (SRA--B fit, SRC fit, SRD fit) that include the relevant control regions as described in Section~\ref{sec:background}. The observed event yields in the various SRA--B, SRC and SRD categories are compared with the post-fit background estimates in
Tables~\ref{tab:srABYields},~\ref{tab:srCYields}, ~\ref{tab:srDYields}, and Figure~\ref{fig:srSummary}. In the SRs optimised for discovery, SRA-TT-Disc and SRC-Disc, 14 and 28 events are observed, respectively, compared with $15.2\pm1.8$ and $28.0\pm4.9$ expected events, respectively. Figure~\ref{fig:SRs} shows the distribution of \objmetsig\ in SRA-TW, \mantikttwelvezero\ in SRB-TT, \rISR\ in SRC, and \htsig\ in SRD0, SRD1 and SRD2. The background predictions are scaled to the values determined from the background-only fits.
 
\begin{table}[htpb]
\caption{Observed event yields in SRA and SRB compared with the expected SM background yields in each signal region after the SRA--B background-only fit. The uncertainties include MC statistical uncertainties, detector-related systematic uncertainties, and theoretical uncertainties.}
\begin{center}
{\renewcommand{\arraystretch}{1.2}
% The next lines are included from the .//tex/SRABYieldTable_QCDInOther.tex input file
\begin{tabular}{
S[table-alignment=left]||
S[table-number-alignment=left]
S[table-number-alignment=left]
S[table-number-alignment=left]
S[table-number-alignment=left]
S[table-number-alignment=left]
S[table-number-alignment=left]
S[table-number-alignment=left]
S[table-number-alignment=left]
S[table-number-alignment=left]
S[table-number-alignment=left]
S[table-number-alignment=left]
S[table-number-alignment=left]
}
\hline\hline
& \multicolumn{2}{c}{\textbf{SRA-TT}} & \multicolumn{2}{c}{\textbf{SRA-TW}} & \multicolumn{2}{c}{\textbf{SRA-T0}} & \multicolumn{2}{c}{\textbf{SRB-TT}} & \multicolumn{2}{c}{\textbf{SRB-TW}} & \multicolumn{2}{c}{\textbf{SRB-T0}}\\ \hline\hline
{Observed} & \multicolumn{1}{l}{$4$} &  & \multicolumn{1}{l}{$8$} &  & \multicolumn{1}{l}{$11$} &  & \multicolumn{1}{l}{$67$} &  & \multicolumn{1}{l}{$84$} &  & \multicolumn{1}{l}{$292$} & \\ \hline
{Total SM} & \multicolumn{2}{l}{$3.2\phantom{00} \pm 0.5\phantom{00}$} & \multicolumn{2}{l}{$5.6\phantom{0} \pm 0.7\phantom{0}$} & \multicolumn{2}{l}{$17.3\phantom{0} \pm 1.7\phantom{0}$} & \multicolumn{2}{l}{$46\phantom{0.} \pm 7\phantom{0.}$} & \multicolumn{2}{l}{$81\phantom{0.} \pm 7\phantom{0.}$} & \multicolumn{2}{l}{$276\phantom{0.} \pm 24\phantom{0.}$}\\ \hline
{\Zjets} & \multicolumn{2}{l}{$1.35\phantom{0} \pm 0.28\phantom{0}$} & \multicolumn{2}{l}{$3.2\phantom{0} \pm 0.4\phantom{0}$} & \multicolumn{2}{l}{$10.5\phantom{0} \pm 1.3\phantom{0}$} & \multicolumn{2}{l}{$15.6 \pm 3.3$} & \multicolumn{2}{l}{$28.7 \pm 3.4$} & \multicolumn{2}{l}{$117\phantom{0.} \pm 14\phantom{0.}$}\\
{Single top} & \multicolumn{2}{l}{$0.50\phantom{0} \pm 0.31\phantom{0}$} & \multicolumn{2}{l}{$0.59 \pm 0.29$} & \multicolumn{2}{l}{$\phantom{1}1.9\phantom{0} \pm 0.8\phantom{0}$} & \multicolumn{2}{l}{$\phantom{4}3.5 \pm 1.2$} & \multicolumn{2}{l}{$\phantom{8}7.0 \pm 3.0$} & \multicolumn{2}{l}{$\phantom{2}31\phantom{0.} \pm 15\phantom{0.}$}\\
{\ttbar} & \multicolumn{2}{l}{$0.08\phantom{0} \pm 0.07\phantom{0}$} & \multicolumn{2}{l}{$0.16 \pm 0.10$} & \multicolumn{2}{l}{$\phantom{1}0.4\phantom{0} \pm 0.4\phantom{0}$} & \multicolumn{2}{l}{$10\phantom{0.} \pm 5\phantom{0.}$} & \multicolumn{2}{l}{$20\phantom{0.} \pm 6\phantom{0.}$} & \multicolumn{2}{l}{$\phantom{2}72\phantom{0.} \pm 19\phantom{0.}$}\\
{\ttZ} & \multicolumn{2}{l}{$1.05\phantom{0} \pm 0.29\phantom{0}$} & \multicolumn{2}{l}{$0.74 \pm 0.17$} & \multicolumn{2}{l}{$\phantom{1}1.50 \pm 0.34$} & \multicolumn{2}{l}{$\phantom{4}9.9 \pm 1.9$} & \multicolumn{2}{l}{$12.5 \pm 2.5$} & \multicolumn{2}{l}{$\phantom{2}22\phantom{0.} \pm \phantom{2}4\phantom{0.}$}\\
{\Wjets} & \multicolumn{2}{l}{$0.16\phantom{0} \pm 0.07\phantom{0}$} & \multicolumn{2}{l}{$0.53 \pm 0.20$} & \multicolumn{2}{l}{$\phantom{1}1.7\phantom{0} \pm 0.6\phantom{0}$} & \multicolumn{2}{l}{$\phantom{4}4.9 \pm 1.9$} & \multicolumn{2}{l}{$\phantom{8}8\phantom{0.} \pm 4\phantom{0.}$} & \multicolumn{2}{l}{$\phantom{2}22\phantom{0.} \pm \phantom{2}9\phantom{0.}$}\\
{Other} & \multicolumn{2}{l}{$0.080 \pm 0.020$} & \multicolumn{2}{l}{$0.34 \pm 0.24$} & \multicolumn{2}{l}{$\phantom{1}1.32 \pm 0.22$} & \multicolumn{2}{l}{$\phantom{4}2.4 \pm 0.4$} & \multicolumn{2}{l}{$\phantom{8}4.7 \pm 2.3$} & \multicolumn{2}{l}{$\phantom{2}10.4 \pm \phantom{2}1.3$}\\
\hline\hline
\end{tabular}
% End of text imported from the .//tex/SRABYieldTable_QCDInOther.tex input file
}
\end{center}
\label{tab:srABYields}
\end{table}
 
\begin{table}[htpb]
\caption{Observed event yields in SRC compared with expected SM background yields in each signal region after the background-only fit.
The uncertainties include MC statistical uncertainties, detector-related systematic uncertainties, and theoretical uncertainties.}
\begin{center}
{\renewcommand{\arraystretch}{1.2}
% The next lines are included from the .//tex/SRCYieldTable_QCDInOther.tex input file
\begin{tabular}{
S[table-alignment=left]||
S[table-number-alignment=left]
S[table-number-alignment=left]
S[table-number-alignment=left]
S[table-number-alignment=left]
S[table-number-alignment=left]
S[table-number-alignment=left]
S[table-number-alignment=left]
S[table-number-alignment=left]
S[table-number-alignment=left]
S[table-number-alignment=left]
}
\hline\hline
& \multicolumn{2}{c}{\textbf{SRC1}} & \multicolumn{2}{c}{\textbf{SRC2}} & \multicolumn{2}{c}{\textbf{SRC3}} & \multicolumn{2}{c}{\textbf{SRC4}} & \multicolumn{2}{c}{\textbf{SRC5}}\\ \hline\hline
{Observed} & \multicolumn{1}{l}{$53$} &  & \multicolumn{1}{l}{$57$} &  & \multicolumn{1}{l}{$38$} &  & \multicolumn{1}{l}{$9$} &  & \multicolumn{1}{l}{$4$} & \\ \hline
{Total SM} & \multicolumn{2}{l}{$46\phantom{00.} \pm 12\phantom{00.}$} & \multicolumn{2}{l}{$52\phantom{00.} \pm 9\phantom{00.}$} & \multicolumn{2}{l}{$32\phantom{00.} \pm 7\phantom{00.}$} & \multicolumn{2}{l}{$11.8\phantom{0} \pm 3.1\phantom{0}$} & \multicolumn{2}{l}{$2.5\phantom{00} \pm 0.7\phantom{00}$}\\ \hline
{\Zjets} & \multicolumn{2}{l}{$\phantom{4}1.2\phantom{0} \pm \phantom{1}0.5\phantom{0}$} & \multicolumn{2}{l}{$\phantom{5}1.4\phantom{0} \pm 0.8\phantom{0}$} & \multicolumn{2}{l}{$\phantom{3}1.2\phantom{0} \pm 0.5\phantom{0}$} & \multicolumn{2}{l}{$\phantom{1}0.67 \pm 0.32$} & \multicolumn{2}{l}{$0.24\phantom{0} \pm 0.12\phantom{0}$}\\
{Single top} & \multicolumn{2}{l}{$\phantom{4}0.90 \pm \phantom{1}0.34$} & \multicolumn{2}{l}{$\phantom{5}2.7\phantom{0} \pm 0.6\phantom{0}$} & \multicolumn{2}{l}{$\phantom{3}1.1\phantom{0} \pm 0.7\phantom{0}$} & \multicolumn{2}{l}{$\phantom{1}0.61 \pm 0.33$} & \multicolumn{2}{l}{$0.25\phantom{0} \pm 0.05\phantom{0}$}\\
{\ttbar} & \multicolumn{2}{l}{$32\phantom{00.} \pm 11\phantom{00.}$} & \multicolumn{2}{l}{$40\phantom{00.} \pm 9\phantom{00.}$} & \multicolumn{2}{l}{$26\phantom{00.} \pm 6\phantom{00.}$} & \multicolumn{2}{l}{$\phantom{1}9.5\phantom{0} \pm 2.7\phantom{0}$} & \multicolumn{2}{l}{$1.7\phantom{00} \pm 0.6\phantom{00}$}\\
{\ttZ} & \multicolumn{2}{l}{$\phantom{4}0.74 \pm \phantom{1}0.32$} & \multicolumn{2}{l}{$\phantom{5}0.50 \pm 0.31$} & \multicolumn{2}{l}{$\phantom{3}0.24 \pm 0.12$} & \multicolumn{2}{l}{$\phantom{1}0.09\;_{-\;0.09}^{+\;0.11}$} & \multicolumn{2}{l}{$0.010\;_{-\;0.010}^{+\;0.030}$}\\
{\Wjets} & \multicolumn{2}{l}{$\phantom{4}1.3\phantom{0} \pm \phantom{1}0.6\phantom{0}$} & \multicolumn{2}{l}{$\phantom{5}1.7\phantom{0} \pm 0.6\phantom{0}$} & \multicolumn{2}{l}{$\phantom{3}1.5\phantom{0} \pm 0.7\phantom{0}$} & \multicolumn{2}{l}{$\phantom{1}0.4\phantom{0}\;_{-\;0.4}^{+\;0.6\phantom{0}}$} & \multicolumn{2}{l}{$0.23\phantom{0} \pm 0.09\phantom{0}$}\\
{Other} & \multicolumn{2}{l}{$\phantom{4}9.7\phantom{0} \pm \phantom{1}2.9\phantom{0}$} & \multicolumn{2}{l}{$\phantom{5}5.6\phantom{0} \pm 1.5\phantom{0}$} & \multicolumn{2}{l}{$\phantom{3}1.8\phantom{0} \pm 0.5\phantom{0}$} & \multicolumn{2}{l}{$\phantom{1}0.51 \pm 0.10$} & \multicolumn{2}{l}{$0.08\phantom{0}\;_{-\;0.08}^{+\;0.19\phantom{0}}$}\\
\hline\hline
\end{tabular}
% End of text imported from the .//tex/SRCYieldTable_QCDInOther.tex input file
}
\end{center}
\label{tab:srCYields}
\end{table}
 
\begin{table}[htpb]
\caption{Observed and expected yields after the background-only fit, for SRD.
The uncertainties include MC statistical uncertainties, detector-related systematic uncertainties, and theoretical uncertainties.}
\begin{center}
{\renewcommand{\arraystretch}{1.2}
% The next lines are included from the .//tex/SRDYieldTable_QCDInOther.tex input file
\begin{tabular}{
S[table-alignment=left]||
S[table-number-alignment=left]
S[table-number-alignment=left]
S[table-number-alignment=left]
S[table-number-alignment=left]
S[table-number-alignment=left]
S[table-number-alignment=left]
}
\hline\hline
& \multicolumn{2}{c}{\textbf{SRD0}} & \multicolumn{2}{c}{\textbf{SRD1}} & \multicolumn{2}{c}{\textbf{SRD2}}\\ \hline\hline
{Observed} & \multicolumn{1}{l}{$5$} &  & \multicolumn{1}{l}{$4$} &  & \multicolumn{1}{l}{$10$} & \\ \hline
{Total SM} & \multicolumn{2}{l}{$6.9\phantom{00} \pm 1.3\phantom{00}$} & \multicolumn{2}{l}{$3.1\phantom{00} \pm 1.0\phantom{00}$} & \multicolumn{2}{l}{$12.2\phantom{0} \pm 1.5\phantom{0}$}\\ \hline
{\Zjets} & \multicolumn{2}{l}{$4.2\phantom{00} \pm 0.8\phantom{00}$} & \multicolumn{2}{l}{$1.07\phantom{0} \pm 0.25\phantom{0}$} & \multicolumn{2}{l}{$\phantom{1}3.5\phantom{0} \pm 0.6\phantom{0}$}\\
{Single top} & \multicolumn{2}{l}{$0.020\;_{-\;0.020}^{+\;0.030}$} & \multicolumn{2}{l}{$0.10\phantom{0}\;_{-\;0.10}^{+\;0.16\phantom{0}}$} & \multicolumn{2}{l}{$\phantom{1}0.84 \pm 0.31$}\\
{\ttbar} & \multicolumn{2}{l}{$0.36\phantom{0} \pm 0.29\phantom{0}$} & \multicolumn{2}{l}{$0.56\phantom{0} \pm 0.31\phantom{0}$} & \multicolumn{2}{l}{$\phantom{1}5.1\phantom{0} \pm 1.0\phantom{0}$}\\
{\ttZ} & \multicolumn{2}{l}{$0.02\phantom{0}\;_{-\;0.02}^{+\;0.04\phantom{0}}$} & \multicolumn{2}{l}{$0.010\;_{-\;0.010}^{+\;0.010}$} & \multicolumn{2}{c}{$<0.01$}\\
{\Wjets} & \multicolumn{2}{l}{$1.9\phantom{00} \pm 0.6\phantom{00}$} & \multicolumn{2}{l}{$0.9\phantom{00} \pm 0.6\phantom{00}$} & \multicolumn{2}{l}{$\phantom{1}2.0\phantom{0} \pm 0.7\phantom{0}$}\\
{Other} & \multicolumn{2}{l}{$0.44\phantom{0} \pm 0.13\phantom{0}$} & \multicolumn{2}{l}{$0.45\phantom{0} \pm 0.21\phantom{0}$} & \multicolumn{2}{l}{$\phantom{1}0.76 \pm 0.20$}\\
\hline\hline
\end{tabular}
% End of text imported from the .//tex/SRDYieldTable_QCDInOther.tex input file
}
\end{center}
\label{tab:srDYields}
\end{table}
 
Observed event yields are in good agreement with the background estimates in all the signal regions. The significance of a data excess with respect to the background predictions can be quantified by the probability ($p$) of a background-only hypothesis to be more signal-like than what is observed. To evaluate these probabilities in each signal region category, alternative fit configurations (discovery fits) are defined. Each discovery likelihood function is defined as the product of the Poisson probability density function describing the numbers of events of a single signal region category and the background-only likelihood function associated with that signal region.
An additional parameter, the signal strength, defined for positive values and corresponding to the signal normalisation in the signal region, is included and free-floating in the fit.
The smallest $p$-value, assuming the background-only hypothesis, is 0.03, corresponding to 1.87$\sigma$, in SRB-TT. In this signal region, 67 events are observed compared with $46.7\pm6.7$ expected events. The largest deficit in the data is found in SRA-T0 where 11 events are observed compared with $17.3\pm1.7$ expected events.
 
Model-independent upper limits set at 95\% CL on the number of beyond the SM (BSM) events in each signal region are derived using the CL$_\mathrm{s}$ prescription~\cite{Read:2002hq} and neglecting any possible signal contamination in the control regions. Normalising these by the integrated luminosity of the data sample, they are interpreted as upper limits on the visible BSM cross section, $\sigma_{\mathrm{vis}}$, where $\sigma_{\mathrm{vis}}$ is defined as the product of the acceptance, reconstruction efficiency and production cross section. The results from the discovery fits are shown in Table~\ref{tab:upLimits}, with SRA-TT having the lowest upper bound on the visible cross section of new physics phenomena of 0.04~fb.
 
{
\def\arraystretch{1.3}
% The next lines are included from the .//tex/UpperLimits.tex input file

\begin{table}
\centering
\caption[Breakdown of upper limits.]{
Left to right: 95\% CL upper limits on the visible cross section
($\langle\epsilon\sigma\rangle_{\mathrm{obs}}^{95}$) and on the number of
signal events ($S_{\mathrm{obs}}^{95}$ ).  The third column
($S_{\mathrm{exp}}^{95}$) shows the 95\% CL upper limit on the number of
signal events, given the expected number (and $\pm 1\sigma$
excursions on the expectation) of background events.
The last two columns
indicate the $\textrm{CL}_{\textrm{B}}$ value, i.e.\ the confidence level observed for
the background-only hypothesis, the discovery $p$-value ($p(s = 0)$), and the significance (Z). In cases where the observed data yields are less than the total expected background yields, the $p$-value is truncated to 0.5.
\label{tab:upLimits}}
\begin{tabular}{l||ccrcc}
\hline\hline
{\textbf {Signal Region}}                        & \textbf{ $\mathbf{\langle\epsilon{\mathrm \sigma}\rangle_{\mathrm{obs}}^{95}}$~[fb]}  &  \textbf{$S_{\mathbf{\mathrm{obs}}}^{95}$}  &  \textbf{$S_{\mathbf{\mathrm{exp}}}^{95}$} & \textbf{$\mathbf{\mathrm{CL}_{\mathrm{B}}}$} & \textbf{$p(s=0)$ ($\mathbf{Z}$)}  \\
\hline\hline
 
SRA-TT    & $0.04$ &  $6.0$ &${5.2}^{+2.7}_{-1.7}$& $0.63$ & $ 0.34$~$(0.40)$ \\%
SRA-TW    & $0.06$ &  $8.6$ &${6.5}^{+3.2}_{-1.6}$& $0.78$ & $ 0.18$~$(0.92)$ \\%
SRA-T0    & $0.05$ &  $6.4$ &${10}^{+5}_{-3}$& $0.11$ & $ 0.50$~$(0.00)$ \\%
SRA-TT-Disc    & $0.06$ &  $8.4$ &${9}^{+4}_{-2}$& $0.39$ & $ 0.50$~$(0.00)$ \\%
SRB-TT    & $0.28$ &  $38.5$ &${22}^{+9}_{-6}$& $0.95$ & $ 0.03$~$(1.87)$ \\%
SRB-TW    & $0.21$ &  $28.6$ &${27}^{+10}_{-7}$& $0.57$ & $ 0.42$~$(0.19)$ \\%
SRB-T0    & $0.51$ &  $71.1$ &${60}^{+22}_{-16}$& $0.69$ & $ 0.30$~$(0.53)$ \\%
SRC1    & $0.19$ &  $26.0$ &${22}^{+4}_{-9}$& $0.75$ & $ 0.49$~$(0.01)$ \\%
SRC2    & $0.24$ &  $32.8$ &${27}^{+10}_{-7}$& $0.76$ & $ 0.22$~$(0.77)$ \\%
SRC3    & $0.17$ &  $24.0$ &${20}^{+7}_{-5}$& $0.76$ & $ 0.23$~$(0.75)$ \\%
SRC4    & $0.06$ &  $8.0$ &${9}^{+4}_{-2}$& $0.29$ & $ 0.50$~$(0.00)$ \\%
SRC5    & $0.05$ &  $6.6$ &${5.0}^{+2.8}_{-1.2}$& $0.73$ & $ 0.22$~$(0.78)$ \\%
SRC-Disc    & $0.11$ &  $15.4$ &${15}^{+6}_{-4}$& $0.53$ & $ 0.49$~$(0.02)$ \\%
SRD0    & $0.04$ &  $5.4$ &${6.8}^{+3.3}_{-2.1}$& $0.28$ & $ 0.50$~$(0.00)$ \\%
SRD1    & $0.04$ &  $6.2$ &${5.5}^{+2.7}_{-1.8}$& $0.63$ & $ 0.34$~$(0.40)$ \\%
SRD2    & $0.05$ &  $6.9$ &${8}^{+4}_{-2}$& $0.28$ & $ 0.50$~$(0.00)$ \\%
 
\hline\hline
\end{tabular}
 
\end{table}
% End of text imported from the .//tex/UpperLimits.tex input file
}
 
A profile-likelihood-ratio test is defined in order to set limits on direct pair production of top squarks. A new fit configuration is defined (referred to as an exclusion fit) for each signal region category (SRA--B, SRC or SRD), where the Poisson probability density functions describing the observed and expected numbers of events in all relevant signal region bins are included in the likelihood function, and the signal-strength parameter, defined for positive values, is free-floating in the fit. Signal contamination in the CRs is taken into account. Limits are derived using the CL$_\mathrm{s}$ prescription and calculated from asymptotic formulae~\cite{Cowan:2010js}. The nominal event yield in each set of SRs is set using the background-level estimates obtained from a background-only fit to both the CRs and the SRs to determine the expected limits, while a coloured band that represents the $\pm1\sigma$ of the total uncertainty ($\sigma_{\mathrm{exp}}$) is also evaluated. The observed event yields and the same background estimates are used to determine the observed limits for each set of SRs (SRA--B, SRC and SRD); these are evaluated for the nominal signal cross sections as well as for $\pm1\sigma$ theory uncertainties in those cross sections, denoted by $\sigma^{\mathrm{SUSY}}_{\mathrm{theory}}$.
 
\begin{figure}[htpb]
\begin{center}
\subfloat[]{\includegraphics[width=0.5\textwidth]{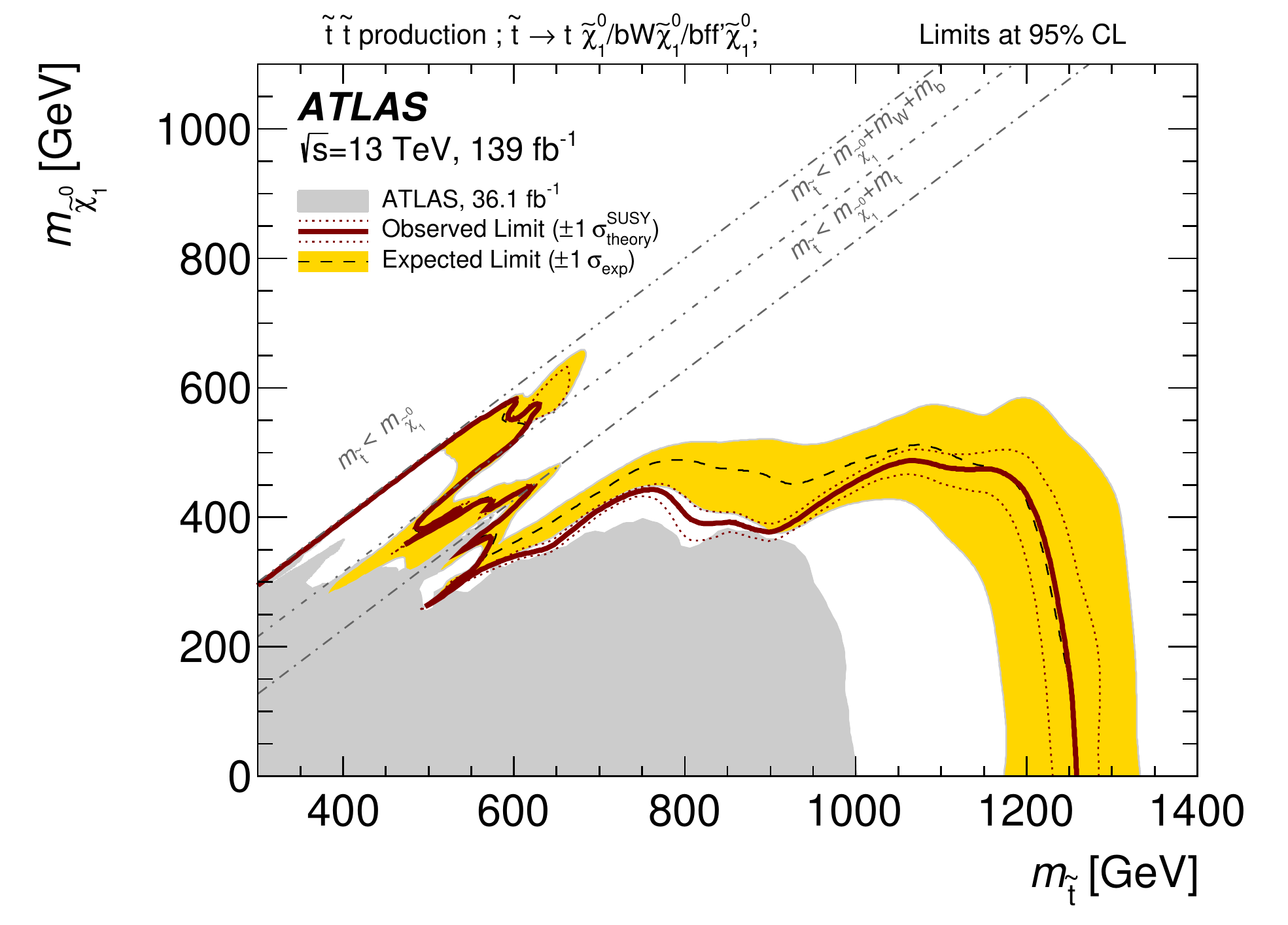}}
\subfloat[]{\includegraphics[width=0.5\textwidth]{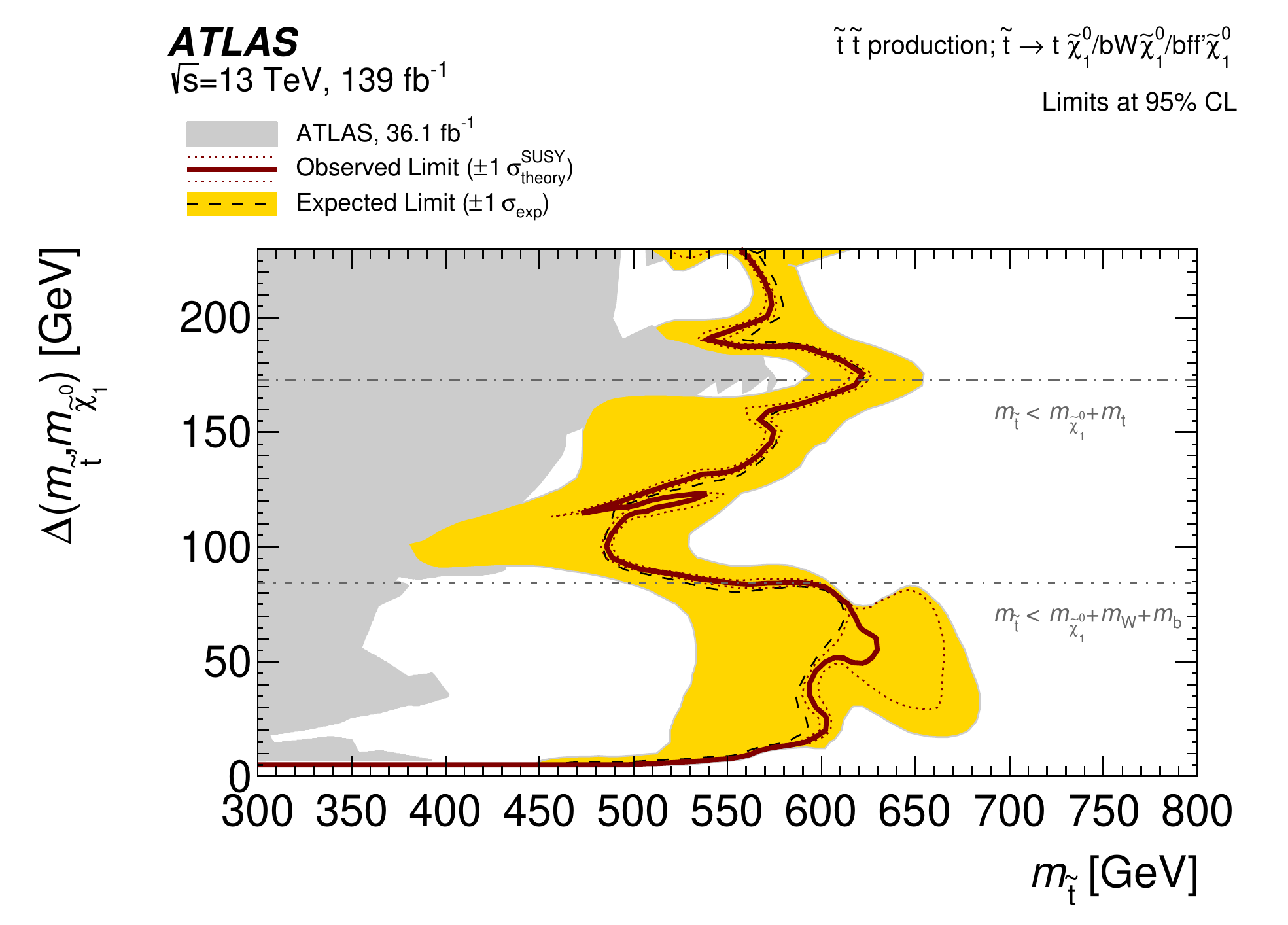}}
\caption{Observed (red solid line) and expected (black dashed line)
exclusion contours at 95\% CL as a function of the (a) \ninoone\ vs $\stop$
masses and (b) $\Delta m(\stop,\ninoone)$ vs $\stop$ mass. Masses that are within the contours are excluded. Uncertainty bands corresponding to the $\pm 1
\sigma$ variation of the expected limit (yellow band) and the
sensitivity of the observed limit to $\pm 1\sigma$ variations of
the signal total cross section (red dotted lines) are also
indicated. Observed limits from previous ATLAS searches~\cite{SUSY-2014-07,SUSY-2016-15,SUSY-2016-16,SUSY-2016-17,EXOT-2016-27,TOPQ-2016-10} based on 36.1 fb$^{-1}$ of $pp$ collision data are provided for comparison in grey.}
\label{fig:SRABCD_exclusion}
\end{center}
\end{figure}

Figure~\ref{fig:SRABCD_exclusion} shows the observed and
expected exclusion contours at 95\% CL as a function of (a) the \ninoone\ mass vs the \stop\ mass and (b) $\Delta m(\stop,\ninoone)$ vs the \stop\ mass. The exclusion contour is obtained by choosing the exclusion fit configuration (SRA--B, SRC or SRD) with the best expected limit for each signal hypothesis.
The data included in the SRA--B fit, together with previous limits from the overlay of zero-, one-, and two-lepton channels and other analyses~\cite{SUSY-2014-07,SUSY-2016-15,SUSY-2016-16,SUSY-2016-17,EXOT-2016-27,TOPQ-2016-10}, exclude top squark masses up to 1250~GeV\ for $\ninoone$ masses below $200$~GeV. Additional constraints are set by the SRC fit in the case where $\Delta m(\stop,\ninoone) \approx m_t$, for which top squark masses in the range \stopLimDiag~GeV\ are excluded. Some structures in the expected exclusion contour are observed in this region and were traced back to the fixed \rISR-binning adopted in SRC. Since the binning of \rISR\ is fixed and does not depend on the signal considered, for some signals the peak is located at the boundary between two bins and therefore leads to a global lower signal over background ratio across all SRC bins. Finally, limits are set by the SRD fit in the case where $\mstop-\mLSP < m_W+m_b$ (with $\Delta m(\stop,\ninoone)\ge 5$~GeV), for which top squark masses in the range \stopLimFourBody~GeV\ are excluded. The sharp structure in the middle of Figure~\ref{fig:SRABCD_exclusion} (b) is an artifact of stitching together the exclusion contours of SRC and SRD. Signals with $\Delta m(\stop,\ninoone)=5$~GeV, which is the smallest $\Delta m(\stop,\ninoone)$ value considered, are excluded for $\mstop<490$~GeV. This is the first time that an ATLAS all-hadronic search reaches exclusion sensitivity in the four-body region. This is due to the newly introduced and dedicated SRD, which takes advantage of track-jet $b$-tagging to discriminate between signal and background. The exclusion limits shown in Figure~\ref{fig:SRABCD_exclusion} are derived for unpolarised top squarks.~\footnote{The polarisation of the top squarks refers to the fraction of right-handed vs left-handed components in the $\stopone$ mass eigenstate.} The exclusion limit stays within $\pm 25$~GeV\ of top squark mass when varying the top squark polarisation, which is within the $\sigma^{\mathrm{SUSY}}_{\mathrm{theory}}$ uncertainty band; hadronic final states are less sensitive to polarisation effects than final states with one or more leptons.

\begin{figure}[htpb]
\begin{center}
\subfloat[]{\includegraphics[width=0.47\textwidth]{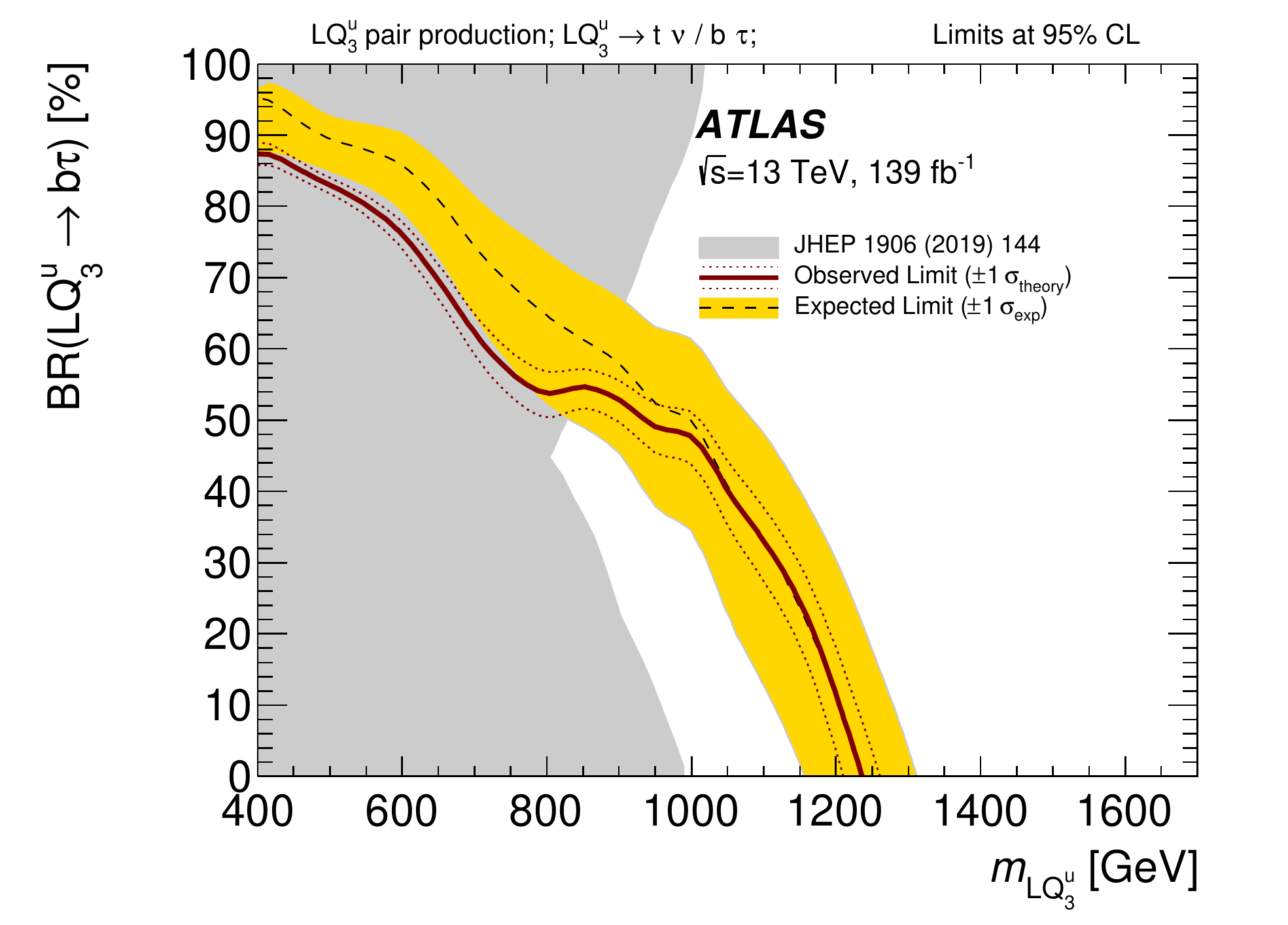}}
\subfloat[]{\includegraphics[width=0.47\textwidth]{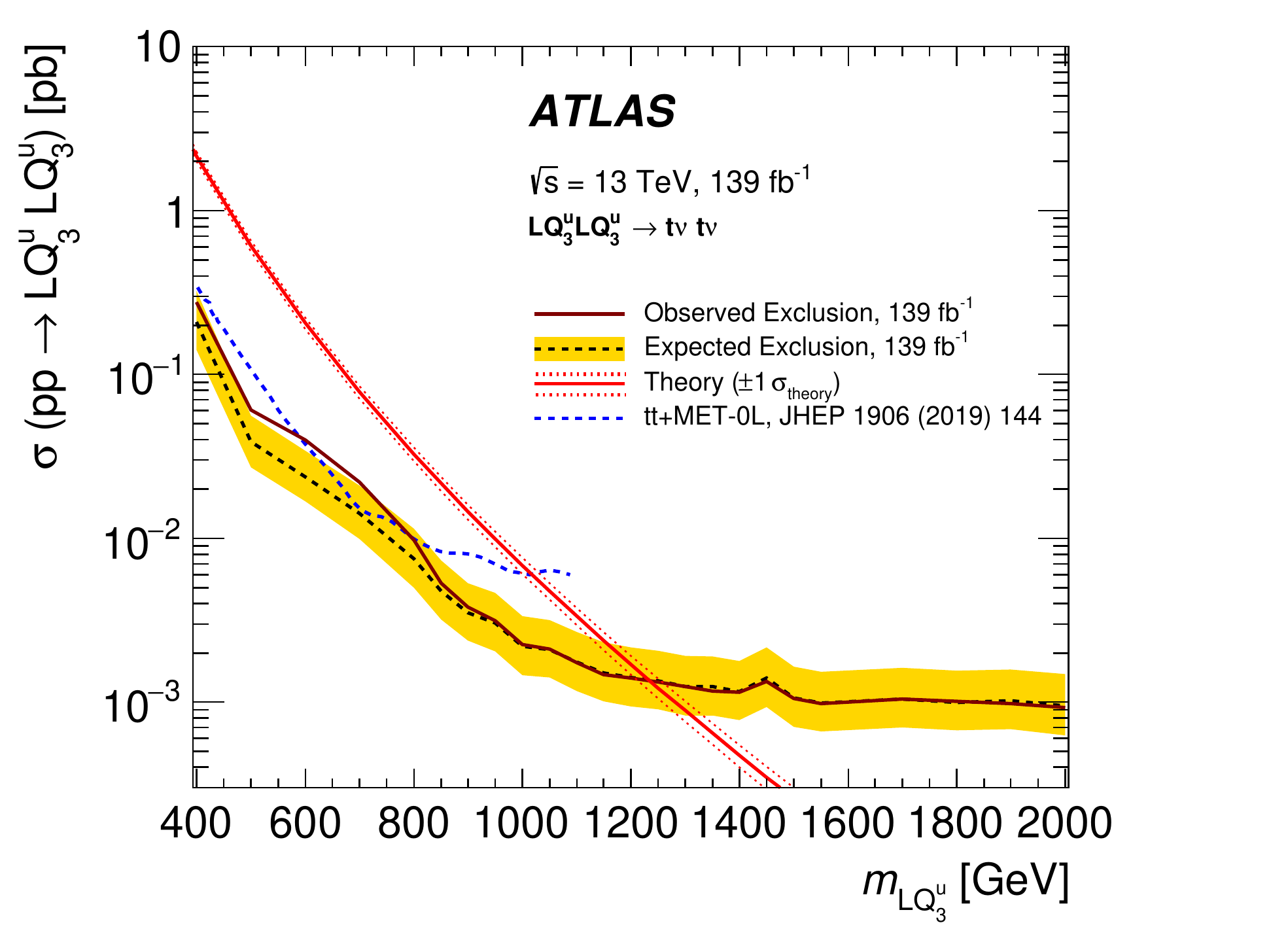}}
\caption{Observed (red solid line) and expected (black dashed line) limits on up-type, third-generation leptoquarks. (a) Limits as a function of the branching ratio of leptoquarks decaying into $b\tau$ (with the only other decay allowed being into $t\nu$) vs leptoquark mass. (b) Limits on the production cross section at 95\% CL as a function of leptoquark mass assuming that all leptoquarks decay into $t\nu$. Uncertainty bands corresponding to the $\pm 1\sigma$ variation of the expected limit (yellow band) and the sensitivity of the observed limit are also indicated. Observed limits from previous searches with the ATLAS detector at $\sqrt{s}=13$~TeV~\cite{EXOT-2017-30} are overlaid (a) in grey and (b) as a blue dashed line.}
\label{fig:lq3Limits}
\end{center}
\end{figure}

The SRA--B exclusion fit was repeated considering the analysis sensitivity to the production of up-type, third-generation scalar leptoquarks. The results are shown in Figure~\ref{fig:lq3Limits}(a) as a function of the leptoquark branching ratio to charged leptons, ${\cal B}\left(\lqThreeUp\to b\tau\right)$ in this scenario, vs the leptoquark mass. For ${\cal B}\left(\lqThreeUp\to b\tau\right) = 0$, where the leptoquarks decay into $t\nu$ $100\%$ of the time, leptoquarks are excluded up to a mass of \lqThreeLim~GeV\ as shown in Figure~\ref{fig:lq3Limits}(b). The difference in exclusion reach at ${\cal B}\left(\lqThreeUp\to b\tau\right) = 0$ between the leptoquark and top squark interpretations comes from the fact that top squark samples were produced at LO in QCD while the leptoquark samples were produced at NLO, which changes the kinematics slightly. This difference, however, is covered by the signal acceptance uncertainty (12\% in SRA--B).
 
% End of text imported from the .//tex/Results.tex input file
\FloatBarrier
 
% The next lines are included from the .//tex/Conclusions.tex input file
\section{Conclusions}
 
Results from a search for a scalar partner of the top quark based on an integrated luminosity of \intlumi\ of $\rts = 13$~\TeV\ proton--proton collision data recorded by the ATLAS experiment at the LHC from 2015 to 2018 are presented. Final states with high-\pT\ jets, large missing transverse momentum, and no electrons or muons are selected. The sensitivity of the search is enhanced by the analysis of the full LHC Run~2 dataset, improvements achieved in the detector performance by the end of the LHC Run~2, and new analysis techniques such as the use of $b$-tagged track-jets, which extend sensitivity to all-hadronic four-body decays.
 
Direct top squark pair production is considered, assuming both top squarks decay via $\stop\rightarrow t^{(*)} \ninoone$, and considering a large range of mass differences between the top squark and the neutralino. In particular, fully hadronic final states are used for the first time in an ATLAS analysis to set limits on the scenario where both the top quarks and $\Wboson$ bosons originating from their decays are off-shell, $\Delta m(\stop,\ninoone)<m_W+m_b$, due to improvements in the identification efficiency of low-transverse-momentum $b$-hadrons. The results are also reinterpreted in the context of third-generation, up-type scalar leptoquark pair production, restricted to the scenario where the leptoquarks decay into $t\nu$ or $b\tau$ final states.
 
No significant excess above the expected SM background is observed. Exclusion limits at 95\% confidence level are derived as a function of $\mstop$ and $\mLSP$, resulting in the exclusion of top squark masses that extend up to $1.25$~\TeV\ for $\ninoone$ masses below $200$~GeV. In the case where $\mstop\sim m_t+\mLSP$, top squark masses in the range \stopLimDiag~GeV\ are excluded, while in the scenario where $\mstop< m_W+m_b+\mLSP$ (with $\Delta m(\stop,\ninoone)\ge 5$~GeV), top squark masses in the range \stopLimFourBody~GeV\ are excluded. Exclusion limits for up-type, third-generation scalar leptoquarks are extended to masses below \lqThreeLim~GeV, assuming the leptoquarks can decay only via $t\nu$. Model-independent limits and $p$-values for each signal region are also reported, with 0.04~fb as the lowest upper bound on the visible cross section of new physics phenomena.
% End of text imported from the .//tex/Conclusions.tex input file
\section{Acknowledgements}
% The next lines are included from the .//acknowledgements/Acknowledgements.tex input file

We thank CERN for the very successful operation of the LHC, as well as the
support staff from our institutions without whom ATLAS could not be
operated efficiently.
 
We acknowledge the support of ANPCyT, Argentina; YerPhI, Armenia; ARC, Australia; BMWFW and FWF, Austria; ANAS, Azerbaijan; SSTC, Belarus; CNPq and FAPESP, Brazil; NSERC, NRC and CFI, Canada; CERN; CONICYT, Chile; CAS, MOST and NSFC, China; COLCIENCIAS, Colombia; MSMT CR, MPO CR and VSC CR, Czech Republic; DNRF and DNSRC, Denmark; IN2P3-CNRS and CEA-DRF/IRFU, France; SRNSFG, Georgia; BMBF, HGF and MPG, Germany; GSRT, Greece; RGC and Hong Kong SAR, China; ISF and Benoziyo Center, Israel; INFN, Italy; MEXT and JSPS, Japan; CNRST, Morocco; NWO, Netherlands; RCN, Norway; MNiSW and NCN, Poland; FCT, Portugal; MNE/IFA, Romania; MES of Russia and NRC KI, Russia Federation; JINR; MESTD, Serbia; MSSR, Slovakia; ARRS and MIZ\v{S}, Slovenia; DST/NRF, South Africa; MINECO, Spain; SRC and Wallenberg Foundation, Sweden; SERI, SNSF and Cantons of Bern and Geneva, Switzerland; MOST, Taiwan; TAEK, Turkey; STFC, United Kingdom; DOE and NSF, United States of America. In addition, individual groups and members have received support from BCKDF, CANARIE, Compute Canada and CRC, Canada; ERC, ERDF, Horizon 2020, Marie Sk{\l}odowska-Curie Actions and COST, European Union; Investissements d'Avenir Labex, Investissements d'Avenir Idex and ANR, France; DFG and AvH Foundation, Germany; Herakleitos, Thales and Aristeia programmes co-financed by EU-ESF and the Greek NSRF, Greece; BSF-NSF and GIF, Israel; CERCA Programme Generalitat de Catalunya and PROMETEO Programme Generalitat Valenciana, Spain; G\"{o}ran Gustafssons Stiftelse, Sweden; The Royal Society and Leverhulme Trust, United Kingdom.
 
The crucial computing support from all WLCG partners is acknowledged gratefully, in particular from CERN, the ATLAS Tier-1 facilities at TRIUMF (Canada), NDGF (Denmark, Norway, Sweden), CC-IN2P3 (France), KIT/GridKA (Germany), INFN-CNAF (Italy), NL-T1 (Netherlands), PIC (Spain), ASGC (Taiwan), RAL (UK) and BNL (USA), the Tier-2 facilities worldwide and large non-WLCG resource providers. Major contributors of computing resources are listed in Ref.~\cite{ATL-GEN-PUB-2016-002}.
 
% End of text imported from the .//acknowledgements/Acknowledgements.tex input file
 
\printbibliography

\clearpage % ATLAS Collaboration author list
% Reference date of SUSY-2018-12 is 2020-01-27
% Author list last updated on date 29-APR-20
% Data extracted on 29-Apr-2020 for paper reference SUSY-2018-12
% at 10:07am
 
\begin{flushleft}
\hypersetup{urlcolor=black}
{\Large The ATLAS Collaboration}

\bigskip

\AtlasOrcid[0000-0002-6665-4934]{G.~Aad}$^\textrm{\scriptsize 102}$,    
\AtlasOrcid[0000-0002-5888-2734]{B.~Abbott}$^\textrm{\scriptsize 128}$,    
\AtlasOrcid{D.C.~Abbott}$^\textrm{\scriptsize 103}$,    
\AtlasOrcid[0000-0002-2788-3822]{A.~Abed~Abud}$^\textrm{\scriptsize 36}$,    
\AtlasOrcid{K.~Abeling}$^\textrm{\scriptsize 53}$,    
\AtlasOrcid{D.K.~Abhayasinghe}$^\textrm{\scriptsize 94}$,    
\AtlasOrcid[0000-0002-8496-9294]{S.H.~Abidi}$^\textrm{\scriptsize 166}$,    
\AtlasOrcid[0000-0002-8279-9324]{O.S.~AbouZeid}$^\textrm{\scriptsize 40}$,    
\AtlasOrcid{N.L.~Abraham}$^\textrm{\scriptsize 155}$,    
\AtlasOrcid[0000-0001-5329-6640]{H.~Abramowicz}$^\textrm{\scriptsize 160}$
% ...
\bigskip
\\

% ...
$^{36}$CERN, Geneva; Switzerland.\\
% ...
$^{40}$Niels Bohr Institute, University of Copenhagen, Copenhagen; Denmark.\\
% ...
$^{53}$II. Physikalisches Institut, Georg-August-Universit\"{a}t G\"ottingen, G\"ottingen; Germany.\\
% ...
$^{94}$Department of Physics, Royal Holloway University of London, Egham; United Kingdom.\\
% ...
$^{102}$CPPM, Aix-Marseille Universit\'e, CNRS/IN2P3, Marseille; France.\\
$^{103}$Department of Physics, University of Massachusetts, Amherst MA; United States of America.\\
% ...
$^{128}$Homer L. Dodge Department of Physics and Astronomy, University of Oklahoma, Norman OK; United States of America.\\
% ...
$^{155}$Department of Physics and Astronomy, University of Sussex, Brighton; United Kingdom.\\
% ...
$^{160}$Raymond and Beverly Sackler School of Physics and Astronomy, Tel Aviv University, Tel Aviv; Israel.\\
% ...
$^{166}$Department of Physics, University of Toronto, Toronto ON; Canada.\\
% ...

\end{flushleft}

% Created with Glance <Atlas.Glance@cern.ch>

\end{document}